\documentclass[prd,preprint,tightenlines,superscriptaddress,floatfix,showpacs,
preprintnumbers,nofootinbib,eqsecnum]{revtex4}

 \usepackage[dvips,final]{graphicx}
     \usepackage{epsfig}
      \usepackage{bm}

\begin{document}

\thispagestyle{empty} \preprint{\hbox{}} \vspace*{-10mm}

\title{$\pi^+ \pi^-$ and $\pi^0 \pi^0$ 
pair production in photon-photon\\ and in ultraperipheral ultrarelativistic heavy ion collisions}

\author{M.~K{\l}usek-Gawenda}
\email{mariola.klusek@ifj.edu.pl}

\affiliation{Institute of Nuclear Physics PAN, PL-31-342 Cracow, Poland}

\author{A.~Szczurek}
\email{antoni.szczurek@ifj.edu.pl}

\affiliation{Institute of Nuclear Physics PAN, PL-31-342 Cracow, Poland}
\affiliation{University of Rzesz\'ow, PL-35-959 Rzesz\'ow, Poland}

\date{\today}

\begin{abstract}
We discuss $\gamma \gamma \to \pi \pi$ reactions, starting from
the two-pion threshold up to about \mbox{$\sqrt{s_{\gamma\gamma}}$ = 6 GeV}. 
Several reaction mechanisms are identified.
We include the dipion continuum due to pion exchange (for $\pi^+\pi^-$) and
$\rho^\pm$ exchange (for $\pi^0\pi^0$) 
as well as the pronounced dipion $\sigma(600)$, $f_0(980)$ and 
$f_2(1270)$ s-channel resonances. 
We discuss also a possible contribution of less pronounced scalar resonances 
$f_0(1500)$ and $f_0(1710)$ being glueball candidates,
as well as the tensor the resonances $f'_2(1525)$, $f_2(1565)$ and $f_2(1950)$. 
While the contribution of $f'_2(1525)$ is rather small, the contribution
of $f_2(1565)$ changes the spectrum around the resonance position.
We find that the inclusion of the spin-four $f_4(2050)$ resonant state
improves the situation at $\sqrt{s_{\gamma\gamma}} \approx$ 2 GeV.
We estimate the relevant diphoton partial decay width to be 
\mbox{$\Gamma_{f_4 \to \gamma \gamma} \approx$ 0.7 keV}.
At higher energies, Brodsky-Lepage and hand-bag mechanisms are included 
in addition.
We nicely describe the world data for $\gamma \gamma \to \pi \pi$ for the first
time both for the total cross section and angular distributions
for $\gamma \gamma \to \pi^+ \pi^-$
and $\gamma \gamma \to \pi^0 \pi^0$ reactions simultaneously at all experimentally 
available energies.

The cross section for the production of two-pions 
in ultraperipheral ultrarelativistic heavy ion collisions is calculated 
in the impact parameter space Equivalent Photon Approximation (EPA).
We obtain the nuclear cross section of
46.7 mb for $\pi^+\pi^-$ and 8.7 mb for $\pi^0\pi^0$ 
at the LHC energy \mbox{$\sqrt{s_{NN}} =$ 3.5 TeV} 
and minimal cut on pion transverse momentum $p_t>$ 0.2 GeV. 
Differential distributions in impact parameter, dipion invariant mass,
single pion and dipion rapidity, pion transverse momentum as well as 
pion pseudorapidity are shown.
The $\gamma \gamma \to \pi^+ \pi^-$ subprocess constitutes a background
to the exclusive $A A \to A \rho^0 (\to \pi^+ \pi^-) A$ process, 
initiated by photon-pomeron or pomeron-photon subprocesses. 
We find that only a part of the dipion invariant mass spectrum 
associated with $\gamma\gamma$-collisions can be potentially visible as 
the cross section for the $A A \to A \rho^0 A$ reaction is very large.

\end{abstract}

       

\maketitle

\section{Introduction}

Ultrarelativistic ultraperipheral production of mesons and elementary
particles is a special category of nuclear reactions \cite{nucl_reac}.
The ultrarelativistic ions provide large fluxes of quasi-real photons which 
can collide leading to different final states. Both total photon-photon
cross sections as well as
cross section for particular simple final states are interesting.
The nuclear cross section is
usually calculated in the Equivalent Photon Approximation in 
momentum space \cite{Serbo, KS_muon}, or in impact parameter space 
\cite{BF90,KS_muon}.
In the latter case the flux of photons depends on the transverse distance
from the heavy ion trajectory. Impact parameter space is very 
convenient to exclude cases when both heavy ions collide, i.e., when they
do not survive the high-energy collision. In the past we have performed 
also full momentum space calculation for $\mu^+ \mu^-$ production 
\cite{KS_muon}.
In the momentum space calculation (EPA or full calculation) the effects 
of nucleus-nucleus collisions and their associated break-up are neglected.
This effect is rather small for light particle production such as $e^+ e^-$
often studied in the literature.

Recently we have studied several processes initiated by 
the photon-photon collisions such as:
$\rho^0 \rho^0$ \cite{KS_rho}, $\mu^+ \mu^-$ \cite{KS_muon}, $Q \bar Q$
\cite{KS_quarks}, $D \bar D$ \cite{LS_DD} and for high mass 
dipions \cite{KS_pQCD}.
We have shown there that the inclusion of realistic charge form factors, 
being Fourier transforms of realistic charge distributions,
is crucial for estimating reliably the nuclear cross sections. 

Here, we shall apply the previously used method for the
exclusive production of $\pi^+ \pi^-$ and $\pi^0 \pi^0$. 
The present analysis is an extension
of the analysis of high-mass dipion production \cite{KS_pQCD} where we
concentrated exclusively on perturbative QCD effects.

The elementary reactions $\gamma \gamma \to \pi^0 \pi^0$ and
$\gamma \gamma \to \pi^+ \pi^-$ are interesting by themselves, since
understanding the mechanism of the $\gamma \gamma \to \pi \pi$ reaction
at low or intermediate energies ($\sqrt{s_{\gamma \gamma}} <$ 1.5 GeV)
is very important for applications of chiral
perturbation theory and pion-pion interaction
\cite{low_energy,Achasov,AS2008,Ko,MNW,Pennington}.
At higher photon-photon energies the Brodsky-Lepage 
\cite{BL81,JA,SS03,QCD_hb,KS_pQCD}
and hand-bag \cite{HB} mechanisms have been discussed in the literature.

In addition, we study the contribution of glueballs or
glueball candidates \cite{ALEPH_glueball}. Only estimates of the total 
cross section were presented in the literature \cite{Machado_glueball} 
recently. 
In general, it is not possible to observe glueballs, even if the
corresponding cross section is sizeable. One should concentrate on a
concrete final state. The $\pi \pi$ channels are a good example.
Estimation of the background and its interplay
with resonant signal should be better understood.
Here we show how the glueball candidates can be seen in reality 
when analyzing e.g. dipion spectra.

In the present paper we construct a hybrid, multi-component 
model for the $\gamma \gamma \to \pi \pi$ reactions 
which describes all experimentally available world data.
We shall extend the range of the photon-photon energies to
the region where the transition from nonperturbative QCD to pQCD may be expected.
It is very important to understand the onset of the pQCD effects.
For example the hand-bag mechanism \cite{HB} is usually fitted to
experimental data for energies bigger than 2.5 GeV.

The $\gamma \gamma \to \pi \pi$ processes are fairly complicated. 
Different mechanisms may contribute in general. 
The present analysis is partially an update of an old analysis in 
Ref.\cite{SS03}.
In the present paper we shall try to understand both 
$\gamma \gamma \to \pi^+ \pi^-$ as well as
$\gamma \gamma \to \pi^0 \pi^0$ processes simultaneously, from the
kinematical threshold ($W = 2 m_{\pi}$) up to the maximal experimentally 
available energy $W_{\gamma\gamma} \approx$ 6 GeV. 
We will include both soft and hard continuum processes
as well as s-channel resonances. 
Here we shall include many more resonances compared to Ref.\cite{SS03}.
In addition, we shall try to describe the new very
precise data of the Belle Collaboration \cite{Belle_Uehara09}, 
including the angular distributions of pions.

Our paper concentrates on realistic predictions 
of the cross sections for dipion
production in ultraperipheral ultrarelativistic heavy ion collisions.
We shall present first realistic predictions for
$Pb Pb \to Pb (\pi \pi) Pb$ reaction at the LHC. Both total cross sections
and differential distributions will be shown and discussed
for the first time in the literature.
The photon-photon induced dipion production in nuclear collision,
interesting by itself,
constitutes a background to another type of nucleus-nucleus reactions
induced by photon-pomeron (pomeron-photon) exchanges leading to a 
coherent production of $\rho^0$ meson \cite{Grube_STAR,BKN2002,FSZ2002} 
and its radial excitations. 
The interplay of the both processes was not discussed so far in 
the literature.

\section{Modelling $\gamma \gamma \to \pi \pi$ reactions}

\subsection{The $\gamma \gamma \to \pi^+ \pi^-$ continuum}
\begin{figure}[!h]             
\centering
\includegraphics[width=0.8\textwidth]{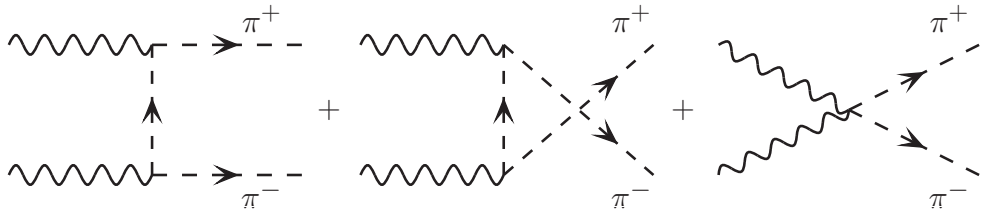}
   \caption{\label{fig:Born}
   \small The Born term matrix elements for non resonant pion pair production.
           }
\end{figure}

%
The Born amplitude for the $\gamma \gamma \to \pi^+ \pi^-$ process
is shown in Fig.\ref{fig:Born}.
%
%
The helicity dependent amplitude for point-like pions is a sum of the three terms
(see e.g. \cite{Brodsky1971}):\\
contact amplitude:
\begin{equation}
\mathcal{M}_{\lambda_1 \lambda_2}^{\pi,c} \left( q_1, q_2, p_{\pi^+}, p_{\pi^-} \right) = 
e^2 \sum 2 g^{\mu\nu} \varepsilon_\mu (\lambda_1) \varepsilon_\nu (\lambda_2) \; ,
\end{equation}
t-channel pion-exchange amplitude:
\begin{equation}
\mathcal{M}_{\lambda_1 \lambda_2}^{\pi,t} \left( q_1, q_2, p_{\pi^+}, p_{\pi^-} \right) = 
e^2 \sum \left(2 p_{\pi^-}^\mu - q_1^\mu\right) \varepsilon_\mu(q_1,\lambda_1) 
\left(2 p_{\pi^+}^\nu - q_2^\nu\right) \varepsilon_\nu(q_2,\lambda_2) \frac{1}{t-m_\pi^2} \; 
\end{equation}
and u-channel pion-exchange amplitude:
\begin{equation}
\mathcal{M}_{\lambda_1 \lambda_2}^{\pi,u} \left( q_1, q_2, p_{\pi^+}, p_{\pi^-} \right) = 
e^2 \sum \left(2 p_{\pi^+}^\mu - q_1^\mu\right) \varepsilon_\mu(q_1,\lambda_1) 
\left(2 p_{\pi^-}^\nu - q_2^\nu\right) \varepsilon_\nu(q_2,\lambda_2) \frac{1}{u-m_\pi^2} \; .
\end{equation}
The full pion-exchange amplitude is a sum of contact + t + u amplitudes
\begin{equation}
\mathcal{M}_{\lambda_1 \lambda_2}^\pi =  
 \mathcal{M}_{\lambda_1 \lambda_2}^{\pi,c} + 
\mathcal{M}_{\lambda_1 \lambda_2}^{\pi,t} + 
\mathcal{M}_{\lambda_1 \lambda_2}^{\pi,u} \; .
\label{eq.Born_gi}
\end{equation}
Such an amplitude can be formally written as:
\begin{equation}
\mathcal{M}_{\lambda_1 \lambda_2}^\pi = 
\varepsilon_\mu (q_1,\lambda_1) \varepsilon_\nu(q_2,\lambda_2) \mathcal{M}^{\mu \nu} \; .
\end{equation}
It is straightforward to show that
\begin{eqnarray}
q_{1\mu} \mathcal{M}^{\pi,\mu \nu} = 0, \nonumber \\
q_{2\mu} \mathcal{M}^{\pi,\mu \nu} = 0 \; ,
\end{eqnarray}
which is consistent with gauge invariance.
The QED Born amplitude for the $\gamma \gamma\to \pi^+ \pi^-$
reaction with point-like particles has been known for a long time.
An interesting problem is to construct the QED amplitude
for real, finite-size, pions.
We use an idea proposed by Poppe \cite{Poppe}, and correct 
the QED amplitude by an overall $t$ and $u$ dependent form factor:
\begin{equation}
\mathcal{M}_{\lambda_1, \lambda_2}^\pi\left(t,u,s\right) = 
\Omega\left(t,u,s\right)
\mathcal{M}_{\lambda_1, \lambda_2}^{\pi,QED}\left(t,u,s\right) \; .
\label{eq.ff_fs}
\end{equation}
It is natural that finite size corrections damp the QED amplitude
for both $t$ and $u$ large.
At sufficiently high energies the following simplification
fulfils this requirement:
\begin{equation}
\Omega\left(t,u,s\right) = 
\frac{F^2\left(t\right)+F^2\left(u\right)} {1+F^2\left(-s\right)} \; .
\label{eq:ff_hadronic}
\end{equation}
In practice one can take 
$F(x) = \exp(\frac{B_{\gamma \pi}} {4}x)$; $B_{\gamma \pi}=$ (4-6) GeV$^{-2}$.
$F(x)$ denotes the standard vertex function 
which provides the convenient normalization $F\left(0\right)$ = 1
and additionally $F\left(t\right) \to $ 0 when $t \to - \infty$.
In the limit of large $s$:
\begin{eqnarray}
\Omega\left(t,u,s\right)& \stackrel{t \to 0}{\longrightarrow} & F^2\left(t\right) \; , \nonumber \\ 
\Omega\left(t,u,s\right)& \stackrel{u \to 0}{\longrightarrow} & F^2\left(u\right) \; .
\end{eqnarray}
This $s$ limit generates standard vertex form factors
and additionally at $z \equiv \cos\theta \approx 0$:
\begin{equation}
\frac{d \sigma^\pi}{dz}\left(z=0,s\right) \propto F^4\left(\frac{-s}{2}\right) \; .
\end{equation}
%

\subsection{$s$-channel $\gamma \gamma \to$  resonances}

\begin{figure}[!h]             
\begin{minipage}[t]{0.46\textwidth}
\centering
\includegraphics[width=0.9\textwidth]{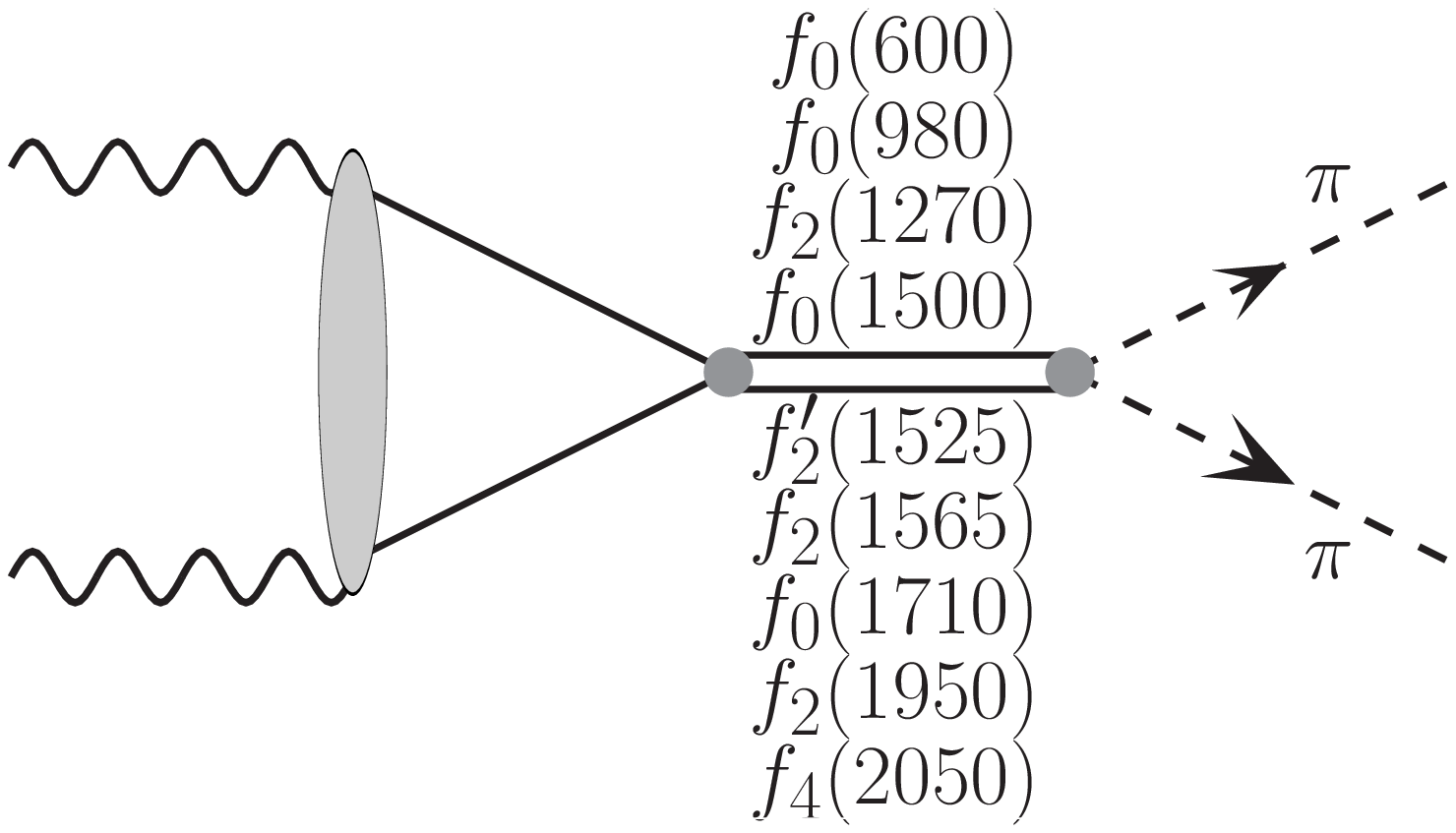}
\end{minipage}
   \caption{\label{fig:resonances}
   \small $\gamma \gamma \to \mbox{resonances} \to \pi^{+/0} \pi^{-/0}$.
           }
\end{figure}

In Fig.\ref{fig:resonances} we show a list of several possible dipion
resonances that could, in principle, contribute to the 
$\gamma \gamma \to \pi \pi$ processes.

%
%
\begin{table}
\caption{
Parameters of resonances used in our calculations.}
\begin{tabular}{||c||c|c|c|c|c||}
\hline
\hline
Resonance			& $m_R$ (MeV)	&	$\Gamma_R$ (MeV)	&	$\Gamma_{\pi \pi}$ (MeV)	&	$\Gamma_{\gamma \gamma}$ (keV)	&	$\frac{\Gamma_{\gamma \gamma} \Gamma_{\pi \pi}}{\Gamma_R}$ (keV) \\
\hline
\hline
$f_0(600)$ \cite{PDG}		& 600			& 400 			& 400							& 0.5				&	\\
$f_0(980)$ \cite{PDG} 		& 980 			& 50 			& 51.3 \cite{Belle_Mori}		& 0.29 				&	\\
$f_2(1270)$ \cite{PDG}		& 1 275 		& 185 			& 156.9  						& 3.035 			&	\\
$f_0(1500)$ \cite{PDG}		& 1 505 		& 109  			& 38 							& -					& 0.033 \cite{Belle_Uehara08}	\\
$f'_2(1525)$ \cite{PDG}		& 1 525 		& 73 			& 0.6 							& 0.081  			&	\\
$f_2(1565)$ \cite{Schegelsky}	& 1 570 	& 160 			& 25 							& 0.7  				&	\\
$f_0(1710)$ \cite{PDG}		& 1 720  		& 135 			& - 							& - 				&	0.82 \\
$f_2(1950)$ \cite{PDG}		& 1 944 		& 472 			& - 							& -  				& 	1.62\\
$f_4(2050)$ \cite{PDG}		& 2 018			& 237 			&  40.3 						& 0.7 				&	\\
\hline
\hline
\end{tabular}
\label{tab:resonances}
\end{table}
%
In Table \ref{tab:resonances} we have collected these resonances with parameters
known from Particle Data Group book \cite{PDG}. 
In the present analysis we shall take into account the following scalar:
$\sigma(600)$, $f_0(980)$, $f_0(1500)$, $f_0(1710)$, tensor:
$f_2(1270)$, $f'_2(1525)$, $f_2(1565)$, $f_2(1950)$ 
as well as spin-4: $f_4(2050)$ resonances.
While the $\pi \pi$ decay widths are usually well known, 
the $\gamma \gamma$ decay widths are known
only for some of them.  Therefore, studying the 
$\gamma \gamma \to \pi \pi$ data may help in extracting the latter quantities.
Below we shall present both, theoretical estimates of some of them, as well
as a fit to experimental total cross sections and angular distributions.

In most cases, PDG gives values of the resonance parameters: 
$m_R$, $\Gamma_R$, $\Gamma_{\pi\pi}$ 
and $\Gamma_{\gamma\gamma}$. We use somewhat smaller values of decay widths 
for $f_0(600)$ resonance than given in the PDG book.
In the PDG book \cite{PDG} a broad range of parameters is given:
$\Gamma_R$=(600-1000) MeV, $\Gamma_{\gamma \gamma}$ =(1.2-10) keV.
In our case we find: $\Gamma_R$=400 MeV, $\Gamma_{\gamma \gamma}$ =0.5 keV 
as the best fit parameters.
In addition, we assume that Br($f_0(600)\to\pi\pi$)=100\%.
For $f_0(1500)$ PDG gives $\frac{\Gamma_{\gamma \gamma} \Gamma_{\pi \pi}}{\Gamma_R}$=0.033 keV
which was obtained by the Belle Collaboration 
which includes in addition in their fit the $f_0(1370)$ resonance \cite{Belle_Uehara08}. 
Using this value, we obtain: $\Gamma_{\gamma \gamma}$=0.1 keV.
For $f_4(2050)$ the situation is even more complicated.
Different values of the ratio (see last column in Table \ref{tab:resonances}) 
has been given in the literature:
\cite{Belle_Uehara09}: $\frac{\Gamma_{\gamma \gamma} \Gamma_{\pi \pi}}{\Gamma_R}$=0.0231 keV, 
\cite{Oest}: $\frac{\Gamma_{\gamma \gamma} \Gamma_{\pi \pi}}{\Gamma_R}<$ 1.1 keV.
In our fit to the Belle experimental data (to be shown in the Result section) we find: 
\mbox{$\frac{\Gamma_{\gamma \gamma} \Gamma_{\pi \pi}}{\Gamma_R}$ = 0.12 keV}.  

The angular distribution for the s-channel resonances can be written
in the standard (typical for Feynman-diagrams) form:
\begin{equation}
 \frac{d \sigma\left( \gamma \gamma \to \pi \pi \right)}{dz} = 
\sum_{\lambda_1,\lambda_2}
\frac{\sqrt{\frac{W^2}{4}-m_\pi^2}}{\frac{W}{2}} 
\left| \mathcal{M} \left( \lambda_1,\lambda_2 \right) \right|^2
\frac{4\pi}{4\times 64\pi^2 W^2} \; ,
\end{equation}
where the helicity-dependent resonant amplitudes must be modelled,
as we do not have a priori microscopic models of the coupling of two photons
to high-spin resonances. Our parameterizations for the $f_0$, $f_2$ and $f_4$ 
resonances are:
\begin{eqnarray}
\mathcal{M} \left( \lambda_1,\lambda_2 \right) & = &
\frac{\sqrt{64\pi^2 W^2 \times 8\pi \left( 2J+1 \right) \left( \frac{m_R}{W} \right)^2 
\Gamma_R \Gamma_R \left( W \right) 
Br \left( R\to\gamma\gamma \right) Br\left( R\to\pi^{+/0}\pi^{-/0} \right)}}
{W^2-m_R^2+im_R \Gamma_R \left( W \right)} e^{i \varphi_R} \nonumber \\
& \times & \sqrt{2} \delta_{\lambda_1,\lambda_2}
\left\{
\begin{array} {l} 
 Y^0_0; \mbox{for } f_0\\
 Y^2_2; \mbox{for } f_2(1270), f'_2(1525), f_2(1950)\\
 Y^0_2; \mbox{for } f_2(1565)\\
 Y^0_4; \mbox{for } f_4(2050)
\end{array}
\right\}
\times \exp \left( \frac{-\left( W-m_R \right)^2}{\Lambda_R^2} \right) \;
.
\end{eqnarray}
The last (form)factor was introduced to correct the resonance form 
far from the actual resonance position where the simple resonance form 
is incorrect. We expect $\Lambda_R$ to
be much larger than the resonance width $\Gamma_R$. In practice we will
treat it as an extra free parameter to be adjusted to experimental data.
This has importance only for broad resonances with $\Gamma >$ 0.1 GeV.

%
%

The energy-dependent resonance widths are parametrized as:
\begin{equation}
 \Gamma_R(W)= \Gamma_R 
\frac{\sqrt{\frac{W^2}{4}-m_\pi^2}}{\sqrt{\frac{m_R^2}{4}-m_\pi^2}} F^J \left(W,R\right) \; ,
\end{equation}
where the function $F^J\left(W,R\right)$ is the spin dependent Blatt-Weisskopf form factor \cite{BW1952}:
\begin{eqnarray}
 F^{J=0}\left(W,R\right)& = & 1, \nonumber \\
 F^{J=2}\left(W,R\right)& = & \frac
{ \left(R p_R \right)^4
+3\left(R p_R\right)^2+9}
{ \left(R p\right)^4
+3\left(R p\right)^2+9},
 \nonumber \\
 F^{J=4}\left(W,R\right)& = & \frac
    {   \left(R p_R\right)^8 
+ 10    \left(R p_R\right)^6 
+ 135   \left(R p_R\right)^4 
+ 1575  \left(R p_R\right)^2 
+ 11025}
   {    \left(R p\right)^8 
+ 10    \left(R p\right)^6 
+ 135   \left(R p\right)^4 
+ 1575  \left(R p\right)^2 
+ 11025} 						\; .
\end{eqnarray}
Above $p_R=\sqrt{\frac{m_R^2}{4}-m_\pi^2}$,
      $  p=\sqrt{\frac{W^2}  {4}-m_\pi^2}$
and R = 1 fm.
%
%

The decay of $f_2(1565)$ into dipions was studied first in
Ref.\cite{Crystal_Barrel} by the Crystal Barrel Collaboration,
while its decay into two photons was studied in Ref.\cite{Schegelsky}.
The $f_4(2050)$ resonance was not studied so far in 
$\gamma \gamma \to \pi \pi$ reactions. 

We consider two simple models of the amplitude for 
the tensor meson production:
\begin{eqnarray}
{\cal M}_{\lambda_1 \lambda_2}^{J=2} 
&&\propto Y_{2,\lambda_1 - \lambda_2}(\theta,\phi) \cdot
\left( \delta_{\lambda_1-\lambda_2,-2} + \delta_{\lambda_1-\lambda_2, 2}
\right)  \; {\mbox{(type A)}},
\nonumber \\
{\cal M}_{\lambda_1 \lambda_2}^{J=2} 
&&\propto Y_{2,\lambda_1 - \lambda_2}(\theta,\phi) \cdot
\delta_{\lambda_1 - \lambda_2,0}  \; {\mbox{(type B)}} .
\label{model_amplitude_f2}
\end{eqnarray}
We shall call them model A and B, respectively.

For $f_2(1270)$ production the amplitude 
is dominantly of the type A \cite{Poppe}. 
We assume the same for the $f'_2(1525)$ and $f_2(1950)$ 
resonances since they are the same nonet partners of $f_2(1270)$.

The $f_2(1565)$ resonance is not well understood so far.
The Belle Collaboration finds in this region both components (A, B)
in the partial wave analysis 
\cite{Belle_Uehara09, Belle_Uehara08, Belle_Mori}.
We shall try both models
in order to describe the experimental data, taking the 
$\Gamma_{\pi \pi}$ decay width from \cite{Schegelsky} and fitting 
$\Gamma_{\gamma \gamma}$.

For the spin-4 resonance again two simple possibilities come into the game:
\begin{eqnarray}
{\cal M}_{\lambda_1 \lambda_2}^{J=4} 
&&\propto Y_{4,\lambda_1 - \lambda_2}(\theta,\phi) \cdot
\left( \delta_{\lambda_1-\lambda_2,-2} + \delta_{\lambda_1-\lambda_2, 2}
\right) \; {\mbox{(type A)}},
\nonumber \\
{\cal M}_{\lambda_1 \lambda_2}^{J=4} 
&&\propto Y_{4,\lambda_1 - \lambda_2}(\theta,\phi) \cdot
\delta_{\lambda_1 - \lambda_2,0} \; {\mbox{(type B)}} .
\label{model_amplitude_f4}
\end{eqnarray}
While the Belle partial wave analysis 
\cite{Belle_Uehara09, Belle_Uehara08, Belle_Mori} suggests
the dominance of the type A form,
in our analysis we find that type B fits better to experimental data.

\subsection{$\gamma \gamma \to \pi^0 \pi^0$ 
in a simple coupled channel model with
$\rho^{\pm}$ exchange}

\begin{figure}[!h]             
\begin{minipage}[t]{0.46\textwidth}
\centering
\includegraphics[width=0.75\textwidth]{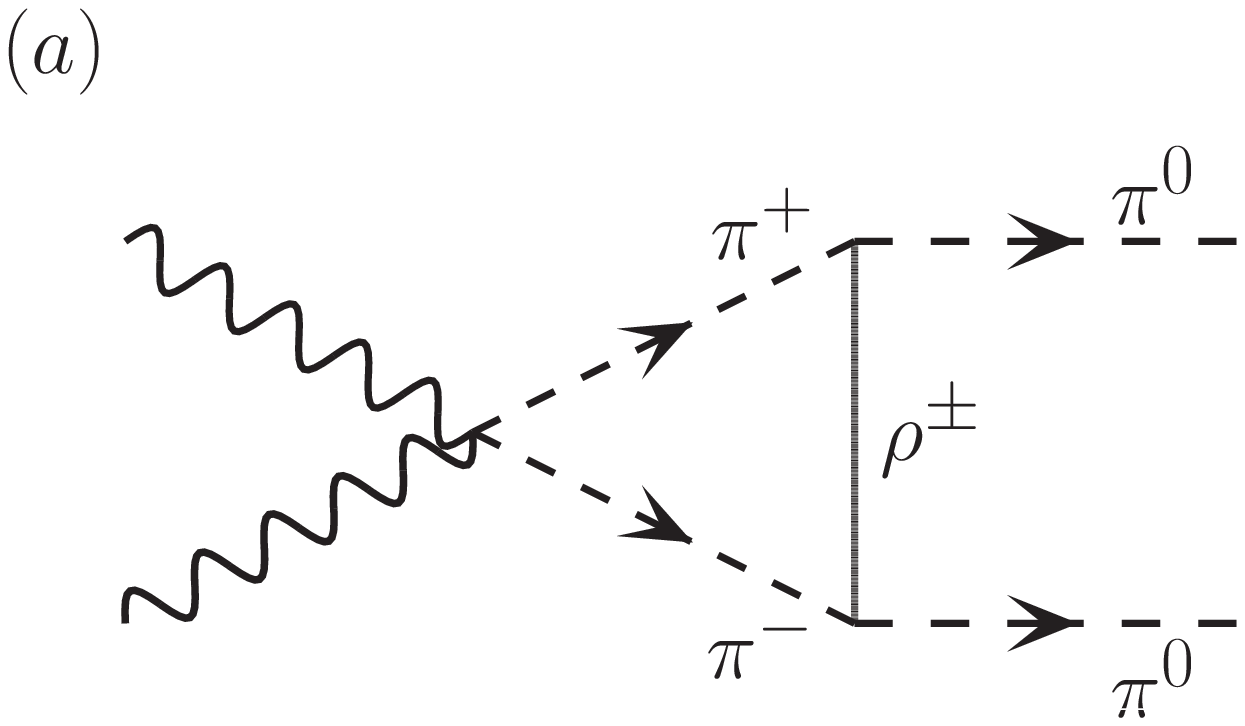}
\end{minipage}
\begin{minipage}[t]{0.46\textwidth}
\centering
\includegraphics[width=0.75\textwidth]{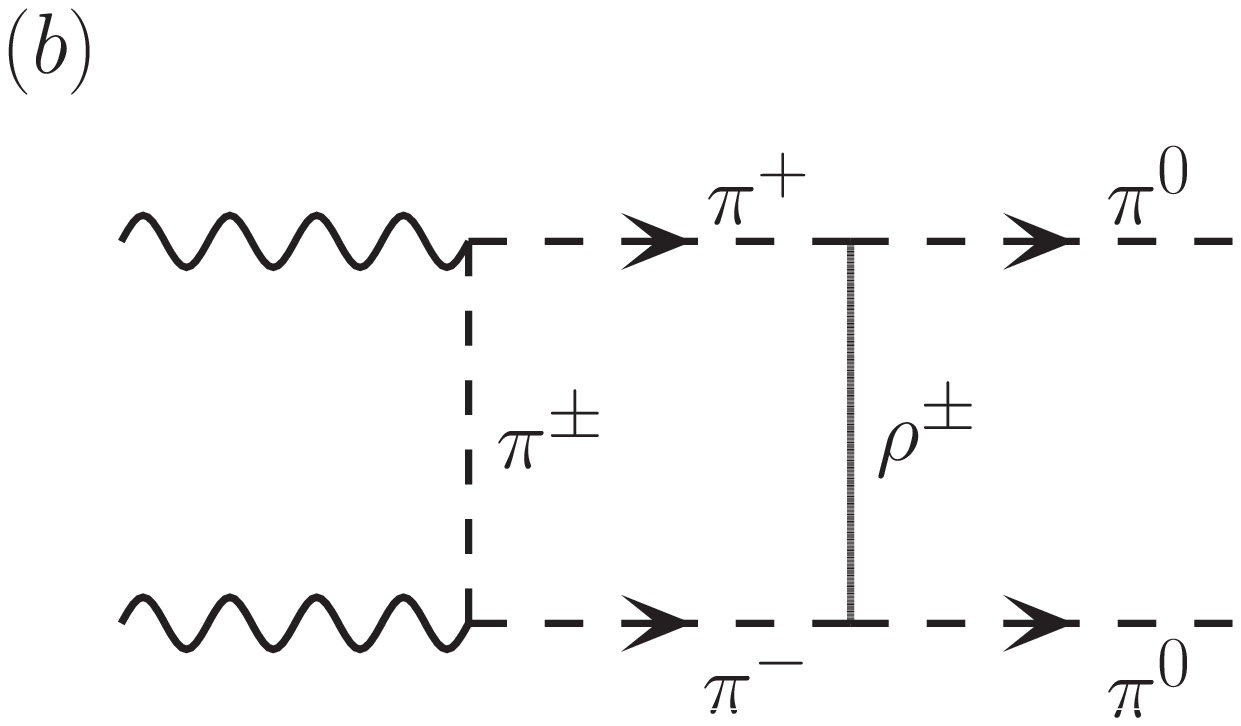}
\end{minipage}
   \caption{\label{fig:gg_pippim_rho_pi0pi0}
   \small $\gamma \gamma \to \pi^+ \pi^- \to \rho \to \pi^0 \pi^0$.
           }
\end{figure}

Since the cross section 
for the $\gamma \gamma \to \pi^+ \pi^-$ reaction is much bigger than that
for the $\gamma \gamma \to \pi^0 \pi^0$ reaction, even a small coupling
between these channels may modify the cross section for the 
$\gamma \gamma \to \pi^0 \pi^0$ reaction.
An example of a process which leads to channel coupling
is shown in Fig.\ref{fig:gg_pippim_rho_pi0pi0}.
%
%

The contact amplitude (see diagram (a) in Fig.\ref{fig:gg_pippim_rho_pi0pi0})
can be written as:
\begin{eqnarray}
\mathcal{M}^c(\lambda_1,\lambda_2) &=& \int 2 e^2 g^{\mu \nu} 
									   \epsilon_\mu(\lambda_1) \epsilon_\nu(\lambda_2)
							  \frac{1}{ k_1^2  -m_\pi^2+i\epsilon} 
							  g_{\pi\pi\to\rho} \left( k_1^\alpha + p_1^\alpha \right) \nonumber \\
 & \times &	\frac{\left(-g_{\alpha \beta} + \frac{{q}_\alpha q_\beta}{m_\rho^2}\right)}
	     {\kappa^2-m_\rho^2+i\Gamma_\rho m_\rho} g_{\pi\pi\to\rho} \left( k_2^\beta + p_2^\beta \right)
							  \frac{1}{ k_2^2  -m_\pi^2+i\epsilon} \nonumber \\
 & \times &					  F(\hat{s},\hat{t},\hat{u})  F^2(\kappa) F^2(k_1) F^2(k_2)  \frac{d^4 \kappa}{\left(2\pi\right)^4} \; 
\end{eqnarray}
and the corresponding t-channel amplitude 
(see diagram (b) in Fig.\ref{fig:gg_pippim_rho_pi0pi0}) as:
\begin{eqnarray}
\mathcal{M}^t(\lambda_1,\lambda_2) &=& \int \frac{e^2}{\kappa_2^2-m_\pi^2+i\epsilon} 
							  \frac{1}{ k_1^2 -m_\pi^2+i\epsilon} 
							  g_{\pi\pi\to\rho} \left( k_1^\alpha + p_1^\alpha \right) \nonumber \\
 & \times &	\frac{\left(-g_{\alpha \beta} + \frac{{q}_\alpha q_\beta}{m_\rho^2}\right)}
	     {\kappa^2-m_\rho^2+i\Gamma_\rho m_\rho}  g_{\pi\pi\to\rho} \left( k_2^\beta + p_2^\beta \right) \frac{1}{ k_2^2  -m_\pi^2+i\epsilon} \nonumber \\
 & \times &					   \epsilon_\mu \left(\lambda_1\right)\left(\kappa_2^\mu+k_1^\mu\right)
							   \epsilon_\nu \left(\lambda_2\right)\left(\kappa_2^\nu-k_2^\nu\right) \nonumber \\
 & \times &					  F(\hat{s},\hat{t},\hat{u})  F^2(\kappa) F^2(k_1) F^2(k_2)  \frac{d^4 \kappa}{\left(2\pi\right)^4} \; .
\end{eqnarray}							 
%
%
The form factors that appear in the above formulas are parametrized as:
\begin{itemize}
\item $F(\hat{s},\hat{t},\hat{u})=\frac{F^2(\hat{t})+F^2(\hat{u})}{1+F^2(\hat{s})}$,
\item $F(x=\hat{s})=\exp\left( \frac{-\left( x-4 m_\pi^2 \right)^2}{\Lambda^4} \right)$,
\item $F(x=\hat{t},\hat{u},k_1,k_2)=\exp\left( \frac{-\left( x^2-m_\pi^2 \right)^2}{\Lambda^4} \right)$,
\item $F(x=\kappa)=\exp\left( \frac{-\left( x^2-m_\rho^2 \right)^2}{\Lambda^4} \right)$,
\end{itemize}
where $\Lambda$ is, in principle, a free parameter.
The resulting amplitude strongly depends on the value of the parameter.
 
The u-channel amplitude can be obtained from the t-channel amplitude
by interchanging $\pi^+$ and $\pi^-$
($\mathcal{M}^u \left(\lambda_1,\lambda_2\right)=
\mathcal{M}^t \left(\lambda_1,\lambda_2\right)$). 
\\
The corresponding distribution in $z=\cos\theta$,
when artificially separating the process, 
can be obtained from the amplitudes as usually:
\begin{equation}
 \frac{d\sigma}{dz} = \sum_{\lambda_1,\lambda_2}
\frac{\sqrt{\frac{W^2}{4}-m_\pi^2}}{\frac{W}{2}} 
\left|  \mathcal{M}^t(\lambda_1,\lambda_2) + 
    	\mathcal{M}^u(\lambda_1,\lambda_2) +
		\mathcal{M}^c(\lambda_1,\lambda_2) \right|^2
\frac{4\pi}{4\times 64\pi^2 W^2} \; .
\end{equation}

\subsection{pQCD mechanisms}

\begin{figure}[!h]             
\begin{minipage}[t]{0.46\textwidth}
\centering
\includegraphics[width=0.65\textwidth]{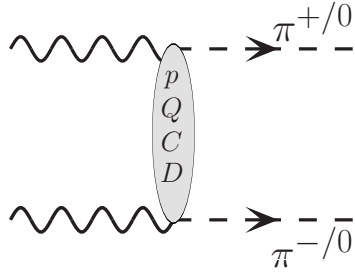}
\end{minipage}
   \caption{\label{fig:pQCD}
   \small The Brodsky-Lepage or hand-bag 
   perturbative mechanisms for large-angle
   $\gamma \gamma \to \pi \pi$ scattering.
           }
\end{figure}

In Ref.\cite{KS_pQCD} we have concentrated on perturbative mechanisms.
In the present paper we shall only summarize the main formulae.
Again, as shown in Fig.\ref{fig:pQCD}, we shall consider simultaneously
the $\pi^+\pi^-$ and $\pi^0\pi^0$ channels.
%

\subsubsection{Brodsky-Lepage mechanism}

\begin{figure}[htb]
\centering
\includegraphics[width=.3\hsize]{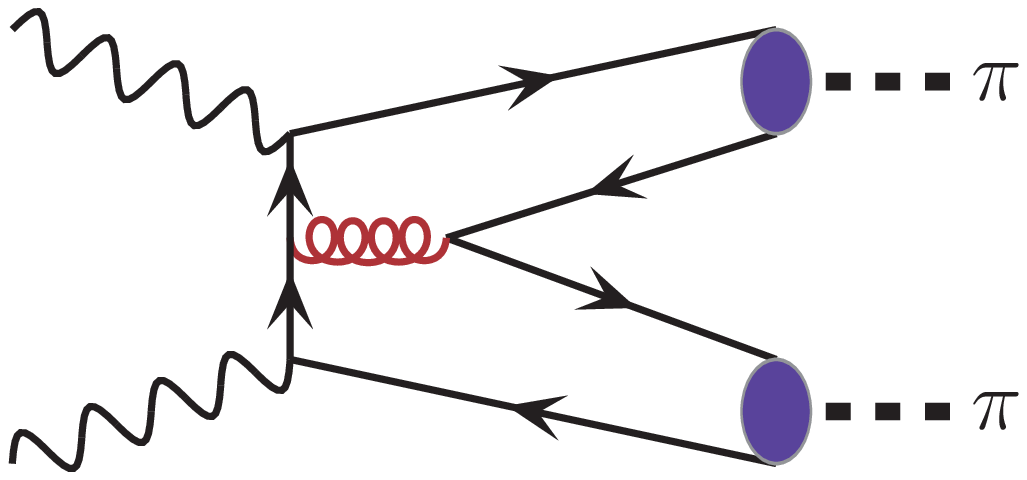}
\includegraphics[width=.3\hsize]{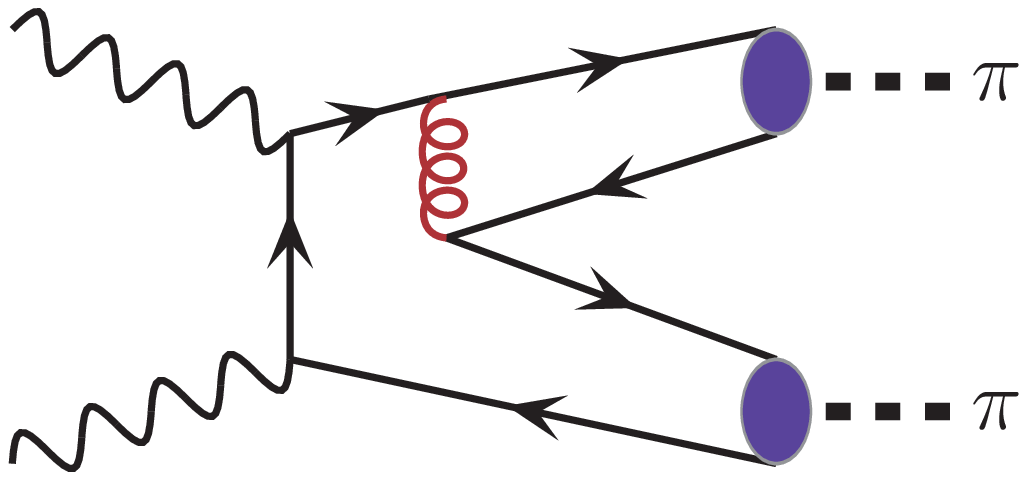}
\includegraphics[width=.3\hsize]{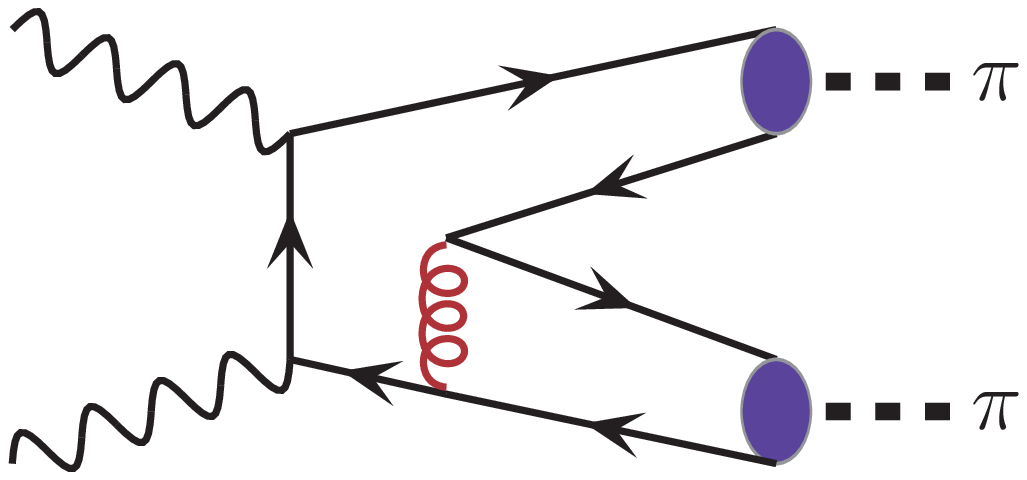}
\includegraphics[width=.3\hsize]{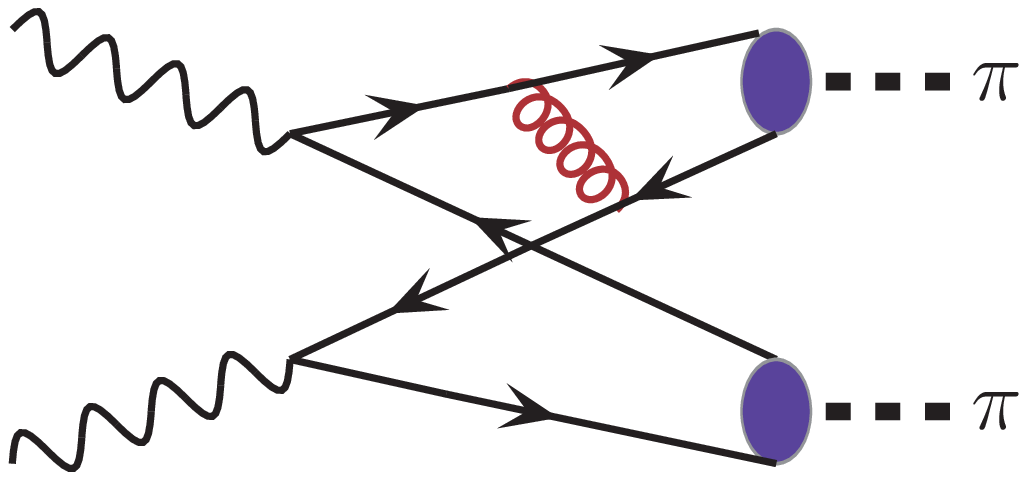}
\includegraphics[width=.3\hsize]{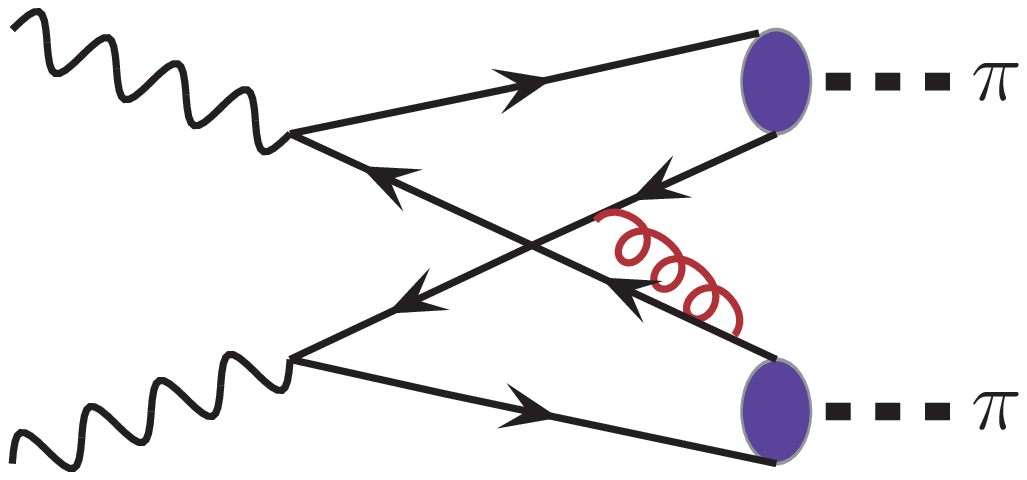}
\vskip -0.3cm
   \caption{\label{fig:pQCD_amplitude}
   \small Feynman diagrams describing the 
$\gamma\gamma  \to \pi\pi$
amplitude in LO pQCD.}
\end{figure}
The basic diagrams of the Brodsky and Lepage formalism are shown 
in Fig.\ref{fig:pQCD_amplitude}.
The invariant amplitude for the initial helicities of two photons 
can be written as:
\begin{eqnarray}
 \mathcal{M}  \left( \lambda_1, \lambda_2 \right)  = 
 \int_0^1 dx \int_0^1 dy \, \phi_\pi \left( x, \mu^2_x \right) 
 T^{\lambda_1 \lambda_2}_H \left( x,y, \mu^2 \right) \phi_\pi \left( y, \mu^2_y \right)
 \times F^{pQCD}_{reg} \left(t,u \right),
\label{eq:pQCDamp}
\end{eqnarray}
where for $\mu_x$ or $\mu_y$:
$\mu_{x_{/y}} = min \left( x_{/y}, 1-x_{/y}  \right) \sqrt{s(1-z^2)}$; 
$z= \cos \theta$ \cite{BL81}.
We take the helicity dependent hard scattering amplitudes from Ref.\cite{JA}.
These scattering amplitudes are different for $\pi^+ \pi^-$
and $\pi^0 \pi^0$. 
The extra form factor in Eq.\ref{eq:pQCDamp} was proposed in Ref.\cite{SS03}:
\begin{equation}
 F^{pQCD}_{reg} \left(t,u \right)= 
 \left[ 1-\exp \left( \frac{t-t_m}{\Lambda_{reg}^2} \right) \right] 
 \left[ 1-\exp \left( \frac{u-u_m}{\Lambda_{reg}^2} \right)  \right],
 \label{eq:ff_pQCD_old}
\end{equation}
where $t_m=u_m$ are the maximal kinematically allowed values of $t$ and $u$. 
This form factor excludes the region of small Mandelstam 
$t$ and $u$ variables which is clearly of nonperturbative nature.
In the present analysis we are even more conservative and try to separate
the phase space into low and high energies and postulate
that the pQCD effects show up only above a certain energy. 
In practice we use the following function which smoothly switches 
off the pQCD contribution at low energies:
\begin{equation}
 F^{pQCD}(s) = 1-\exp\left(\frac{-\left(s-4m_\pi^2\right)^4}{\Lambda^8_{pQCD}}\right) \; .
 \label{eq:ff_pQCD_new}
\end{equation}
%
\begin{figure}[!h]             
\begin{minipage}[t]{0.47\textwidth}
\centering
\includegraphics[width=0.99\textwidth]{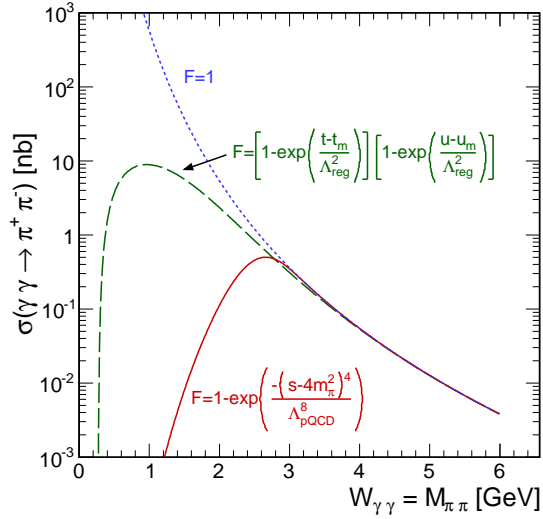}
\end{minipage}
   \caption{\label{fig:pQCD_ff}
   \small Comparison of different forms of the extra form factor.
          We show results for a new separation of low and high energy processes
          given by Eq.(\ref{eq:ff_pQCD_new}) and represented by the red solid line (\mbox{$\Lambda_{pQCD}=$ 2.5 GeV}).
          In Ref.\cite{KS_pQCD} we used the form given by Eq.(\ref{eq:ff_pQCD_old}) 
          represented in the figure by the green dashed line ($\Lambda_{reg}=$ 1 GeV).
          The dotted line shows the BL pQCD cross section without any extra form factor.}
\end{figure}
%
%
Fig.\ref{fig:pQCD_ff} illustrates the role of the extra form factors discussed above.

The distribution amplitudes are subjected to the ERBL pQCD evolution \cite{ER, BL79}.
The scale dependent quark distribution amplitude of the pion \cite{Muller,AB}
can be expanded in terms of Gegenbauer polynomials:
\begin{equation}
\phi_\pi \left( x, \mu^2 \right) = 
\frac{f_\pi}{2 \sqrt{3}}
6x \left( 1-x \right) \sum_{n=0}^{\infty '} C_n^{3/2} 
\left(2x-1 \right) a_n \left(\mu^2 \right)  ,
\end{equation}
where the expansion coefficients $a_n \left(\mu^2 \right)$
depend on the form of the distribution amplitude
$\phi_\pi \left( x, \mu_0^2 \right)$.
Wu and Huang \cite{WH} proposed recently a new form 
of the distribution amplitude:
\begin{eqnarray}
 \phi_\pi \left( x, \mu_0^2 \right) & = 
 & \frac{\sqrt{3}A \, m_q \beta}{2 \sqrt{2} \pi^{3/2} f_\pi} 
 \sqrt{x \left( 1-x \right)} 
 \left( 1+B \times C_2^{3/2} \left(2x-1 \right) \right) \nonumber \\
 &\times& \left( \mbox{Erf} \left[ \sqrt{\frac{m_q^2+\mu_0^2}{8 \beta^2 x \left(1-x \right)}} \right] 
 - \mbox{Erf} \left[ \sqrt{\frac{m_q^2}{8 \beta^2 x \left(1-x \right)}} \right]\right).
 \label{eq.WH}
\end{eqnarray}
This pion distribution amplitude at the initial scale 
($\mu_0^2=$ 1 GeV$^2$) is controlled by 
the parameter B. This simple model better describes
recent BABAR data \cite{BABAR}.

\begin{figure}[!h]             
\begin{minipage}[t]{0.46\textwidth}
\centering
\includegraphics[width=0.8\textwidth]{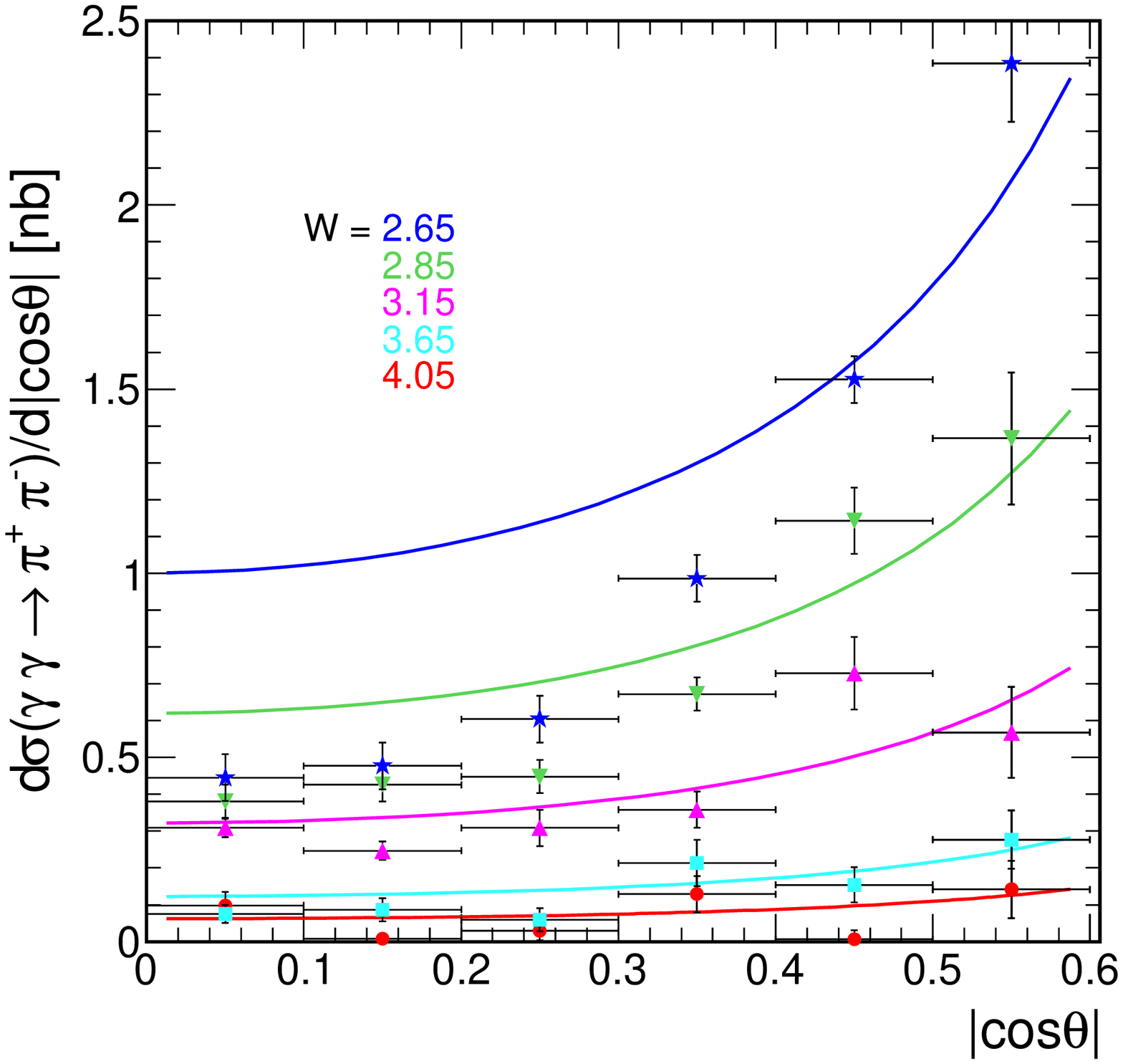}
\end{minipage}
\begin{minipage}[t]{0.46\textwidth}
\centering
\includegraphics[width=0.8\textwidth]{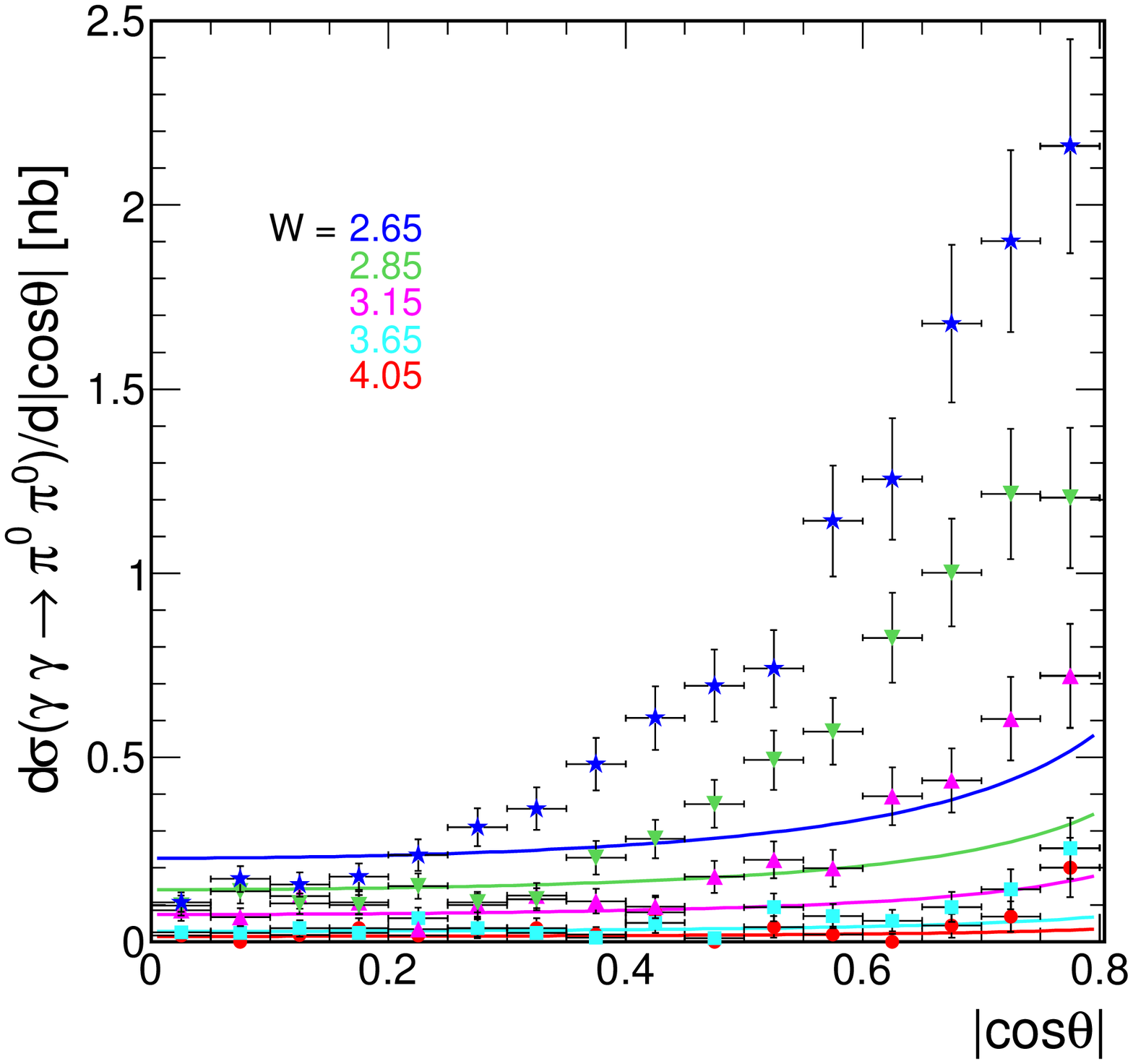}
\end{minipage}
   \caption{\label{fig:dsig_dz_pQCD}
   \small Comparison of BL pQCD angular distributions 
   with Belle data \cite{Belle_Nakazawa, Belle_Uehara09}.
           }
\end{figure}

In Fig.\ref{fig:dsig_dz_pQCD} we compare the results of the Brodsky-Lepage (BL)
approach and experimental data for different energies. As can be seen
from the figures the BL pQCD approach is not enough to describe
the ''high energy'' data, especially for the $\gamma\gamma\to\pi^0\pi^0$ reaction.
%
%
\subsubsection{Hand-bag mechanism}
%
The hand-bag approach was proposed in Ref.\cite{HB}.
In this approach the only nonzero helicity-dependent amplitudes read
\begin{equation}
 \mathcal{A}_{+-}=\mathcal{A}_{-+}=-4 \pi \alpha_{em} \frac{s^2}{tu} R_{2\pi}\left(s\right) \; .
\end{equation}
The form factors $R_{2\pi}(s)$ are of nonperturbative nature and are in
principle unknown. In practice they were fitted to the experimental total
(integrated over experimentally measured region)
cross section for $\gamma \gamma \to \pi^+ \pi^-$ \cite{HB},
assuming that the mechanism exhausts the measured cross section at 
high energy. This is not a necessary condition. In our analysis we
shall relax this rather restrictive assumption.
In the present paper we discuss also respective angular distributions.
We shall use the parametrization proposed in Ref.\cite{HB}:
\begin{equation}
 R_{2\pi}\left(s\right) = 
\frac{5}{9s} a_u \left(\frac{s_0}{s}\right)^{n_u} + 
\frac{5}{9s} a_s \left(\frac{s_0}{s}\right)^{n_s} \; .
\end{equation}

The values of $a_u$, $n_u$, $a_s$ and $n_s$ are taken from Ref.\cite{HB}.
We have averaged the values for different energies: 
$a_u=$ 1.375 GeV$^2$, $n_u=0.4175$, 
$a_s=$ 0.5025 GeV$^2$ and $n_s=1.195$.
%
\begin{figure}[!h]             
\begin{minipage}[t]{0.46\textwidth}
\centering
\includegraphics[width=0.8\textwidth]{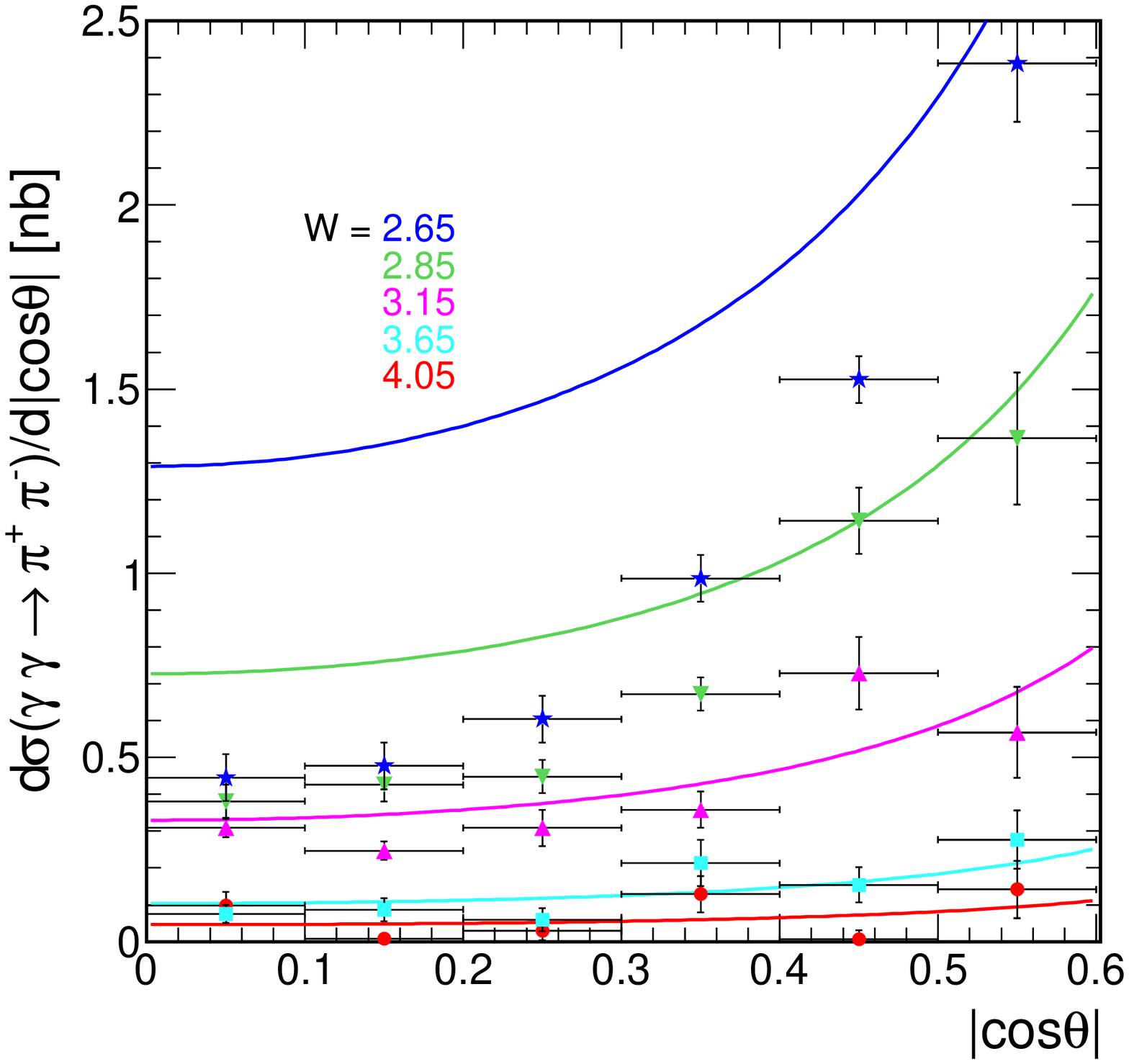}
\end{minipage}
\begin{minipage}[t]{0.46\textwidth}
\centering
\includegraphics[width=0.8\textwidth]{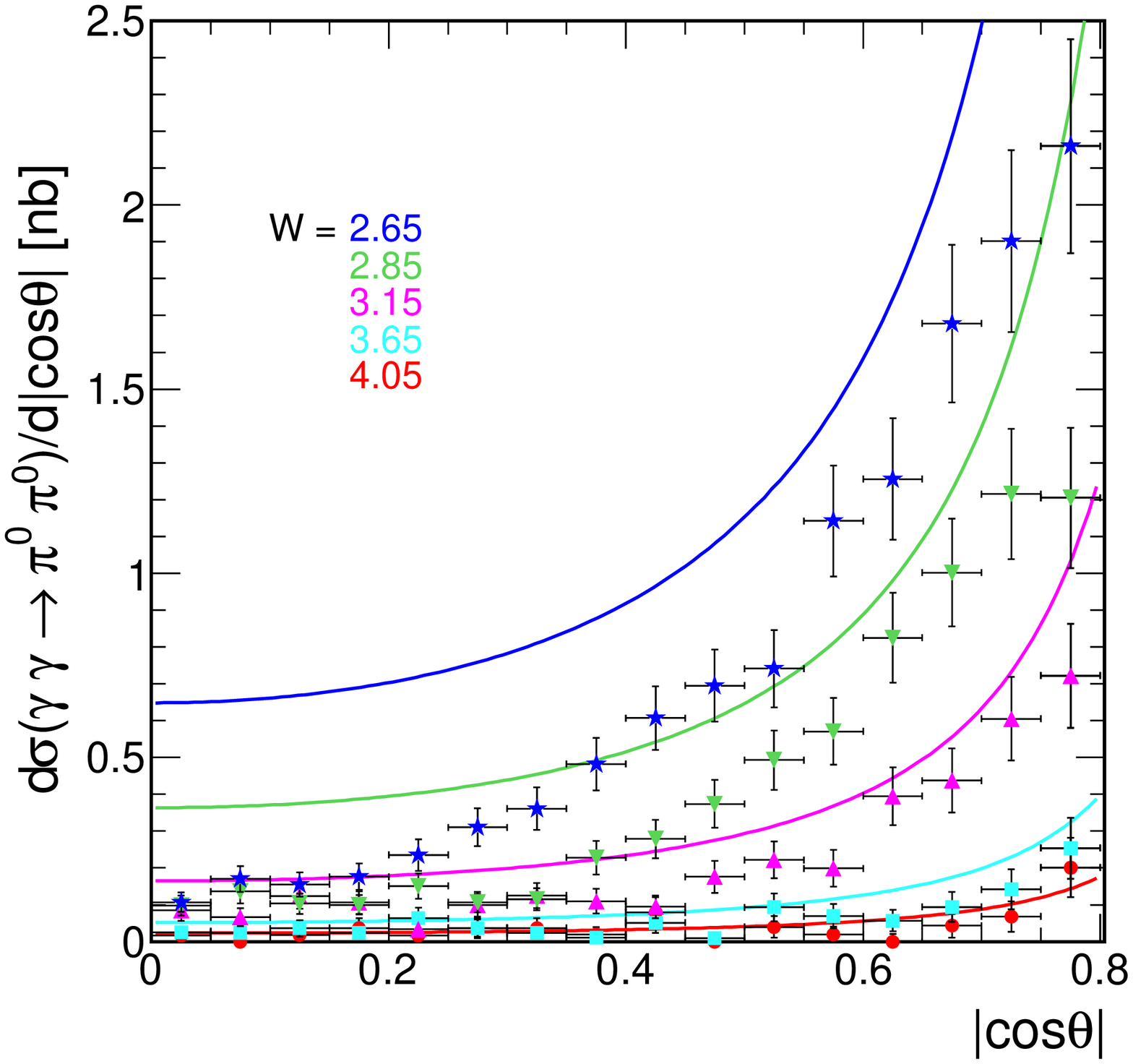}
\end{minipage}
   \caption{\label{fig:dsig_dz_hb}
   \small Comparison of hand-bag angular distributions 
   with Belle data \cite{Belle_Nakazawa, Belle_Uehara09}.
           }
\end{figure}
%

In Fig.\ref{fig:dsig_dz_hb} we show angular distributions
calculated in the hand-bag approach together
with the Belle experimental data \cite{Belle_Nakazawa, Belle_Uehara09}. 
Only at the highest energies
the hand-bag parametrization looks reasonable, however, does not describe properly
the ratio of the $\gamma\gamma\to\pi^0\pi^0$ and $\gamma\gamma\to\pi^+\pi^-$ 
cross section (see Fig.\ref{fig:ratio}).
At somewhat lower energies the hand-bag and experimental distributions
differ significantly. The onset of the hand-bag mechanism was not
discussed so far in the literature.
We shall return to the problem in the Result section where other
mechanisms will be discussed too.

\subsection{Equivalent photon approximation for $AA\to AA\pi\pi$}

How to calculate the production of pairs of particles in nucleus-nucleus 
collisions was explained in detail in Ref.\cite{KS_muon}.
Here we only sketch the formalism.
The total nuclear cross section can be expressed by folding
the $\gamma\gamma\to\pi\pi$ subprocess cross section  
with equivalent photon fluxes as:
\begin{eqnarray}
\sigma(AA\to AA\pi\pi;s_{AA} ) & = & 
\int \hat{\sigma}(\gamma\gamma\to\pi\pi;W_{\gamma\gamma}) S^2_{abs}({\bf b}) \nonumber \\ 
& \times & N(\omega_1,{\bf b}_1) N(\omega_2,{\bf b}_2) 
d^2{\bf b}_1 d^2{\bf b}_2 d\omega_1 d\omega_2 \; .
\label{eq.first_cross_section}
\end{eqnarray}
In principle, the absorption factor $S_{abs}({\bf b})$ can be calculated in different models. 
At not too high energies the Glauber approach is a reasonable approach.
Here we take a somewhat simpler approach and approximate the absorption 
factor as:
\begin{equation}
S^2_{abs}({\bf b}) = \theta({\bf b}-2R_A) = 
                     \theta(|{\bf b}_1-{\bf b}_2|-2R_A) \; .
\end{equation}
This form excludes the situation at high energies when the colliding
nuclei overlap which at high energies unavoidably leads to their breakup. 

In order to simplify our calculation we shall use the following
transformations:
\begin{equation}
\omega_{1/2} = \frac{W_{\gamma\gamma}}{2} e^{\pm Y_{\pi\pi}}
\mbox{, } 
d\omega_1 d\omega_2 = \frac{W_{\gamma\gamma}}{2} dW_{\gamma\gamma} dY_{\pi\pi} \; .
\end{equation}
Then formula (\ref{eq.first_cross_section}) can be rewritten as:
\begin{eqnarray}
\sigma(AA\to AA\pi\pi;s_{AA} ) & = & 
\int \hat{\sigma}(\gamma\gamma\to\pi\pi;W_{\gamma\gamma}) S^2_{abs}({\bf b}) \nonumber \\ 
& \times & N(\omega_1,{\bf b}_1) N(\omega_2,{\bf b}_2) 
\frac{W_{\gamma\gamma}}{2} d^2{\bf b}_1 d^2{\bf b}_2 d W_{\gamma\gamma} d Y_{\pi\pi} \; .
\label{eq.second_cross_section}
\end{eqnarray}

While in the above approach one can easily calculate only the total nuclear 
cross section, distributions in rapidity of the pair of pions,
and the invariant mass of the dipions (see e.g. \cite{KS_rho,KS_muon}),
experimental constraints can not be easily imposed.

If one wants to calculate in addition kinematical distributions of 
each of the individual particles (transverse momentum, rapidity, pseudorapidity),
or impose corresponding experimental cuts on kinematical quantity of each
of dipions, a more complicated calculation is required. Then instead 
of the one-dimensional array 
$\sigma(\gamma \gamma \to \pi \pi;W_i)$, a two-dimensional array
$\frac{d \sigma(\gamma \gamma \to \pi \pi)}{dz} \left(W_i, z_j\right)$
has to be calculated, stored and passed to the code calculating 
cross sections for nuclear collisions.
Then an extra integration in $z=\cos\theta$ is required in Eqs.~(\ref{eq.first_cross_section})
or (\ref{eq.second_cross_section}), which makes the calculation much more time-consuming. 

Next four-momenta of pions in the $\pi \pi$ center
of mass frame are calculated:
\begin{eqnarray}
E_{\pi} = \sqrt{\hat{s}}/2. \; , \nonumber \\
p_{\pi} = \sqrt{\frac{\hat{s}}{4} - m_{\pi}^2}  \; , \nonumber \\
p_{t,\pi} = \sqrt{1-z^2} p_{\pi}  \; , \nonumber \\
p_l = z p_{\pi} \; .
\label{pipi_cm_kinematics}
\end{eqnarray}
The rapidity of each of the pions can be calculated easily as:
\begin{equation}
y_i = Y_{\pi \pi} + y_{i/\pi \pi} \; ,
\label{rapidities_of_pions}
\end{equation}
where $y_{i/\pi \pi}=y_{i/\pi \pi}(W,z)$ is the rapidity
of one of the pions in the recoil $\pi\pi$ system of reference.
The transverse momenta of pions in both frames of reference, as Lorentz
invariants, are the same.
Other kinematical variables are calculated by adding relativistically
velocities \cite{Hagedorn_book}
\begin{equation}
\vec{v}_i = \vec{V}_{\pi \pi} \oplus \vec{v}_{i/\pi\pi}  \;  ,
\label{adding_velocities}
\end{equation}
where 
\begin{equation}
\overrightarrow{V_{\pi \pi}} = \frac{\overrightarrow{P_{\pi \pi}}}{E_{\pi \pi}}
\label{velocity_of_recoiled_system}
\end{equation}
and from the energy-momentum conservation:
\begin{eqnarray}
E_{\pi \pi}   &=& \omega_1 + \omega_2   \; , \nonumber \\
P_{\pi \pi}^z &=& \omega_1 - \omega_2   \; ,
\end{eqnarray}
the energies of photons can be expressed in terms of our
integration variables as:
\begin{eqnarray}
\omega_1 &=& \frac{W_{\gamma \gamma}}{2} \exp ( Y) \; , \nonumber \\
\omega_2 &=& \frac{W_{\gamma \gamma}}{2} \exp (-Y) \; .
\end{eqnarray}

Now the pion velocities can be converted to four-momenta of pions
in the overall nucleus-nucleus center of mass frame.
Then angles or pseudorapidities of pions can be easily calculated
and experimental cuts can be imposed in addition.

Distributions in kinematical variables are obtained by
corresponding binning into histograms. A smooth distributions in
transverse momenta or pseudorapidities require
many points in $W$ and $z$. If only experimental cuts are imposed the
number of points can be smaller. In the present approach we are 
interested first of all in two-pion invariant mass distribution 
but including cuts of real experiments. 

\subsection{$A A \to A A \rho$}

In nucleus-nucleus collisions another mechanism exists which contributes
to the $\pi^+ \pi^-$ channel -- coherent production of $\rho^0$ mesons
and its subsequent decay into the $\pi^+ \pi^-$ channel 
($\rho^0 \to \pi^+ \pi^-$).
This mechanism was studied theoretically in the classical Glauber \cite{KN99},
quantum Glauber \cite{FSZ}, Glauber-Gribov \cite{FSZ02,RSZ2011} and
Color Glass Condensate \cite{GM11} approaches.
The rapidity distribution of $\rho^0$ 
can be written in an almost model-independent way as:
\begin{equation}
\frac{d \sigma}{d y} 
= \omega_1 \frac{d N_{\gamma}(\omega_1)}{d \omega_1} 
  \sigma_{\gamma A \to V A}
+ \omega_2 \frac{d N_{\gamma}(\omega_2)}{d \omega_2}
  \sigma_{A \gamma \to A V} \; .
\label{V_photoproduction} 
\end{equation}
Essentially the models enter in calculating the cross section for 
$\gamma A \to V A$ processes. While for heavy quarkonium production
the situation is more transparent \cite{CSS12}, for light meson
production the situation is somewhat less certain.
We shall not discuss the issue here, we rather refer the reader to the existing
literature (see e.g. \cite{ultraperiheral}).
In the most rudimentary approach:
\begin{equation}
\sigma_{\gamma A \to \rho^0 A} = \frac{\alpha_{em}}{4 f_{\rho}^2} 
\left( \sigma_{\rho A}^{tot} \right)^2 \int_{t_{min}}^{\infty} dt
F_A^2(t) \; ,
\label{photoproduction_cross_section}
\end{equation}
where $\sigma_{\rho A}^{tot} \approx \pi R_A^2$ and 
$t_{min} = \frac{m_{\rho}^4}{4 q_0^2}$ \cite{FSZ}.

%
\section{Results}

Before we go to the nuclear processes we wish to discuss individual
subprocesses $\gamma \gamma \to \pi^+ \pi^-$ and 
$\gamma \gamma \to \pi^0 \pi^0$ which are interesting by themselves.

\subsection{$\gamma \gamma \to \pi \pi$}


%
\begin{figure}[!h]             
\begin{minipage}[t]{0.45\textwidth}
\centering
\includegraphics[width=1.05\textwidth]{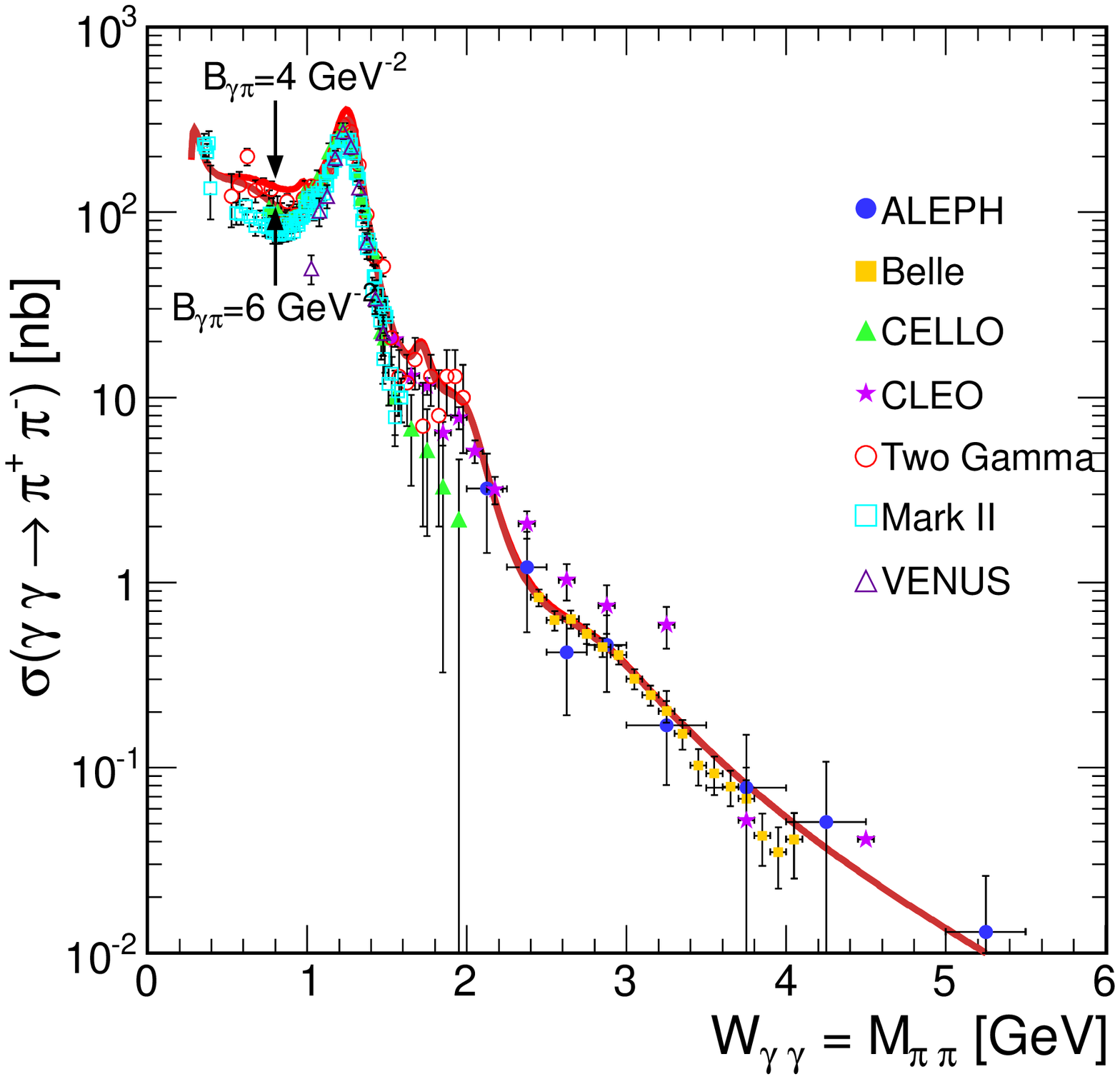}
\end{minipage}
\begin{minipage}[t]{0.45\textwidth}
\centering
\includegraphics[width=1.05\textwidth]{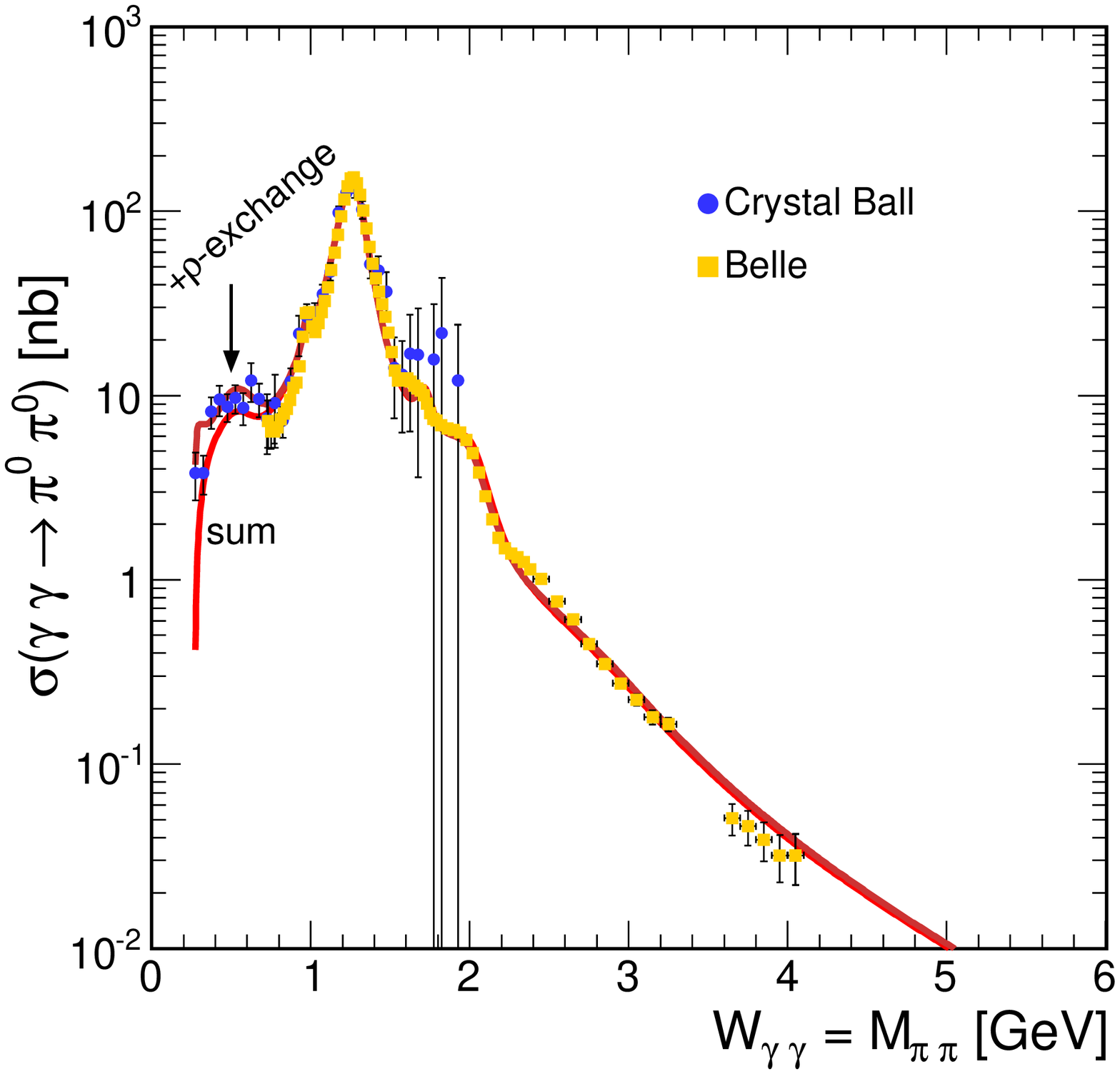}
\end{minipage}
	\caption{\label{fig:sig_W_sum}
	\small Results of the fit: the total cross section
	($\gamma\gamma\to\pi^+\pi^-$ for $|\cos\theta|<$ 0.6, 
	$\gamma\gamma\to\pi^0\pi^0$ for $|\cos\theta|<$ 0.8).}
\end{figure}

\begin{figure}[!h]             
\begin{minipage}[t]{0.45\textwidth}
\centering
\includegraphics[width=1.05\textwidth]{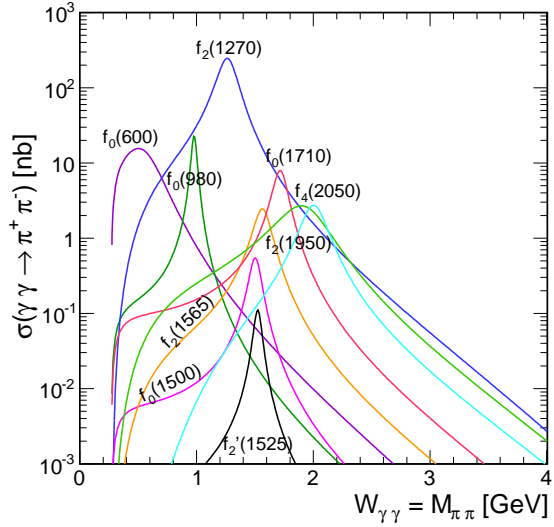}
\end{minipage}
	\caption{\label{fig:sig_W_resonances}
	\small Contributions of different s-channel resonances
               to the $\gamma\gamma\to\pi^+\pi^-$ reaction.}
\end{figure}

\begin{figure}[!h]             
\begin{minipage}[t]{0.45\textwidth}
\centering
\includegraphics[width=1.05\textwidth]{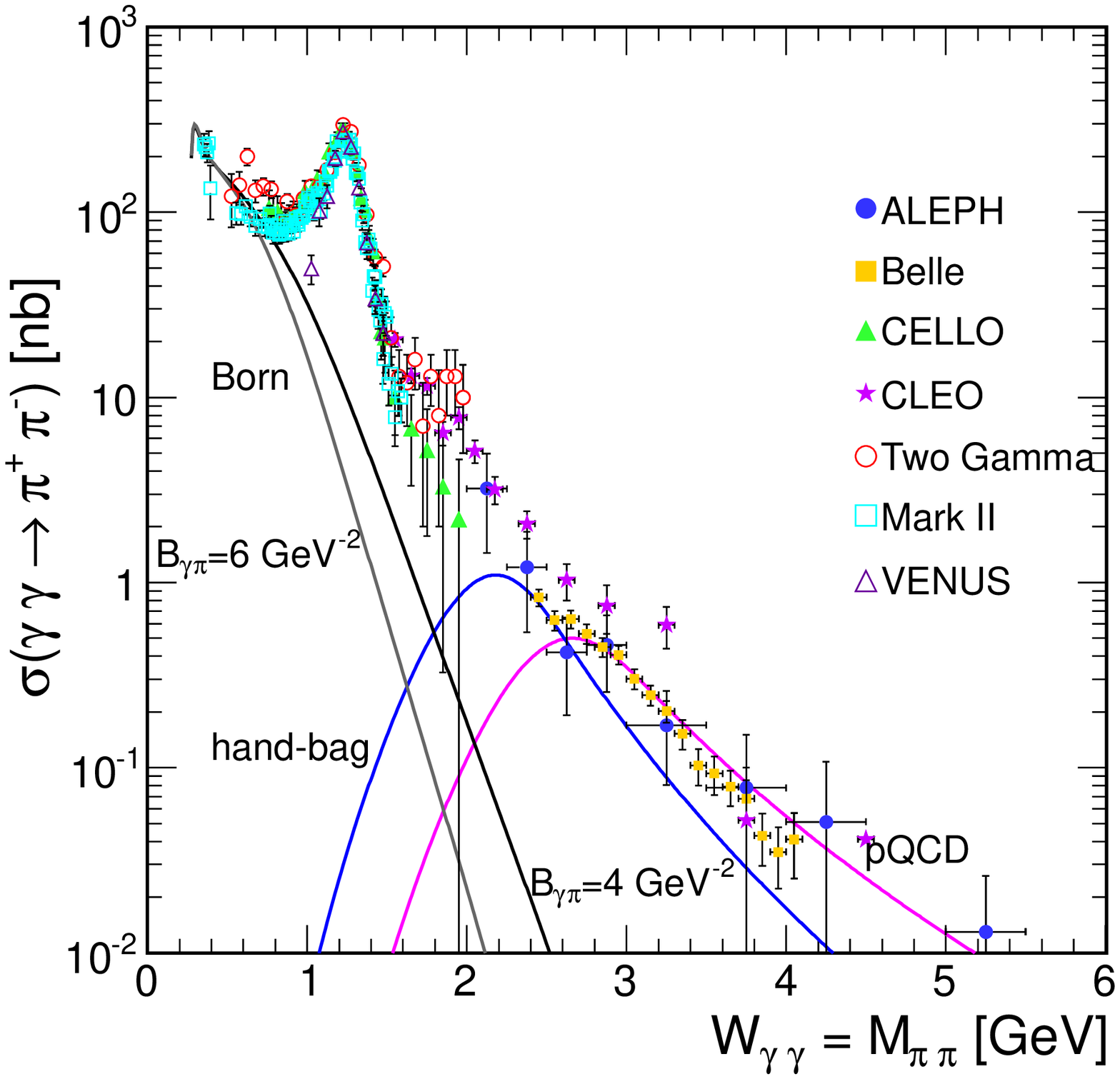}
\end{minipage}
\begin{minipage}[t]{0.45\textwidth}
\centering
\includegraphics[width=1.05\textwidth]{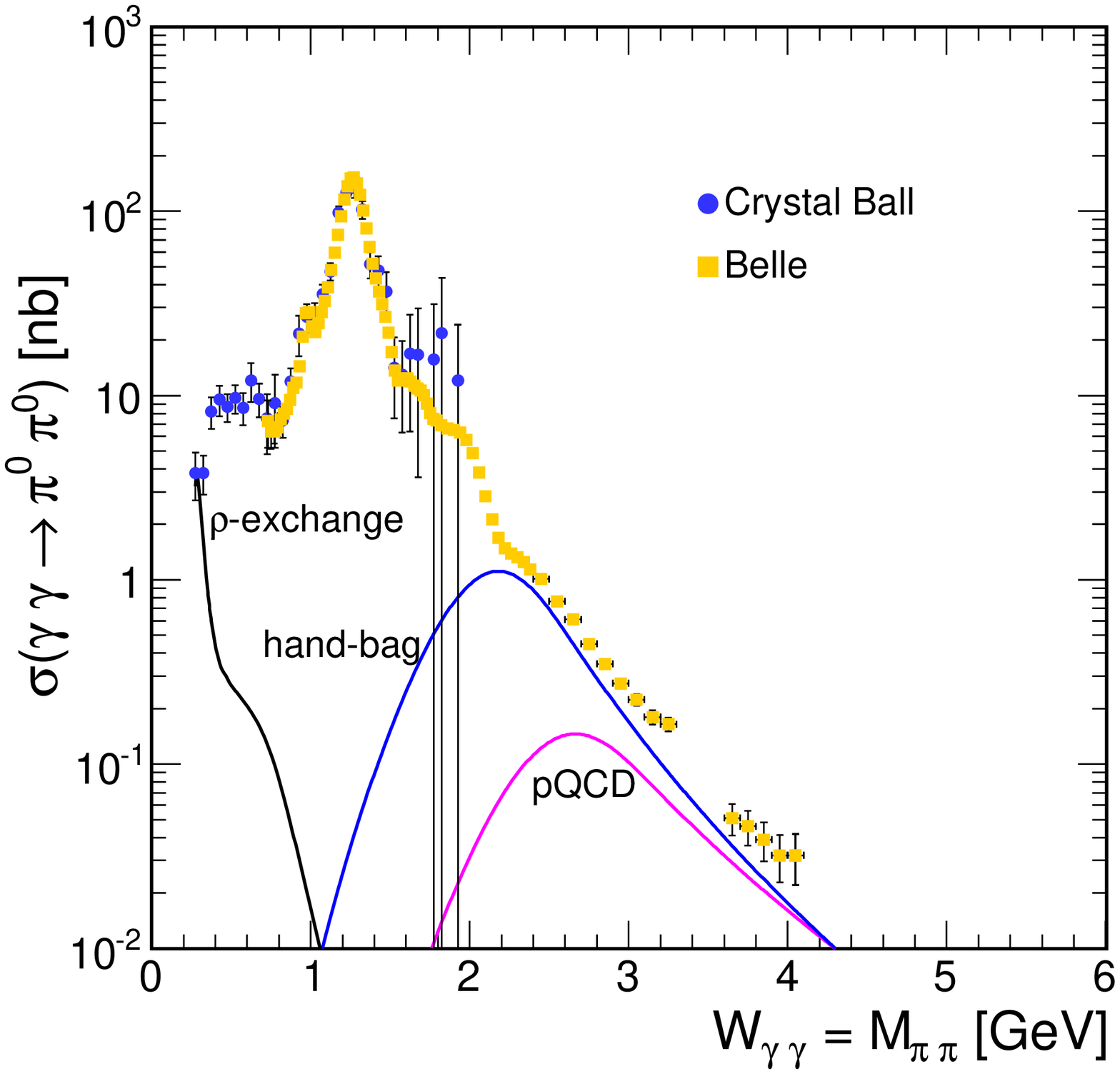}
\end{minipage}
	\caption{\label{fig:sig_W_continuum}
	\small Contributions of different continuum processes.
	The meaning of the curves is explained in the figures.}
\end{figure}
%
We wish to show first cross section for 
$\gamma \gamma \to \pi^+ \pi^-$ and $\gamma \gamma \to \pi^0 \pi^0$
processes as a function of the subprocess energy.
We shall include the pion exchange mechanism for the 
$\gamma \gamma \to \pi^+ \pi^-$ reaction, several resonance states,
pion-pion rescattering important, as will be shown below,
for $\gamma \gamma \to \pi^0 \pi^0$ reaction where in the leading order
the pion exchange continuum vanishes. At somewhat higher energies
we include also the Brodsky-Lepage pQCD and hand-bag mechanisms.
In Fig.\ref{fig:sig_W_sum} we show our model calculations
against world data for $\gamma\gamma\to\pi\pi$ \cite{Aleph, Belle_Mori, Belle_Nakazawa, Cello, Cleo, Gamma, 
Mark_II, Venus, Belle_Uehara09, Crystal_Ball}. We get a good agreement
with all available data points for the first time in such a large range of energies.
This makes our hybrid model for $\gamma \gamma \to \pi \pi$ well suited 
for the predictions of the cross sections for nucleus-nucleus collisions.

Our model amplitude is a sum of many ingredients 
as discussed in the previous section. 
In Fig.\ref{fig:sig_W_resonances} and Fig.\ref{fig:sig_W_continuum}
we show some somewhat artificially \footnote{We add up amplitudes not cross sections.} 
separated ingredients to the final spectrum. 
We show s-channel resonance (Fig.\ref{fig:sig_W_resonances})
and continuum (Fig.\ref{fig:sig_W_continuum}) contributions separately. 
At low energies, the $\sigma(600)$ s-channel resonance as well as
pion rescattering due to $\rho$-meson exchange play crucial role.
At somewhat higher energies the spectra are dominated by
the contribution of the tensor $f_2(1270)$ meson. At still higher
energies the situation is less clear. The smallness of the cross section at
$W \sim$ 1.5 GeV can be understood as an interference of right wing
of the tensor $f_2(1270)$ resonance and other resonances 
($f'_2(1525)$, $f_0(1500)$, $f_2(1565)$). At around $W \sim$ 2 GeV, we observe a new
irregular behaviour. This can be understood as the presence of s-channel 
spin-4 contribution. In order to describe the angular distribution we use
a simple model assuming that the spin-4 amplitude is proportional to
$\mathcal{M}_{\lambda_1 \lambda_2} \sim Y_{4,\lambda_1 - \lambda_2} 
\delta_{\lambda_1-\lambda_2,}$ (type B).
This is similar to what was found from a phenomenological
analysis in Ref.\cite{Belle_Uehara09}.
A much better description of the data is obtained when the $f_4(2050)$
resonance is included. The corresponding radiative decay width is
treated as a free parameter. We get $\Gamma_{f_4 \to \gamma \gamma}$ =
0.7 keV $\pm$ 0.1 keV. This is rather a rough estimate than a
rigorous statistical analysis. The problem is in the background
which is in this range of invariant masses rather difficult to control.
Our analysis is a first trial to extract the radiative
decay width of the $f_4(2050)$ resonance. The found value is much
smaller than that for the $f_2(1270)$ (3.035 keV) and comparable to
that for $f_0(980)$ (0.29 keV) (see Table \ref{tab:resonances}).
We shall not discuss consequences of better understanding of the structure
of $f_4(2050)$ resonance. This certainly goes beyond the scope of the
present paper.

Clearly, the tensor $f_2(1270)$ is the dominant resonance. 
Other resonances give much smaller but sizeable absolute cross section.
However, their separation is not an easy task especially 
from the analysis of total cross section. 
As was discussed recently \cite{Machado_glueball}, 
the cross section for the production of 
glueballs (G) in the nucleus-nucleus $A A \to A G A$ reaction could be large.
However, selecting a final state is in this context a very important 
issue. The $\pi \pi$ final state is an obvious candidate.
Our analysis shows that searching for glueballs in the $\pi \pi$ channels
may be extremely difficult, if not impossible. This makes
the heavy ion processes not the best tool to study glueballs
or other exotic mesons.
We think that a precise fit of the present Belle
angular distribution is probably better than any fit to the future nuclear data.

In Fig.\ref{fig:sig_W_continuum}, we show the contributions of the Brodsky-Lepage 
and hand-bag mechanisms. Please note that our normalization 
of the hand-bag mechanism was adjusted to describe
high-energy experimental data together with the Brodsky-Lepage mechanism. 
In Ref.\cite{HB} only the hand-bag mechanism, ignoring the Brodsky-Lepage mechanism, 
was fitted to the high-energy data.
For the $\pi^+ \pi^-$ channel we show in addition the continuum distribution
corresponding to pion exchange for two different values of slope parameter
and for the $\pi^0 \pi^0$ channel rescattering contribution due to $\rho^{\pm}$ t-channel
exchanges. The corresponding contribution falls quickly 
with increasing energy and plays an important role only up to $W =$ 1 GeV.
In Fig.\ref{fig:sig_W_sum} (right panel) we show the sum 
of resonances, BL pQCD and hand-bag mechanisms - lower line.
The upper curve includes in addition the $\rho^\pm$ exchange contribution.
Adding the $\rho$ exchange contribution 
allows us to better describe experimental data for 
small invariant masses.
%
\begin{figure}[!h]             
\begin{minipage}[t]{0.31\textwidth}
\centering
\includegraphics[width=1.05\textwidth]{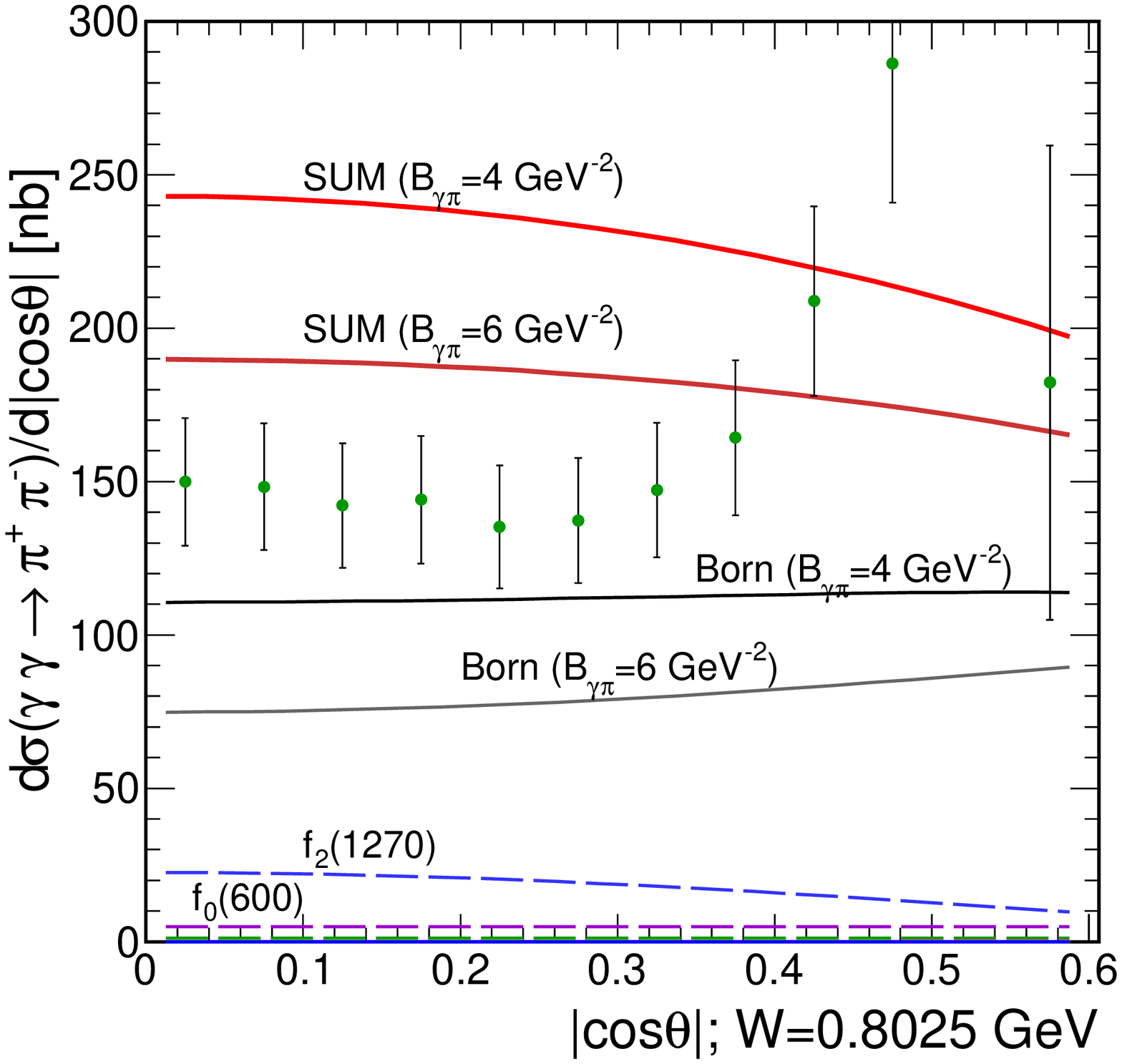}
\end{minipage}
\begin{minipage}[t]{0.31\textwidth}
\centering
\includegraphics[width=1.05\textwidth]{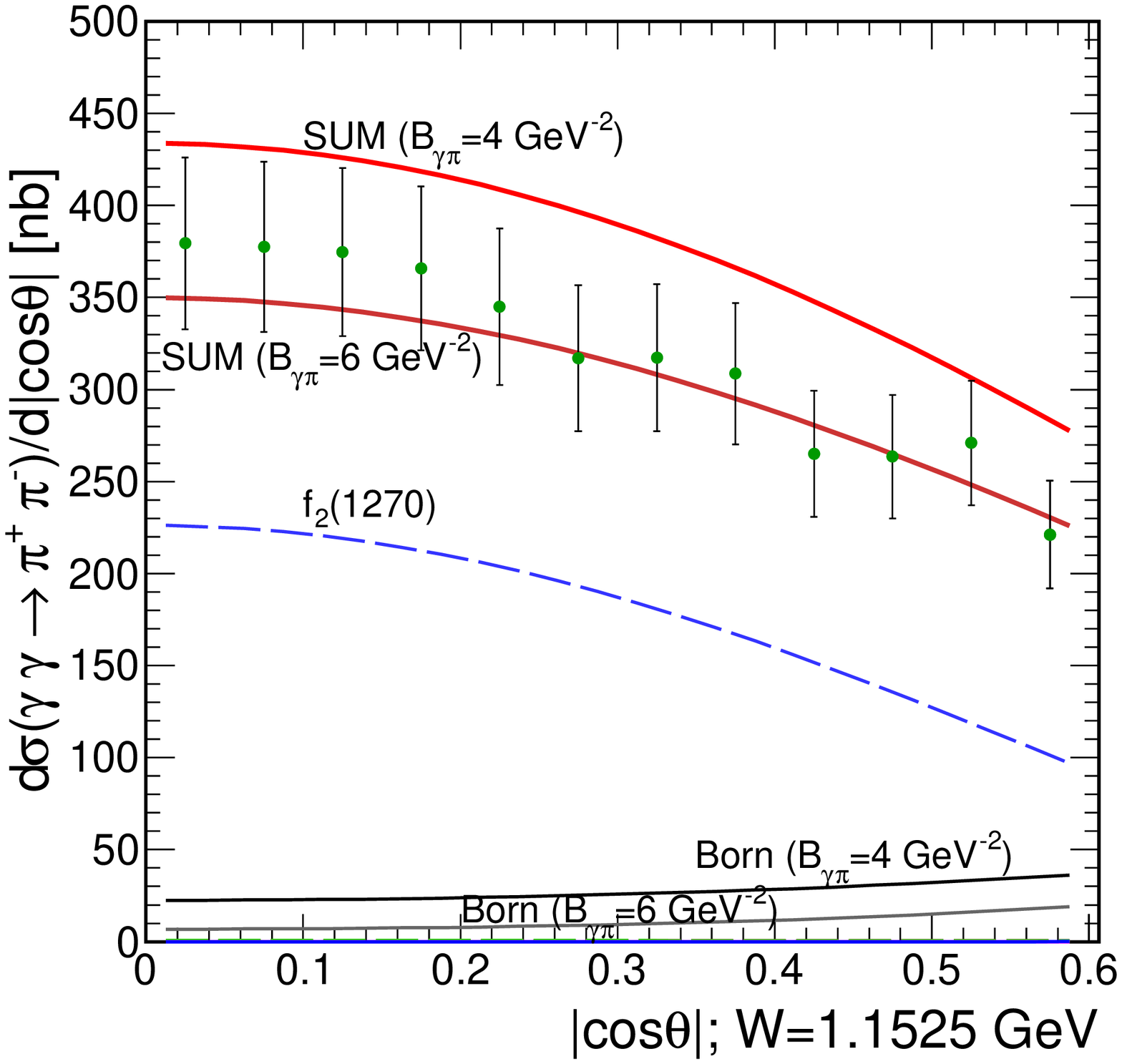}
\end{minipage}
\begin{minipage}[t]{0.31\textwidth}
\centering
\includegraphics[width=1.05\textwidth]{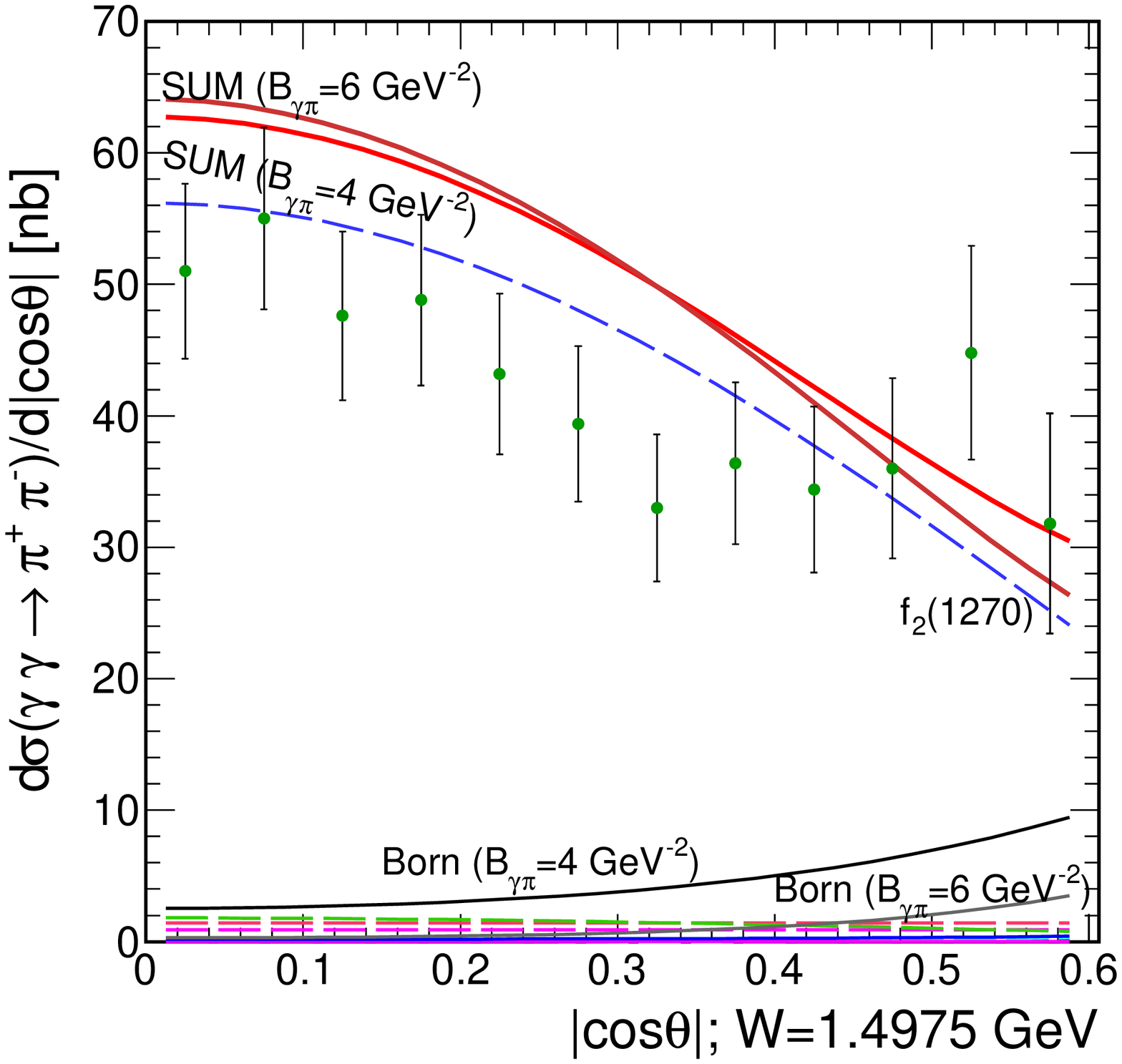}
\end{minipage}
\begin{minipage}[t]{0.31\textwidth}
\centering
\includegraphics[width=1.05\textwidth]{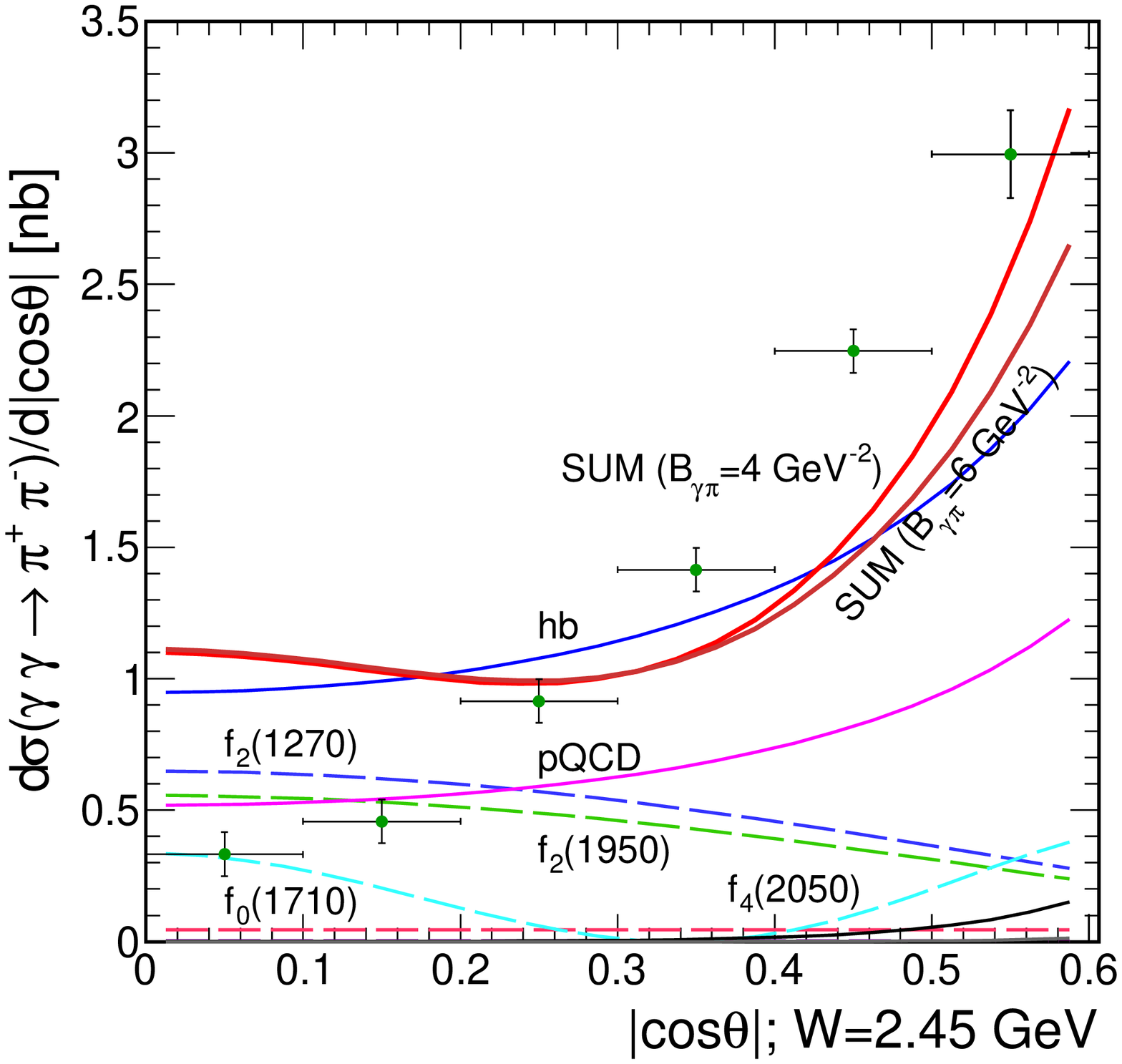}
\end{minipage}
\begin{minipage}[t]{0.31\textwidth}
\centering
\includegraphics[width=1.05\textwidth]{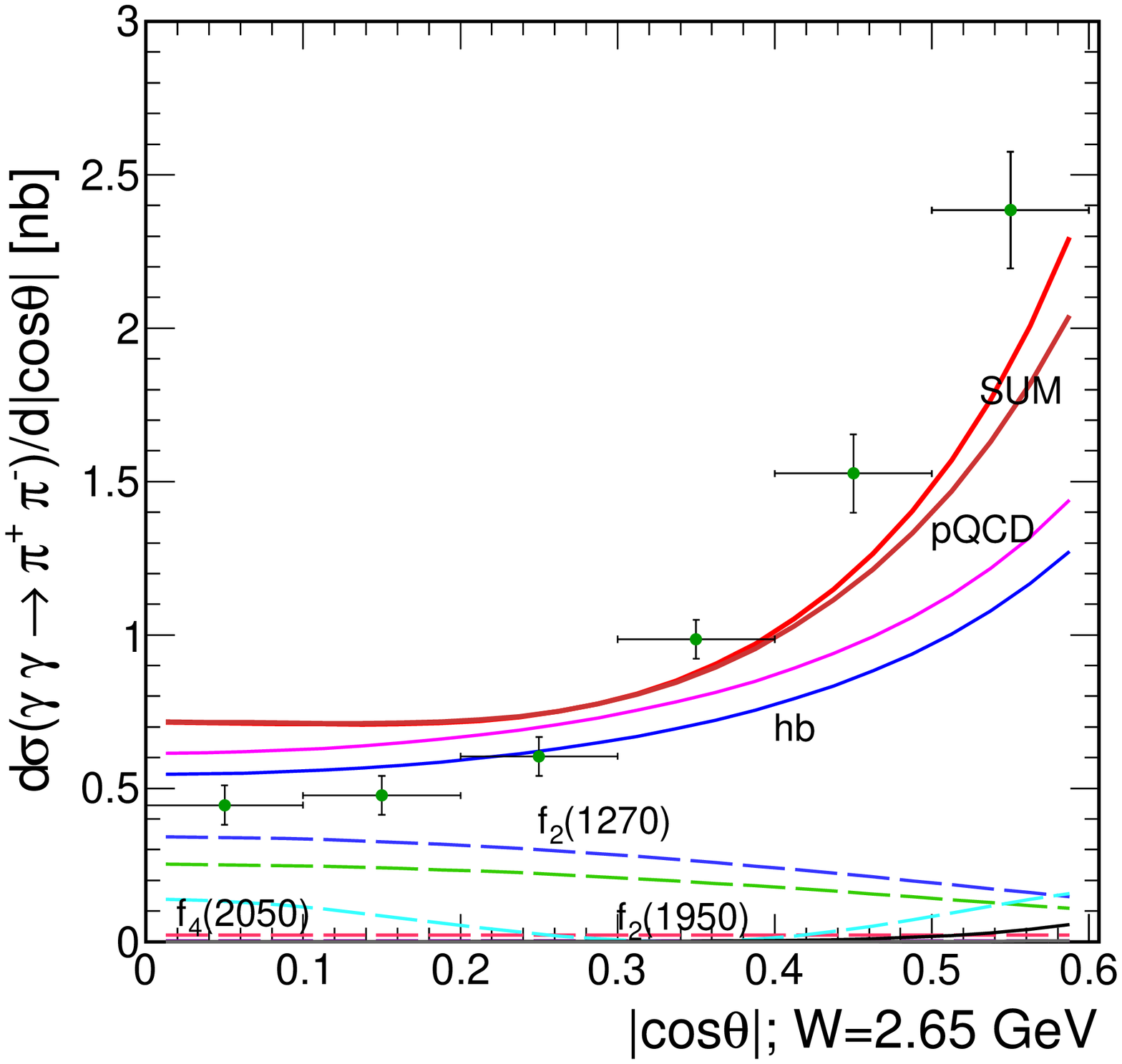}
\end{minipage}
\begin{minipage}[t]{0.31\textwidth}
\centering
\includegraphics[width=1.05\textwidth]{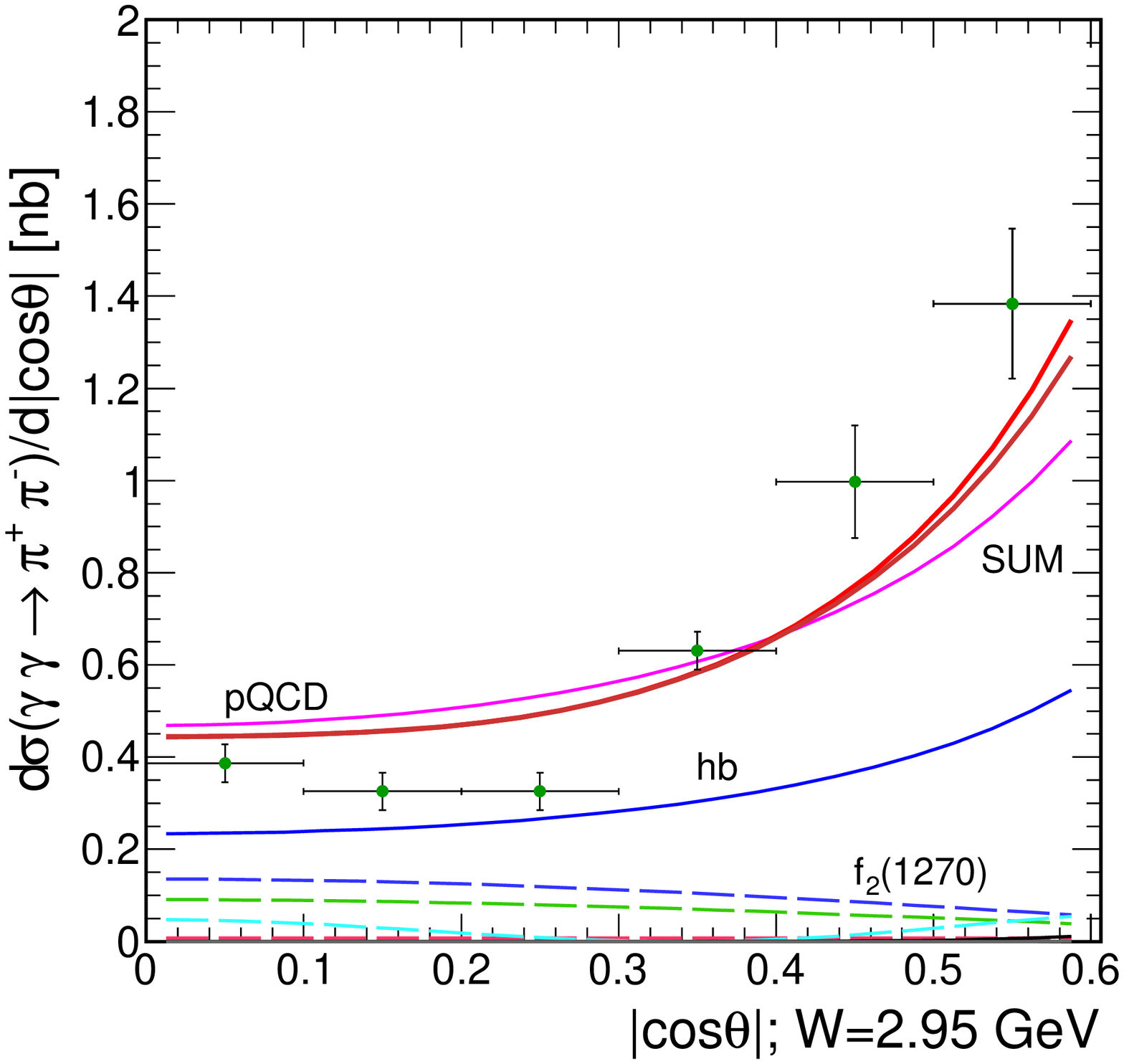}
\end{minipage}
\begin{minipage}[t]{0.31\textwidth}
\centering
\includegraphics[width=1.05\textwidth]{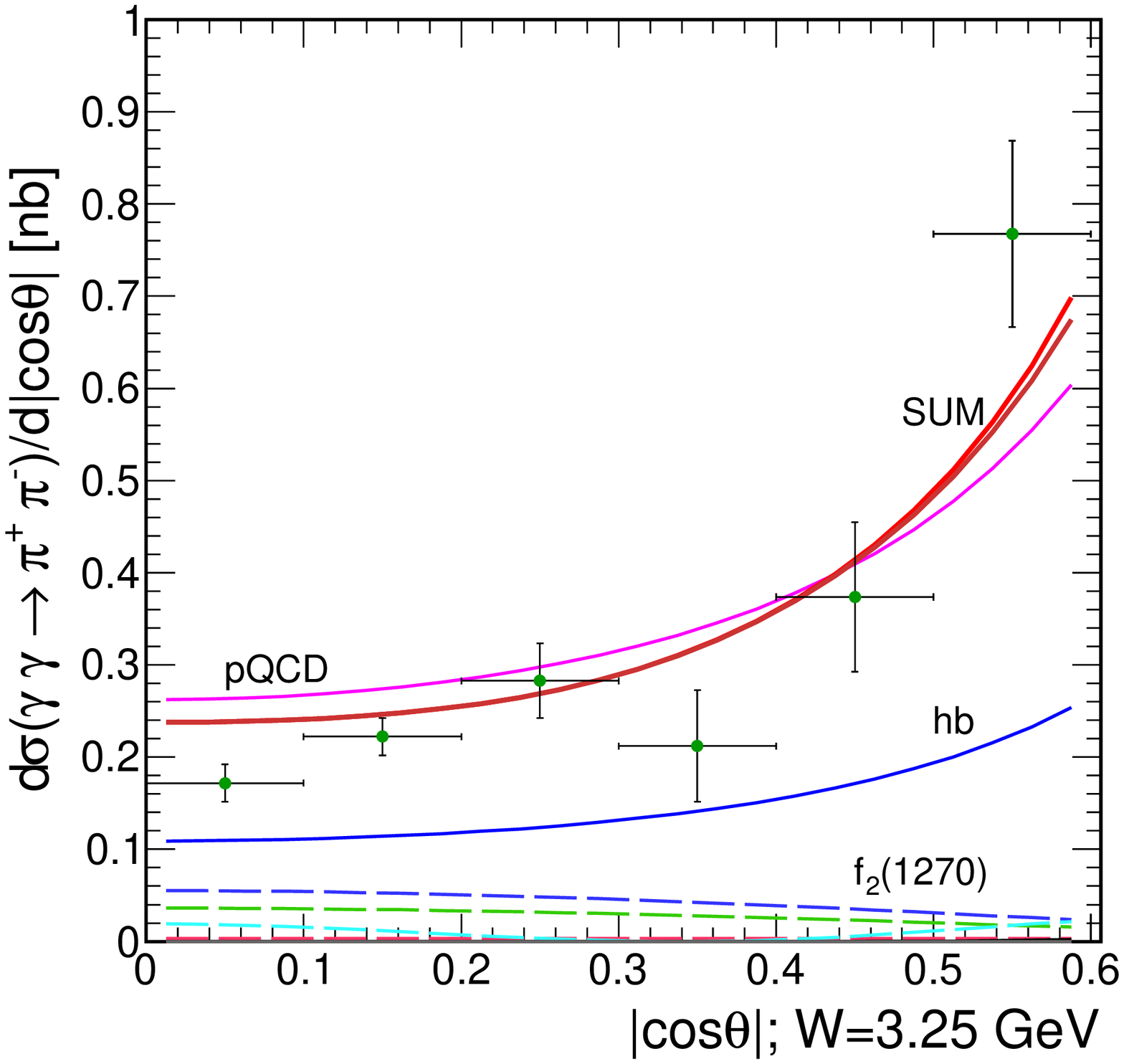}
\end{minipage}
\begin{minipage}[t]{0.31\textwidth}
\centering
\includegraphics[width=1.05\textwidth]{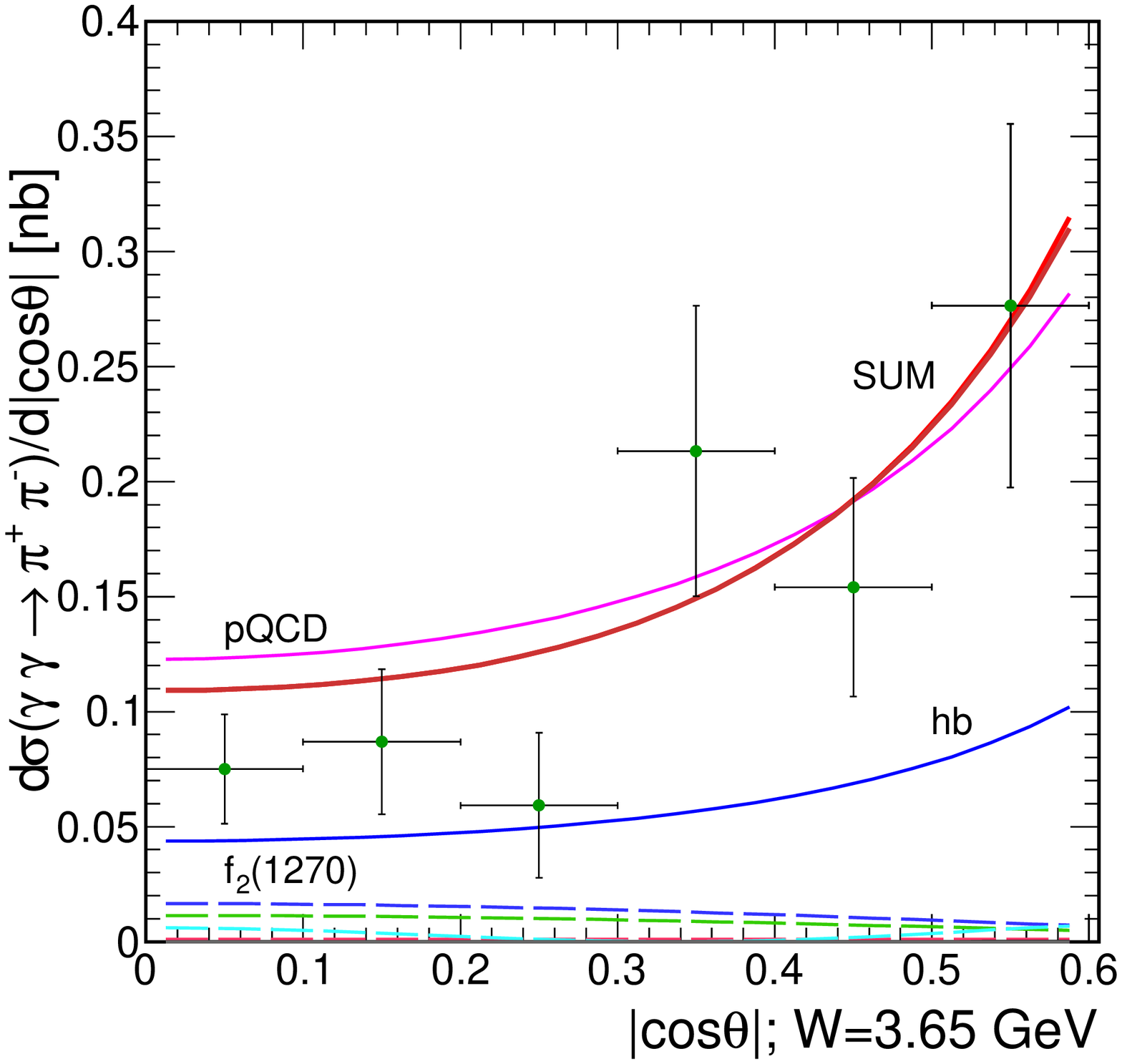}
\end{minipage}
\begin{minipage}[t]{0.31\textwidth}
\centering
\includegraphics[width=1.05\textwidth]{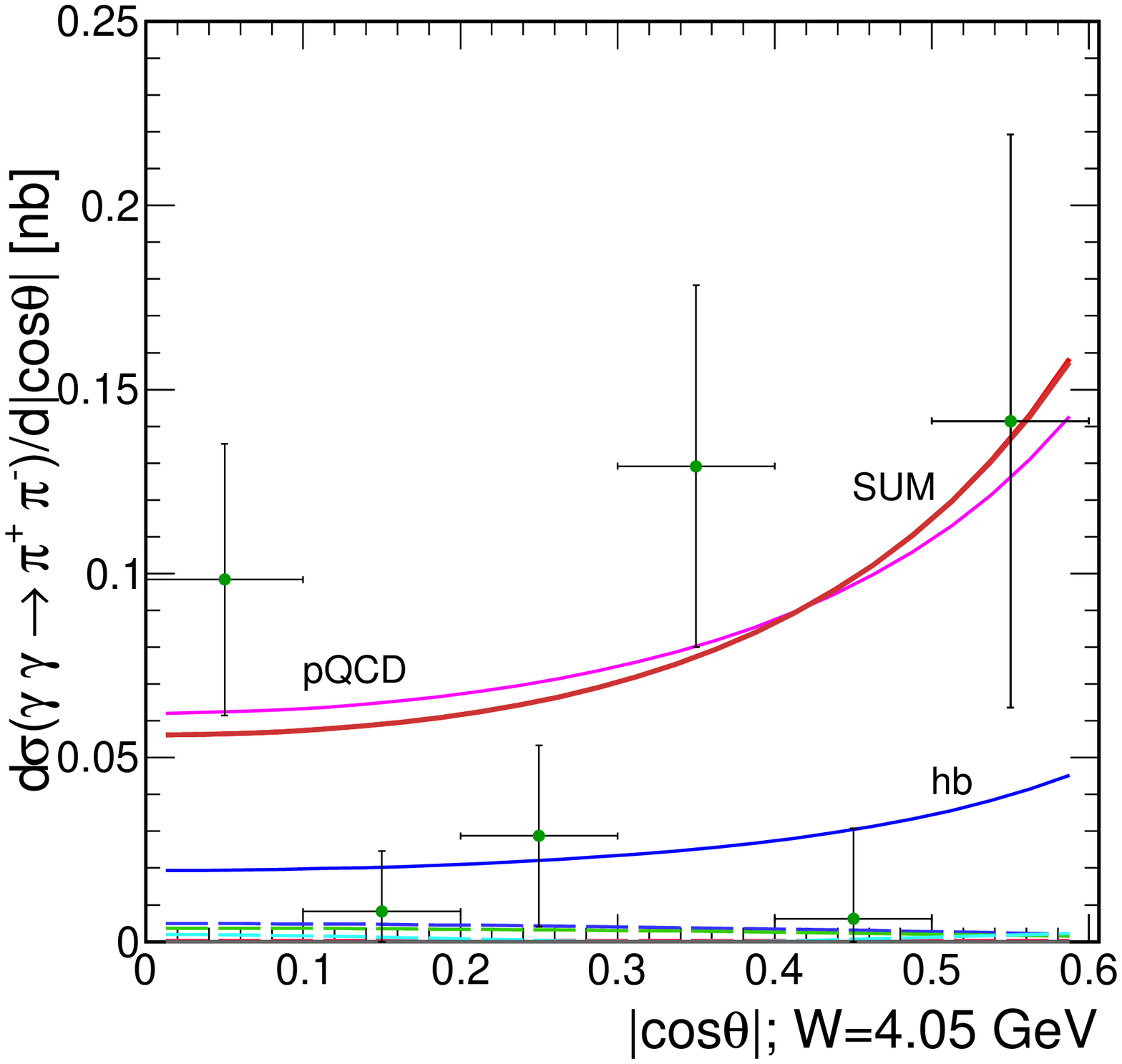}
\end{minipage}
   \caption{\label{fig:dsig_dz_pippim}
   \small Examples of angular distributions for the $\gamma\gamma\to\pi^+\pi^-$ reaction.}
\end{figure}

\begin{figure}[!h]             
\begin{minipage}[t]{0.31\textwidth}
\centering
\includegraphics[width=1.05\textwidth]{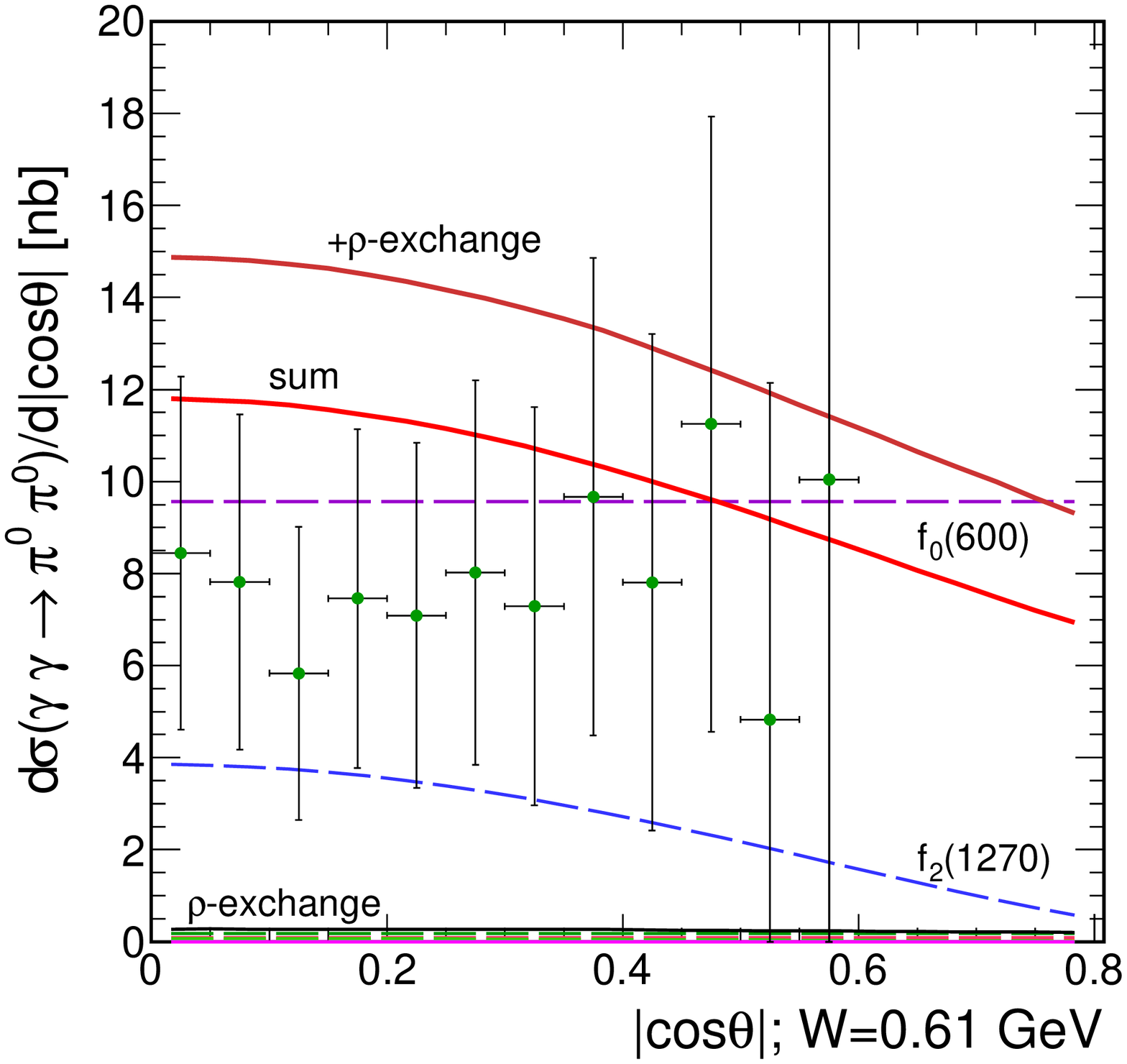}
\end{minipage}
\begin{minipage}[t]{0.31\textwidth}
\centering
\includegraphics[width=1.05\textwidth]{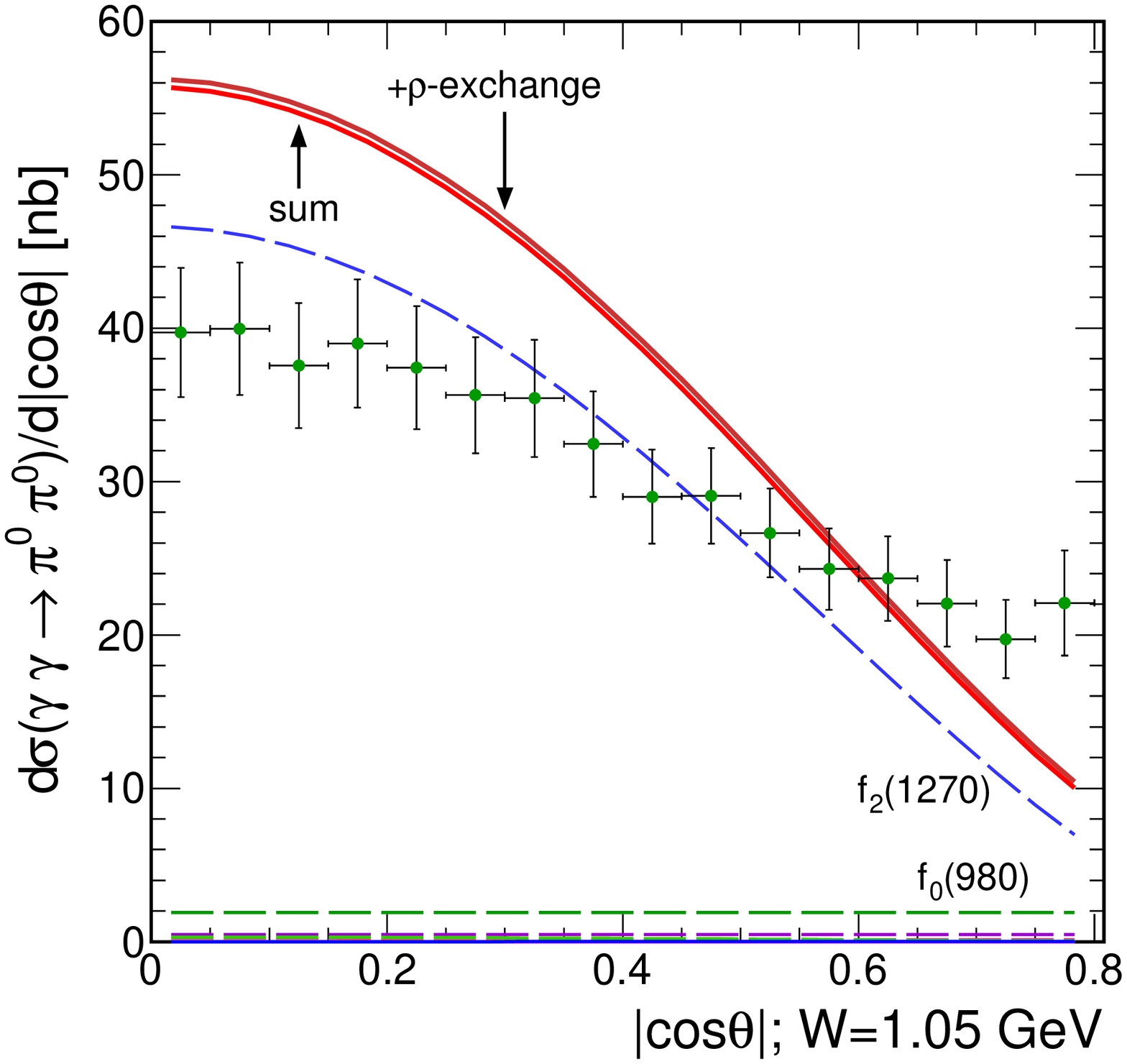}
\end{minipage}
\begin{minipage}[t]{0.31\textwidth}
\centering
\includegraphics[width=1.05\textwidth]{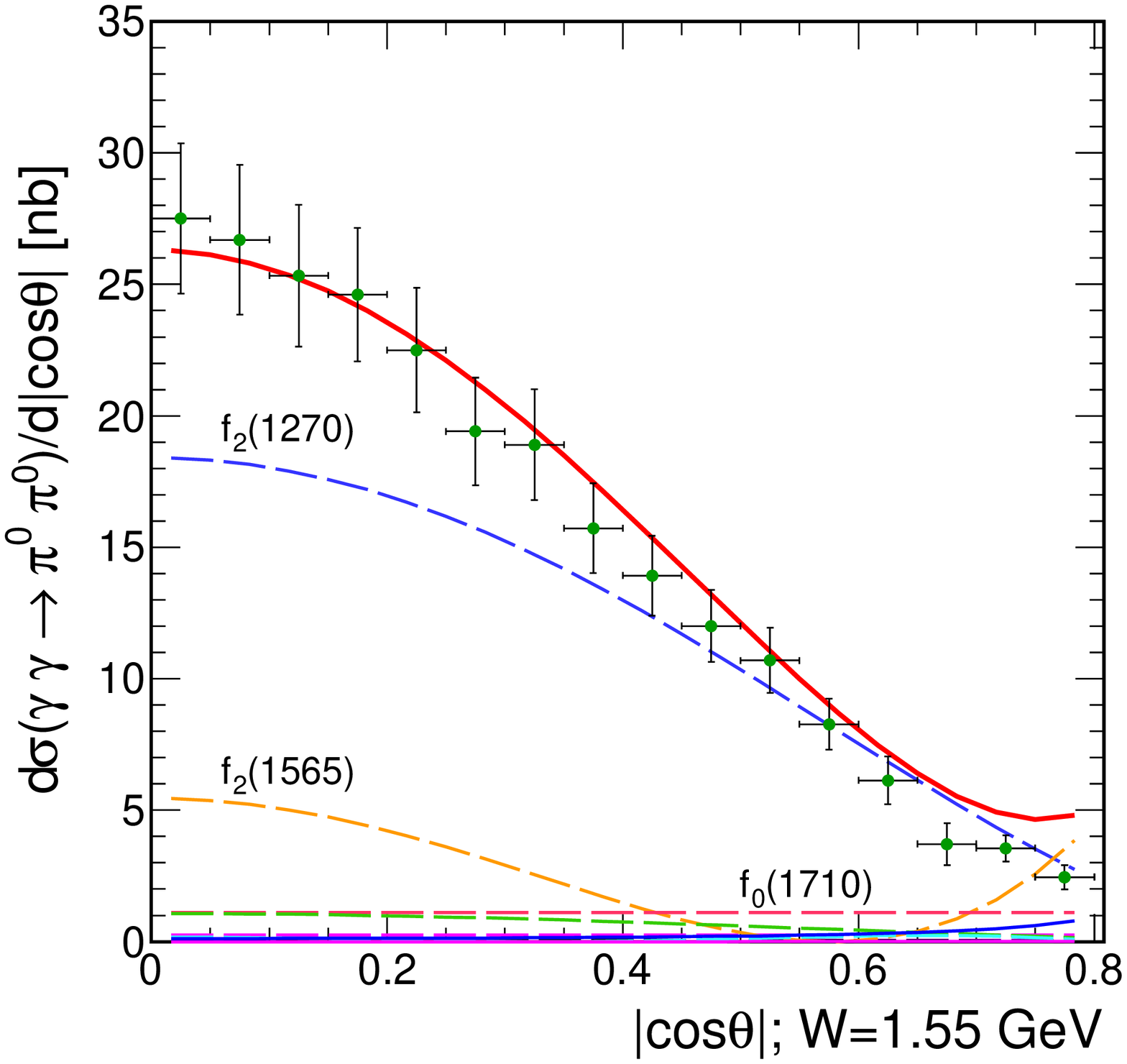}
\end{minipage}
\begin{minipage}[t]{0.31\textwidth}
\centering
\includegraphics[width=1.05\textwidth]{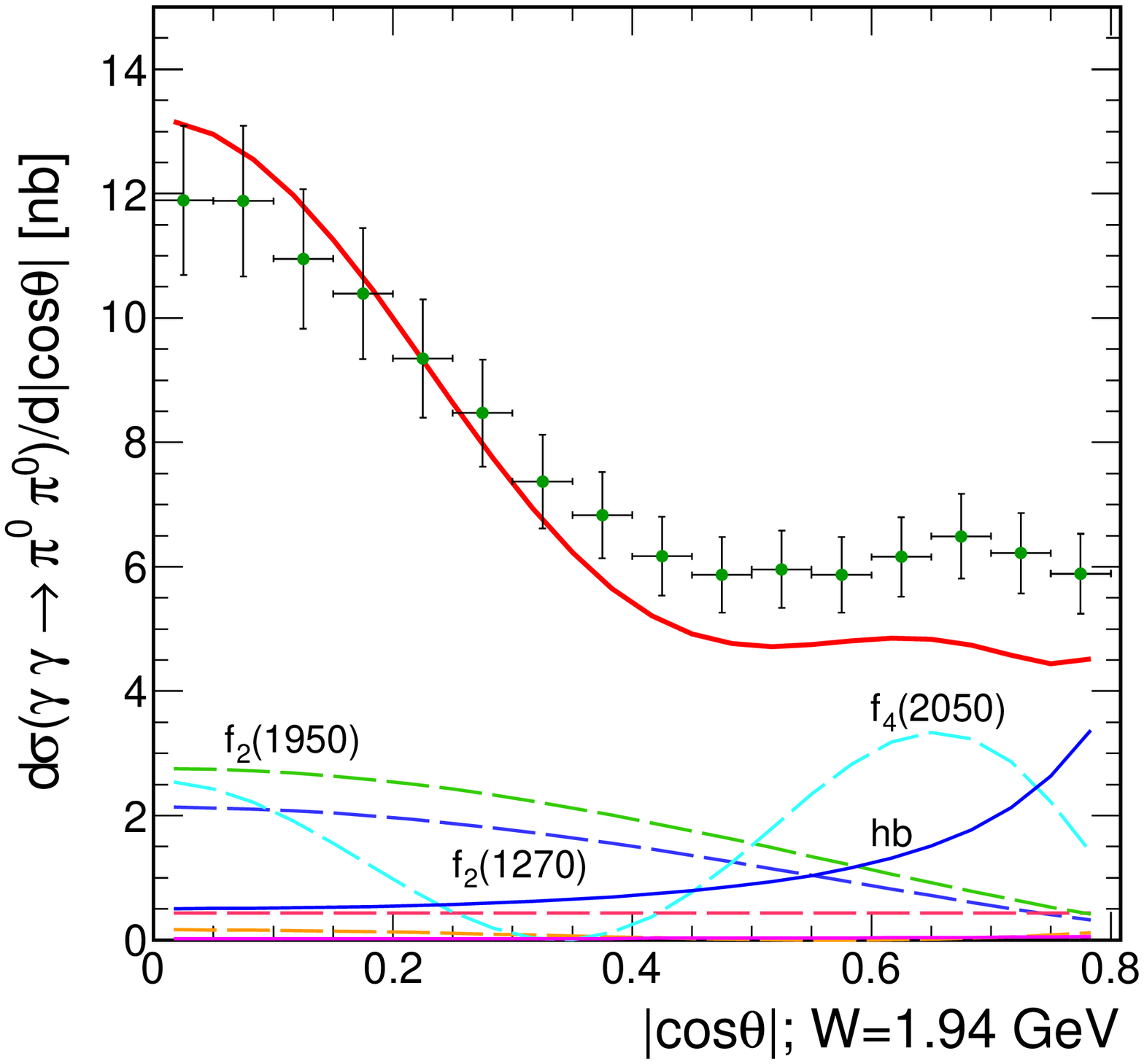}
\end{minipage}
\begin{minipage}[t]{0.31\textwidth}
\centering
\includegraphics[width=1.05\textwidth]{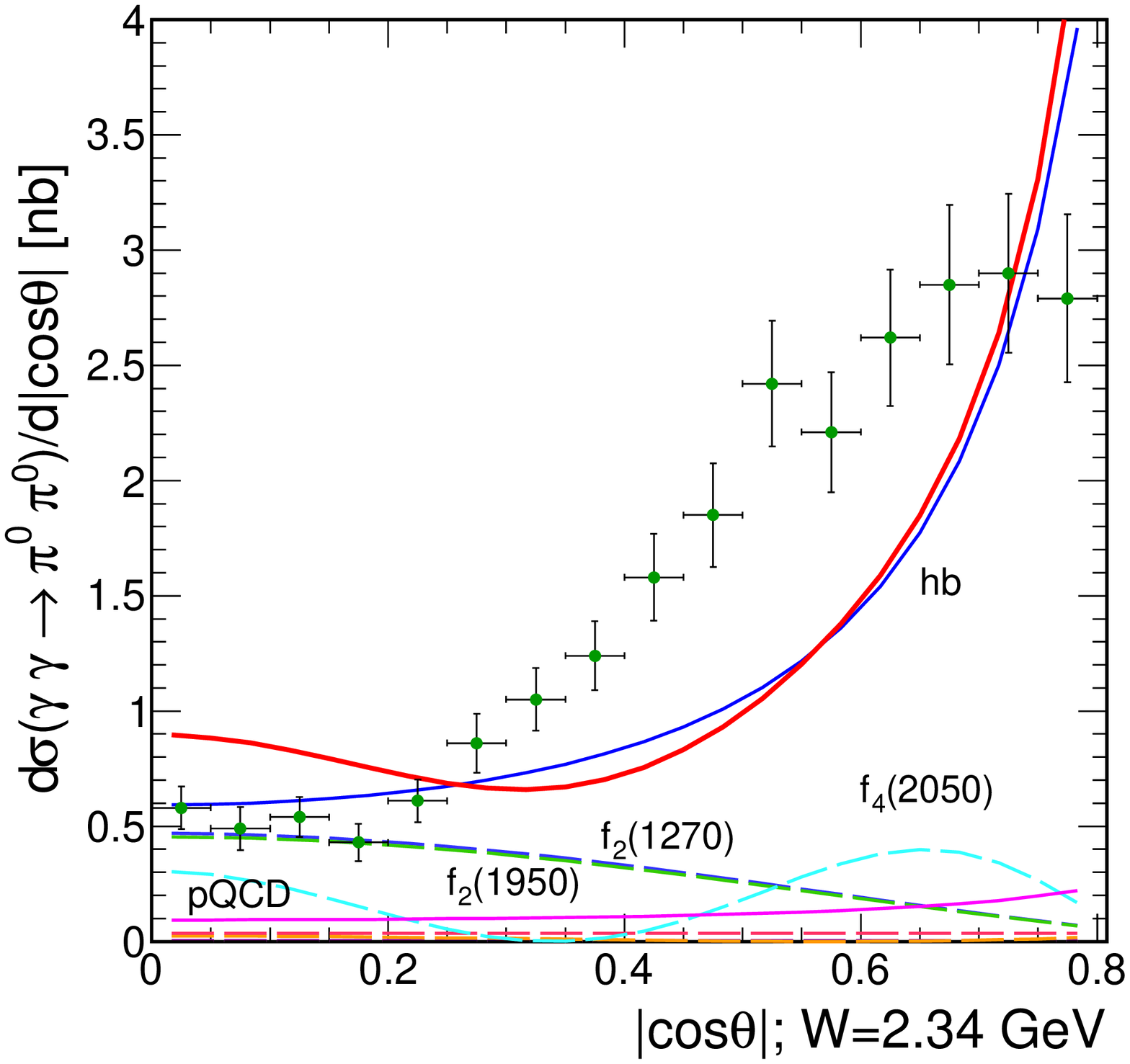}
\end{minipage}
\begin{minipage}[t]{0.31\textwidth}
\centering
\includegraphics[width=1.05\textwidth]{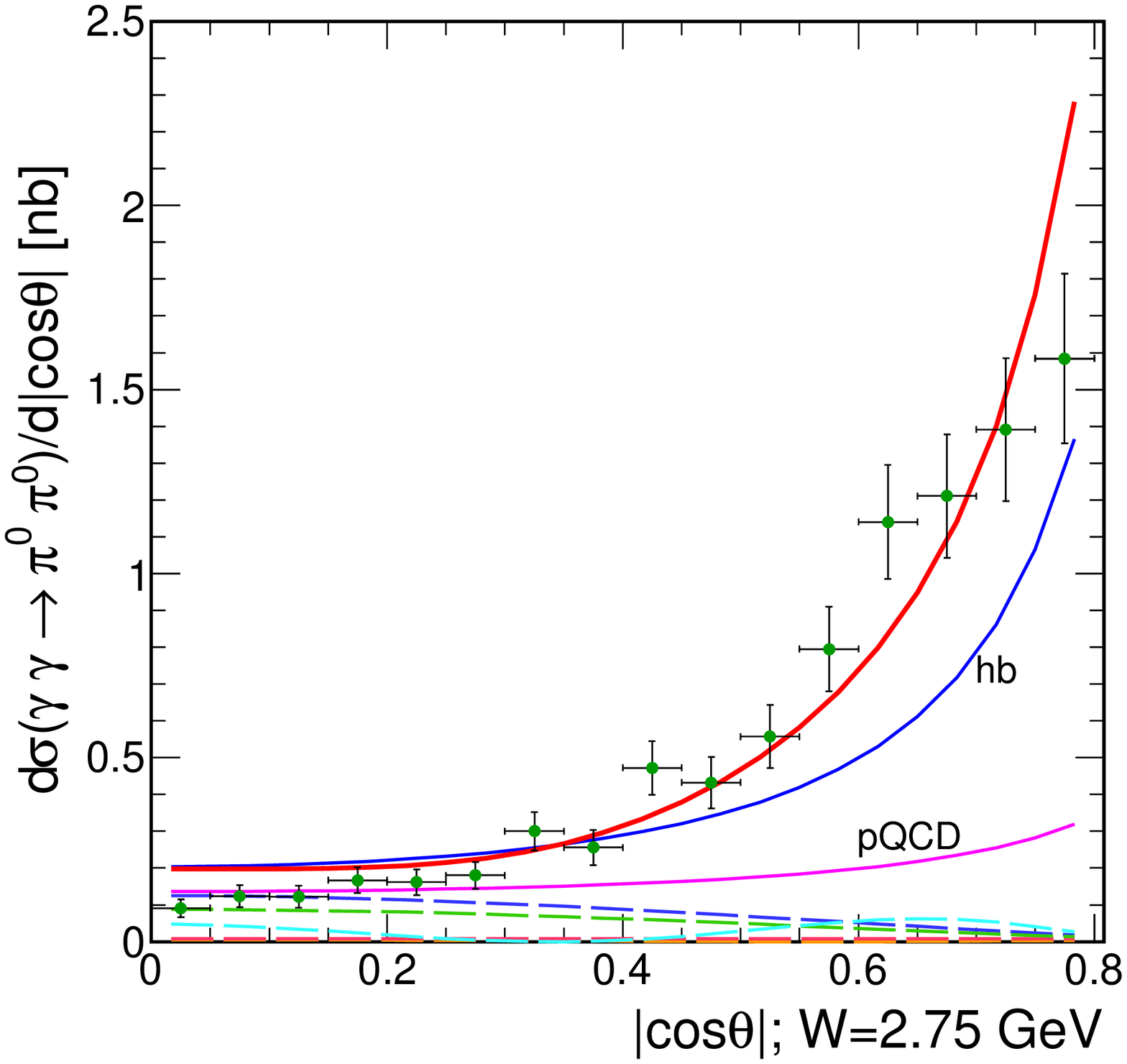}
\end{minipage}
\begin{minipage}[t]{0.31\textwidth}
\centering
\includegraphics[width=1.05\textwidth]{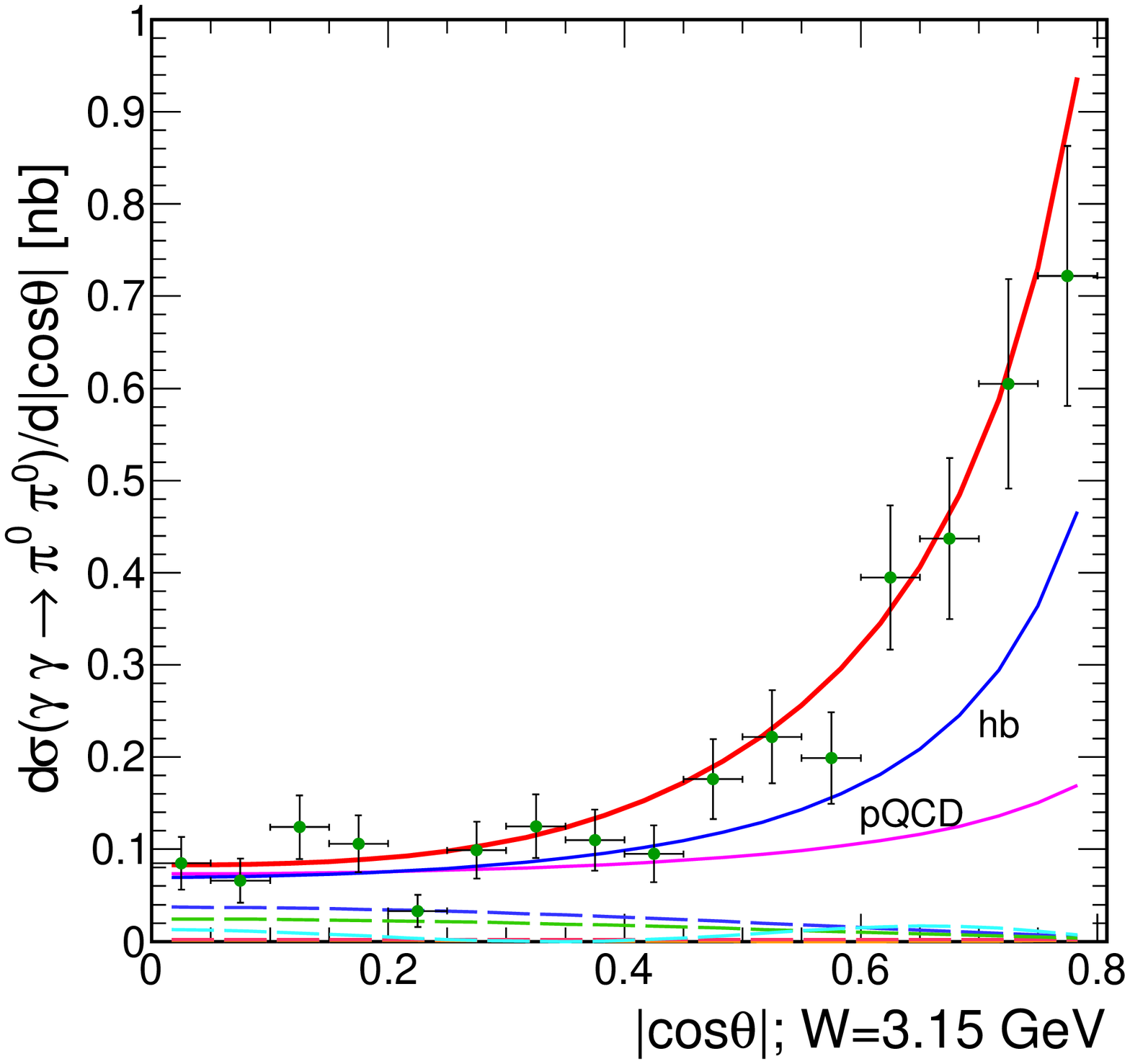}
\end{minipage}
\begin{minipage}[t]{0.31\textwidth}
\centering
\includegraphics[width=1.05\textwidth]{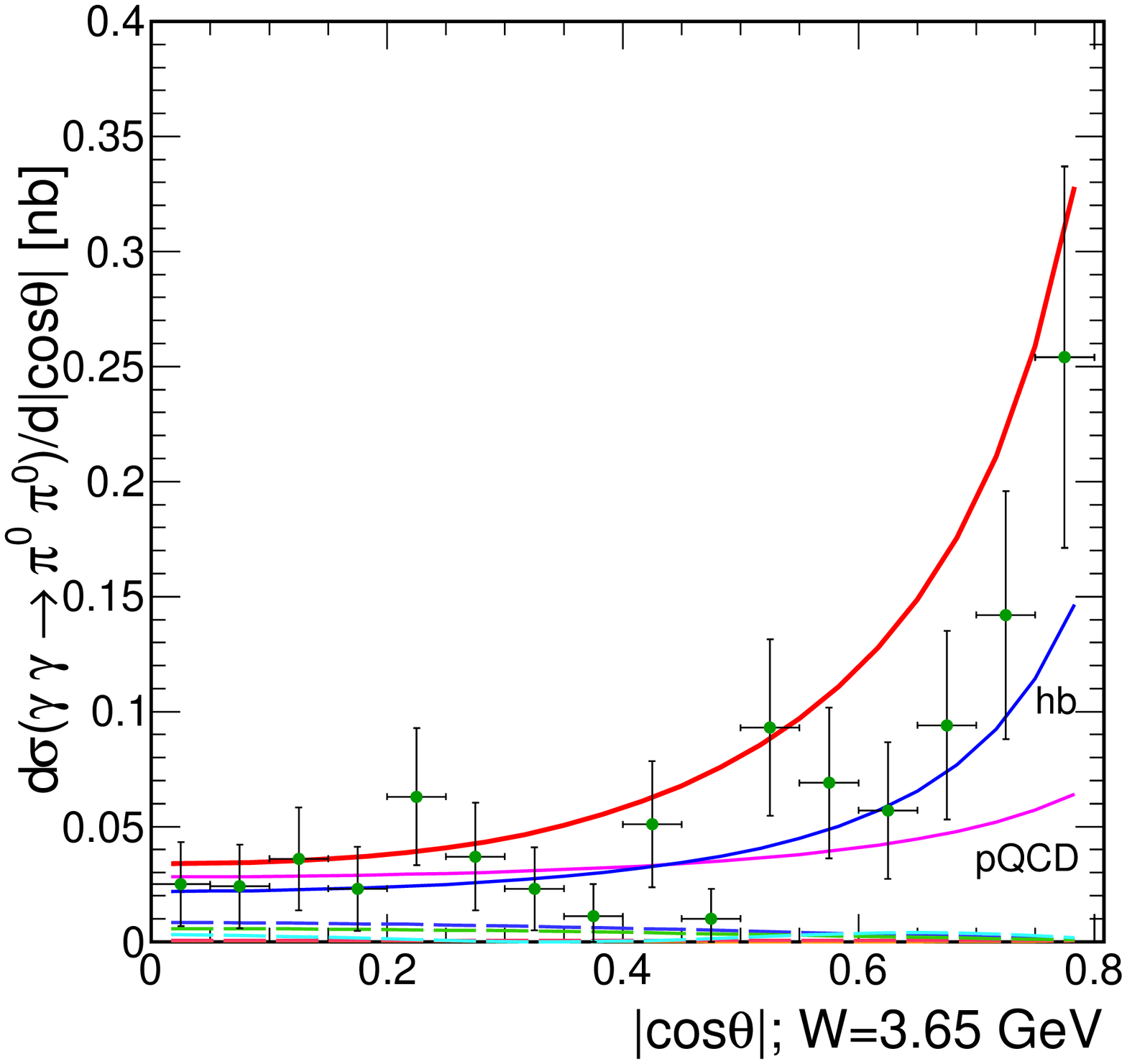}
\end{minipage}
\begin{minipage}[t]{0.31\textwidth}
\centering
\includegraphics[width=1.05\textwidth]{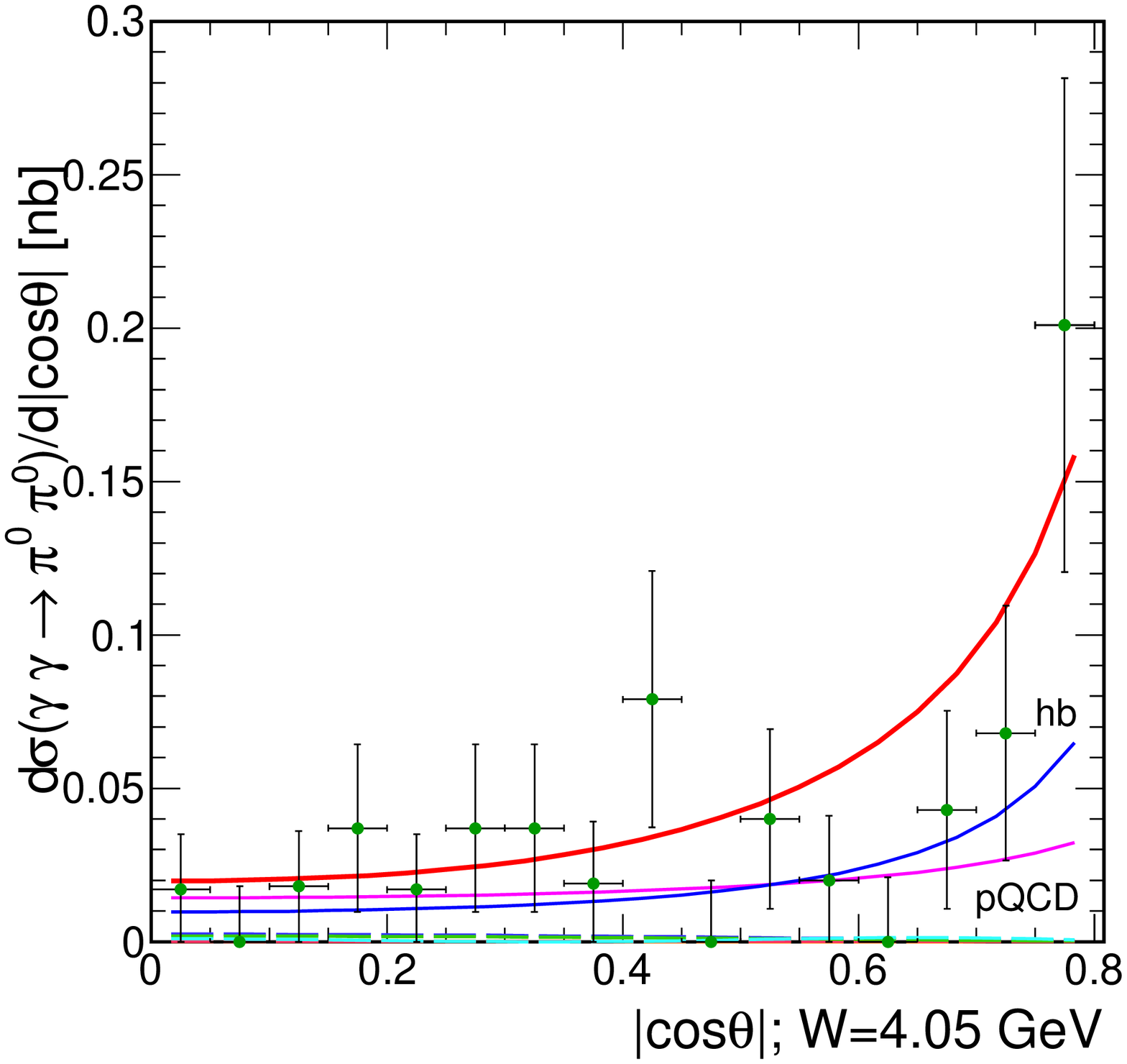}
\end{minipage}
   \caption{\label{fig:dsig_dz_pi0pi0}
   \small Examples of angular distributions for the $\gamma\gamma\to\pi^0\pi^0$ reaction.}
\end{figure}
%
\begin{figure}[!h]             
\begin{minipage}[t]{0.45\textwidth}
\centering
\includegraphics[width=1.05\textwidth]{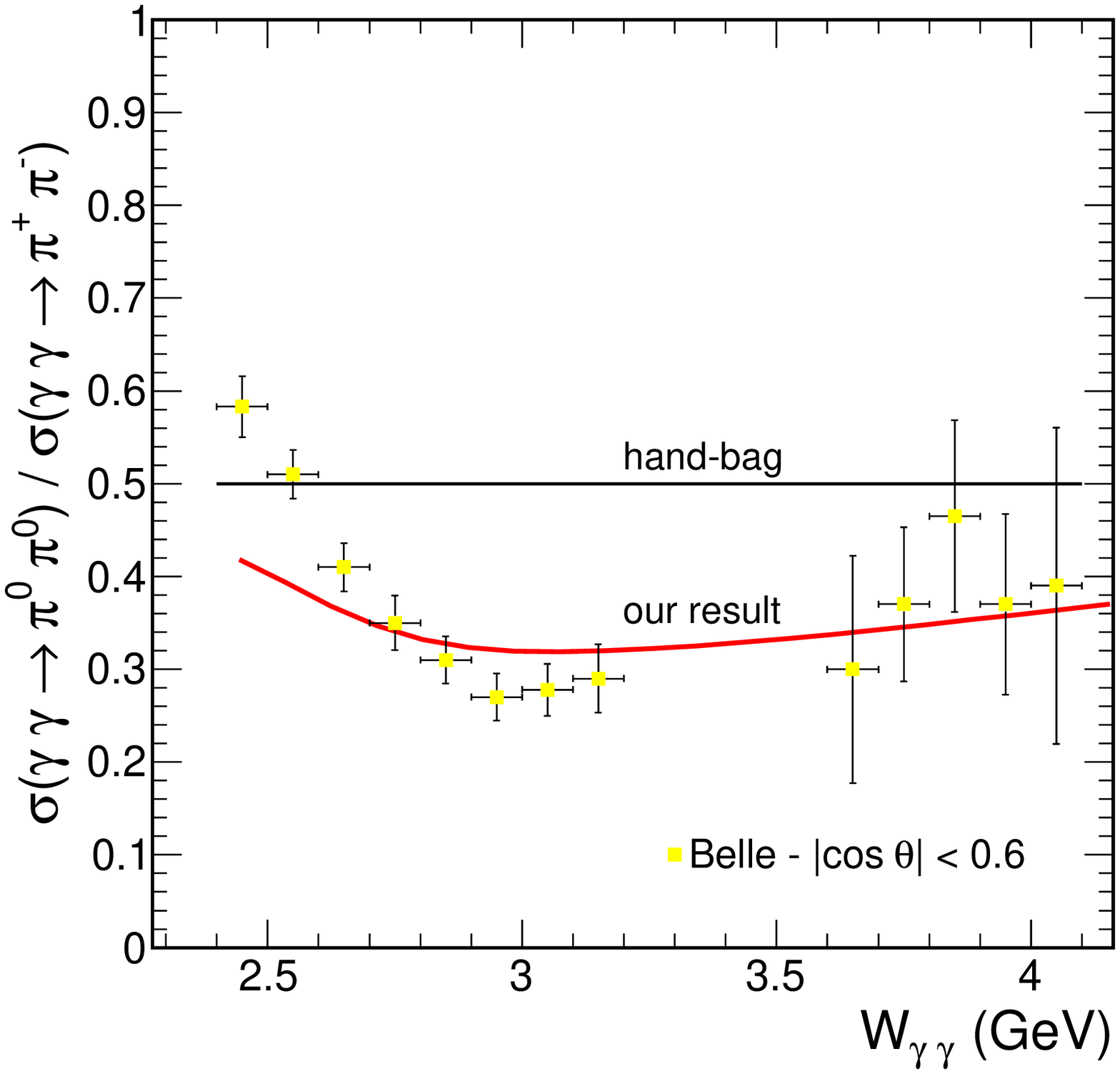}
\end{minipage}
	\caption{\label{fig:ratio}
	\small The ratio of the $\gamma\gamma\to\pi^0\pi^0$
	and $\gamma\gamma\to\pi^+\pi^-$ cross sections as a function of 
	$W_{\gamma\gamma}=\sqrt{s_{\gamma\gamma}}$.}
\end{figure}
%
In Fig.\ref{fig:dsig_dz_pippim} and \ref{fig:dsig_dz_pi0pi0}
we compare our predictions to the angular experimental data 
measured by the Belle Collaboration. 
Experimental data points are taken from Refs.
\cite{Belle_Mori,Belle_Nakazawa} 
for $\gamma\gamma\to\pi^+\pi^-$ 
and from Ref.\cite{Belle_Uehara09} 
for $\gamma\gamma\to\pi^0\pi^0$ reactions.
In Fig.~\ref{fig:dsig_dz_pippim} (the first four panels)
we show the sum of contributions of many mechanisms for 
two different values of the slope parameter 
in the hadronic form factors (see below Eq.~(\ref{eq:ff_hadronic})).
We show how the final result depends on the value of this parameter.
In Ref.\cite{SS03} the results for $B<$ 4 GeV$^{-2}$ were shown, but here
we obtain a better description of the data when we use 
the exponential form factor with $B_{\gamma\pi}$=6 GeV$^{-2}$.
The description for $W=$1.4975 GeV shows that one should include
also $D$ wave, but we have no theoretical guidance how to do it.
For larger $W$, the contribution from pQCD mechanisms starts to dominate.
In Fig.\ref{fig:dsig_dz_pi0pi0} we show the angular distribution 
for $\gamma\gamma\to\pi^0\pi^0$ process for nine different energies
ranging from 0.61 GeV to 4.05 GeV. Here we use the same parameters 
of resonances (masses, widths, phases) as 
for the $\gamma\gamma\to\pi^+\pi^-$ reaction.
Here we have a small problem with good description for two energies 
(1.05 and 2.34 GeV). In addition, we show the 
cross section corresponding to the sum of processes (amplitudes)
with and without 
contribution of the $\rho^\pm$-meson exchange.
We can observe that the inclusion of this contribution is important 
only for $W<$ 0.6 GeV (see. Fig.\ref{fig:sig_W_sum}, right panel).   
In conclusion, we have obtained a reasonable description 
of the angular experimental data, which results in excellent description 
of the total cross section (see Fig.\ref{fig:sig_W_sum}).

In Fig.\ref{fig:ratio} we present the ratio of the neutral
and charged pion pair cross sections. The upper line represents
the hand-bag model result, which is independent of $\theta$
and is equal to $\frac{1}{2}$. Our result, which includes the BL pQCD and 
hand-bag contribution simultaneously, describes the experimental data
measured by the Belle Collaboration \cite{Belle_Nakazawa, Belle_Uehara09}.
We have a problem with correct description of the data 
at $W =$ (2.4-2.7) GeV. We think that this is due to
not exact description of angular experimental data 
for $\gamma\gamma\to\pi^0\pi^0$ reaction in this range of energy.
However, our result shows that for the correct description of
high energy cross section it is necessary to take into account
also the BL pQCD contribution. 

\begin{figure}[!h]             
\begin{minipage}[t]{0.44\textwidth}
\centering
\includegraphics[width=1.05\textwidth]{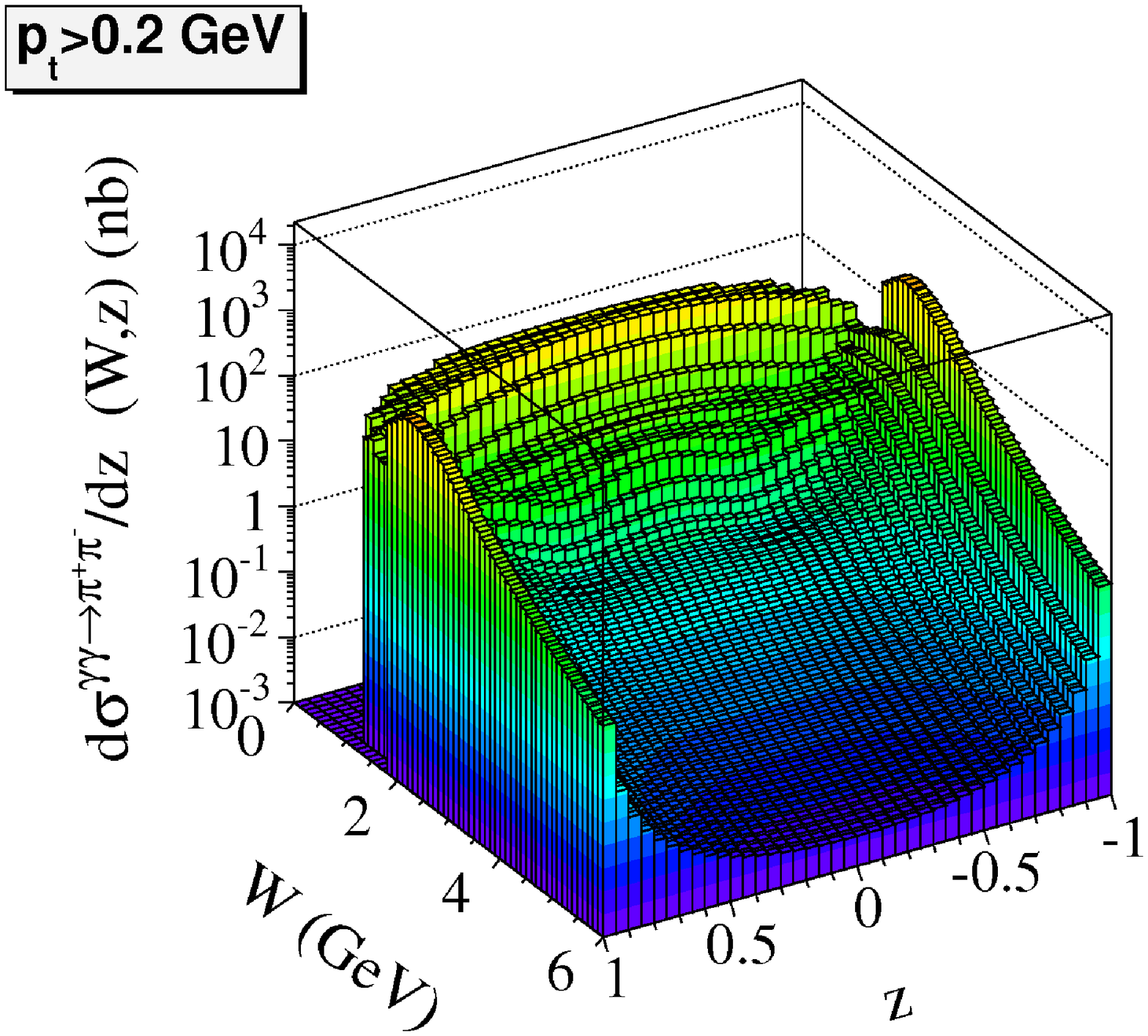}
\end{minipage}
\begin{minipage}[t]{0.44\textwidth}
\centering
\includegraphics[width=1.05\textwidth]{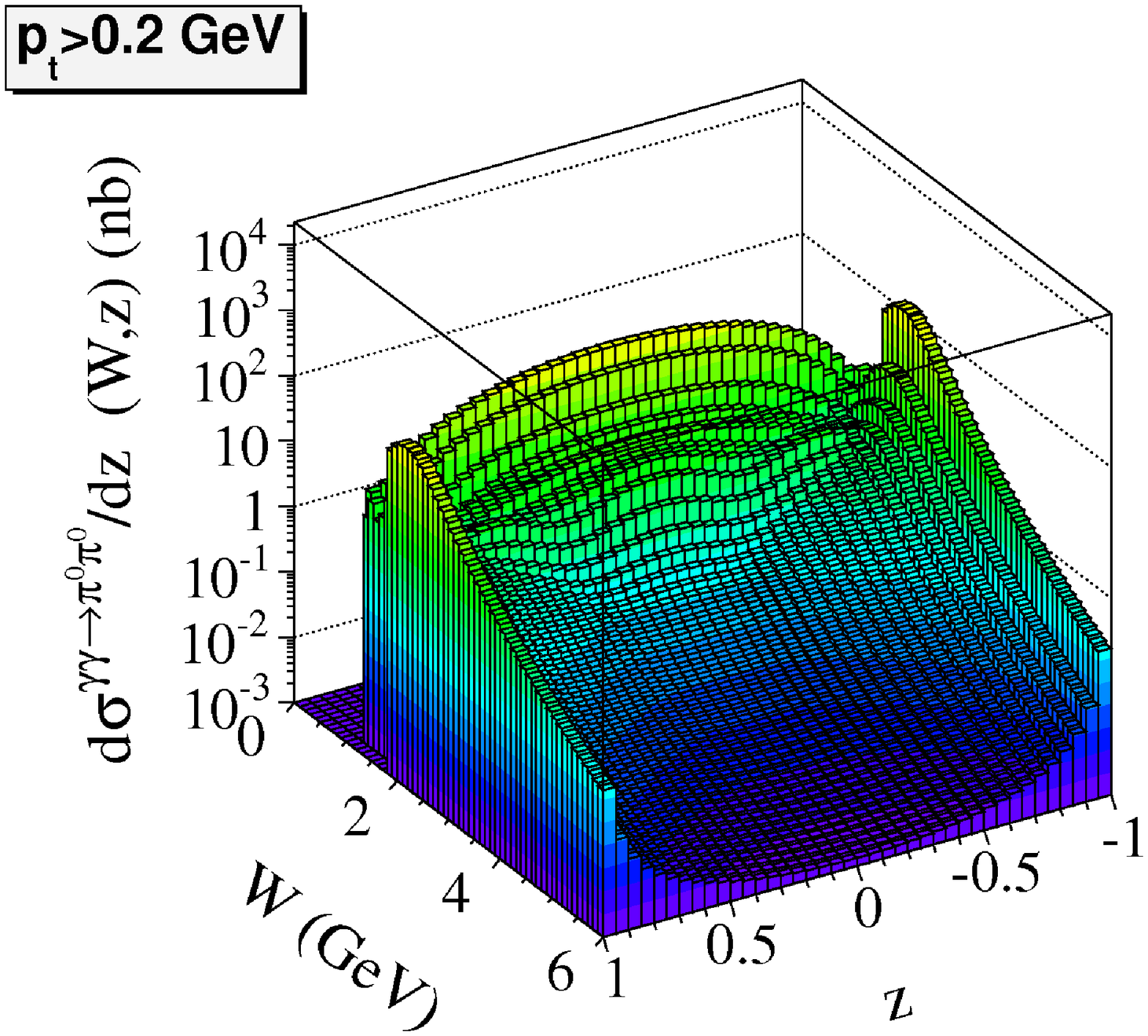}
\end{minipage}
\begin{minipage}[t]{0.44\textwidth}
\centering
\includegraphics[width=1.05\textwidth]{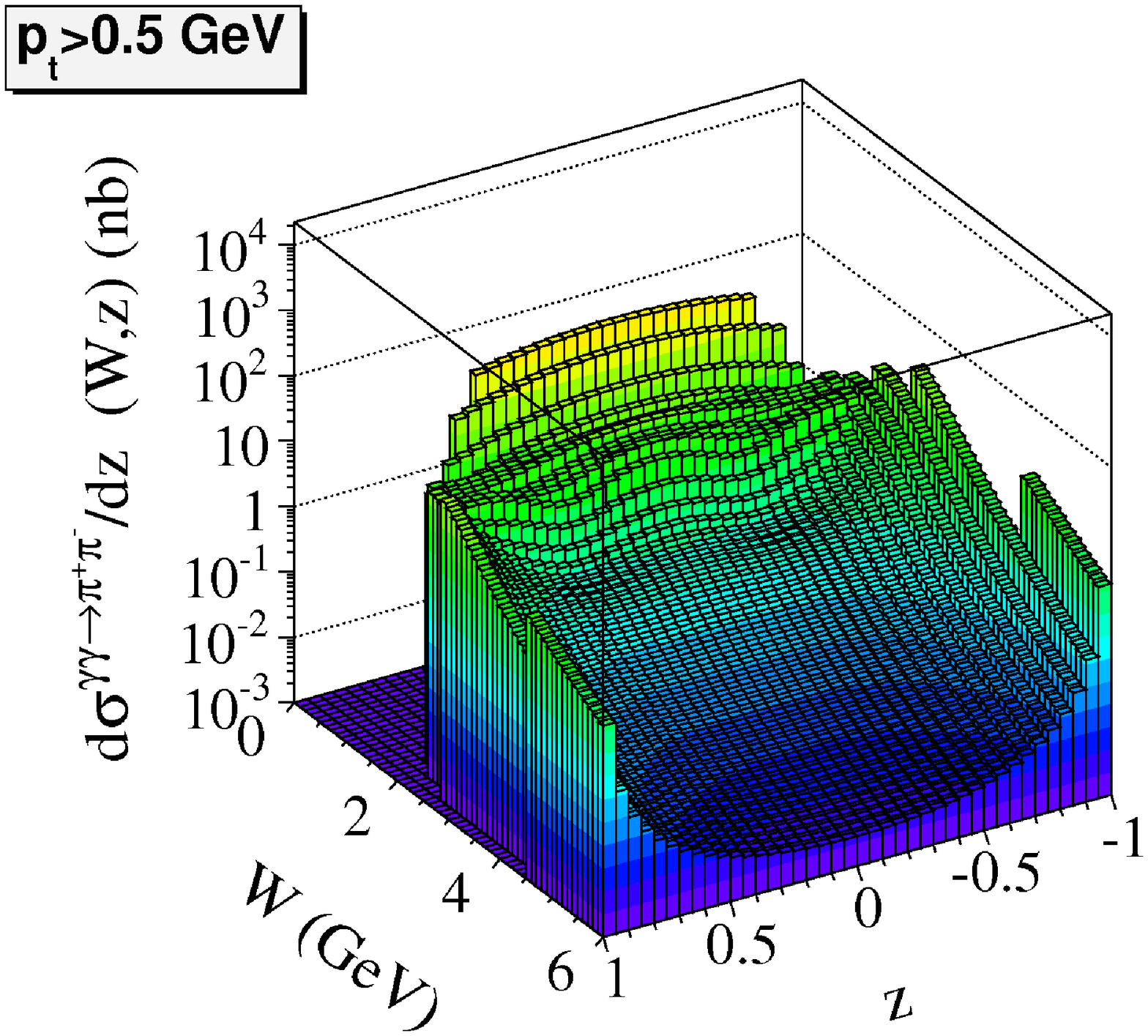}
\end{minipage}
\begin{minipage}[t]{0.44\textwidth}
\centering
\includegraphics[width=1.05\textwidth]{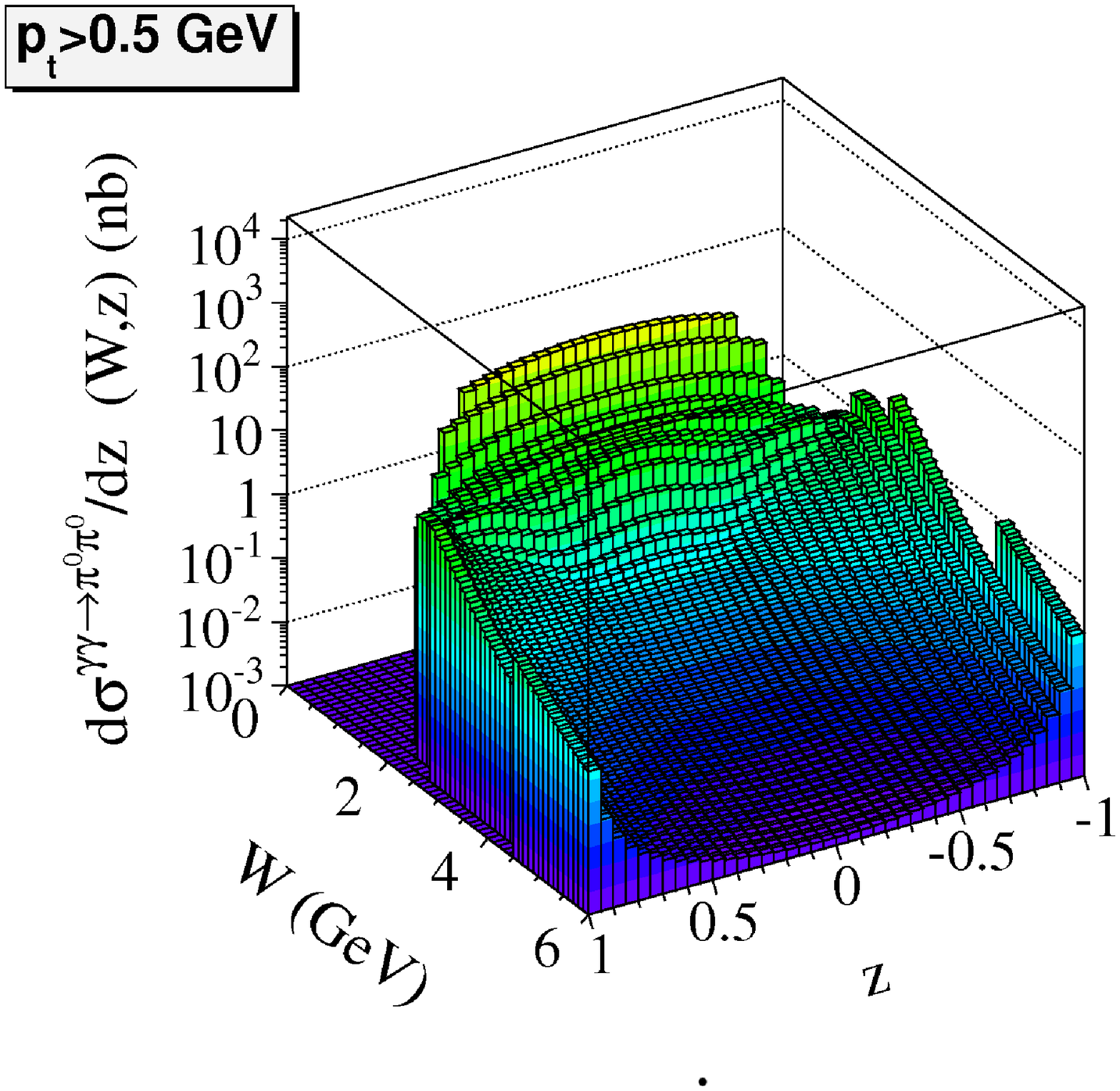}
\end{minipage}
   \caption{\label{fig:paw_Wz}
   \small  $\frac{d \sigma}{dz}(W,z)$ for $\gamma\gamma\to\pi^+\pi^-$ (left) 
    and $\gamma\gamma\to\pi^0\pi^0$ (right) for two different
    cuts on pion transverse momenta: $p_{t,\pi} >$ 0.2 GeV (top) and $p_{t,\pi} >$ 0.5 GeV (bottom).}
\end{figure}

In Fig.\ref{fig:paw_Wz} we summarize our elementary cross sections
as a function of $\sqrt{s}_{\gamma \gamma}$ and $\cos\theta$.
Here transverse momentum cuts were imposed as explained in 
the figure caption. The lower cut roughly corresponds to the lower cut
of the ALICE experiment. It is not clear for us if such a cut
is sufficient to regularize the calculations of the BL pQCD and hand-bag
mechanisms. Therefore we present also a result with slightly bigger cut.
Another option would be to cut in $\cos\theta$ as is usually done
at $e^+ e^-$ colliders.
The irregularities observed at z $\approx \pm$ 1 are caused
by small number of bins in the figure.
Such two-dimensional distributions as shown in Fig.\ref{fig:paw_Wz} are then used 
in nuclear calculations discussed in the next subsection.

%
\subsection{$Pb Pb \to Pb Pb \pi \pi$}
\begin{figure}[!h]             
\begin{minipage}[t]{0.45\textwidth}
\centering
\includegraphics[width=1.05\textwidth]{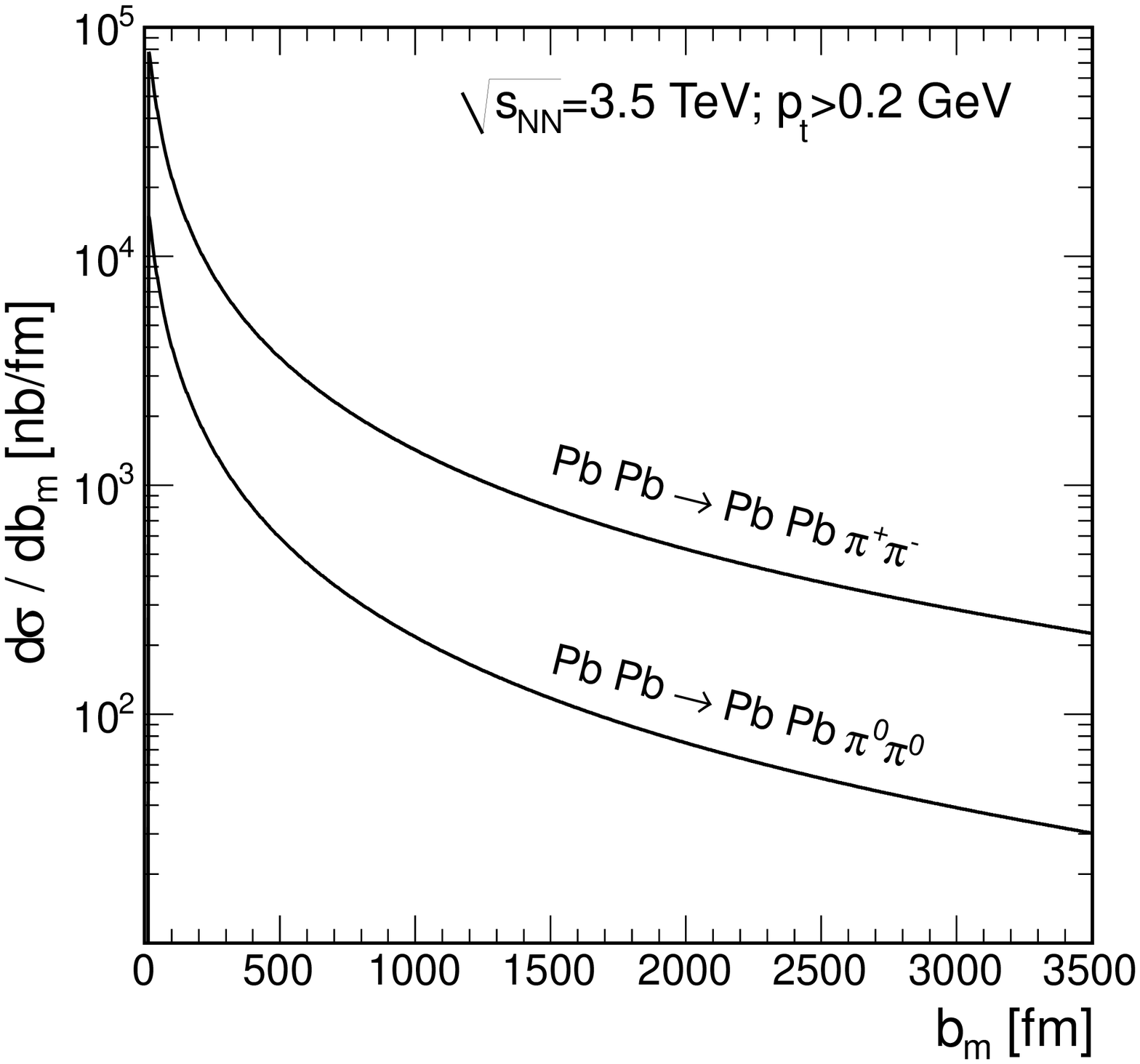}
\end{minipage}
	\caption{\label{fig:dsig_dbm}
	\small The impact parameter distribution for 
the $Pb Pb \to Pb Pb \pi^+ \pi^-$ and $Pb Pb \to Pb Pb \pi^0 \pi^0$
reactions at $\sqrt{s_{NN}} = $ 3.5 TeV.}
\end{figure}

%
In order to check our nuclear calculation, in Fig.\ref{fig:dsig_dbm}
we show interesting distribution in impact parameter between
the two lead nuclei. The distribution is different from zero 
starting from the distance
of $b = 2 R_A \approx$ 14 fm and extends till ''infinity''. 
This means that a big part of the cross section comes from the situations 
when the two $^{208}Pb$ nuclei fly by very far one from each other. This
figure clearly demonstrates the ultraperipheral character of the discussed 
process of exclusive dipion production via photon-photon fusion.

%
%
\begin{table}[h!]
\caption{Cross sections for different lower cuts on pion transverse
momenta at \mbox{$\sqrt{s_{NN}}$ = 3.5 TeV}.}
\begin{center}
\begin{tabular}{|c|r|r|}
\hline
$p_{t,min}$ [GeV]  &  $\pi^+ \pi^-$ [mb]	&  $\pi^0 \pi^0$ [mb] \\
\hline
    0.2            &   46.7       		&  8.7    \\
    0.5            &   12.1       		&  5.1    \\
    1.0            &    0.08       		&  0.05    \\
\hline
\end{tabular}
\end{center}
\label{tab:total_cross_section}
\end{table}
%

The total cross sections for $\sqrt{s_{NN}}$ = 3.5 TeV are collected
for both $\pi^+ \pi^-$ and $\pi^0 \pi^0$ channels for different
lower cuts on pion transverse momentum in Table \ref{tab:total_cross_section}.
%
\begin{figure}[!h]             
\centering
\includegraphics[width=8cm]{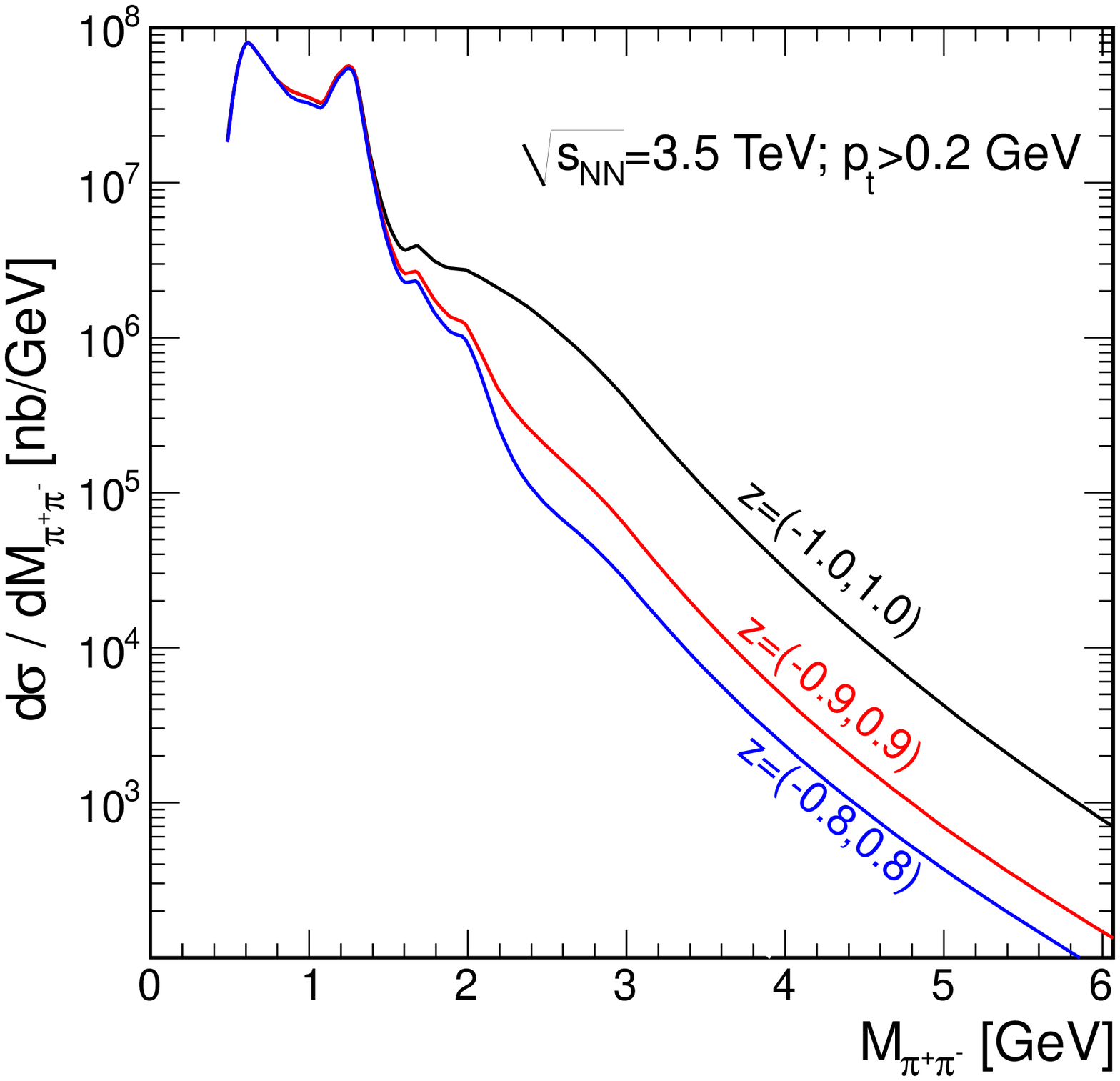}
\includegraphics[width=8cm]{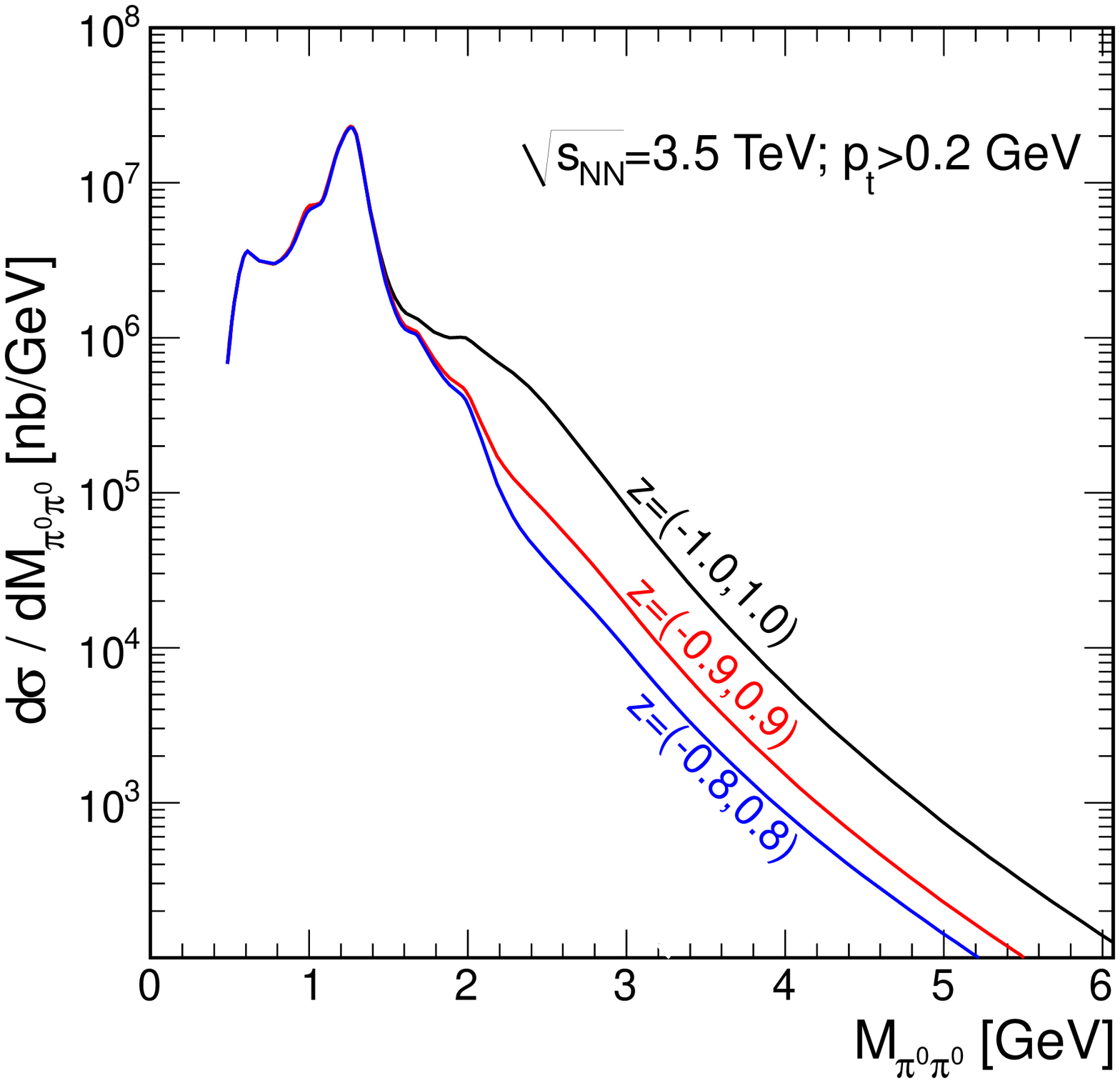}
   \caption{
   \small  Total cross section for $\gamma \gamma \to \pi^+ \pi^-$ (left
  panel) and $\gamma \gamma \to \pi^0 \pi^0$ (right panel) as a function
of the system energy at the LHC energy $\sqrt{s_{NN}} =$ 3.5 TeV
and $p_{t,\pi} >$ 0.2 GeV. We show in addition results with extra cuts on $z$.}
\label{fig:dsig_dW}
\end{figure}
%

The distribution in invariant dipion mass shown in Fig.\ref{fig:dsig_dW}
is the most interesting one. We show somewhat theoretical distribution 
for the full phase space (upper solid line).
We show distributions 
for the $PbPb \to PbPb\pi^+\pi^-$ (left panel) and 
for the $PbPb \to PbPb\pi^0\pi^0$ (right panel) reactions
for the full phase space
(upper lines), for $|\cos\theta|<$ 0.9 (middle lines) and for the angular
range corresponding to the experimental limitations usually used
for the $\pi\pi$ production in $e^+e^-$ collisions ($|\cos\theta|<$ 0.8).
At lower energies ($W<$ 1.5 GeV) the result does not depend on the angular cuts.
The big differences start in the region 
where the elementary cross section is described by the BL pQCD and hand-bag mechanisms.
%
\begin{figure}[!h]             
\centering
\includegraphics[width=8cm]{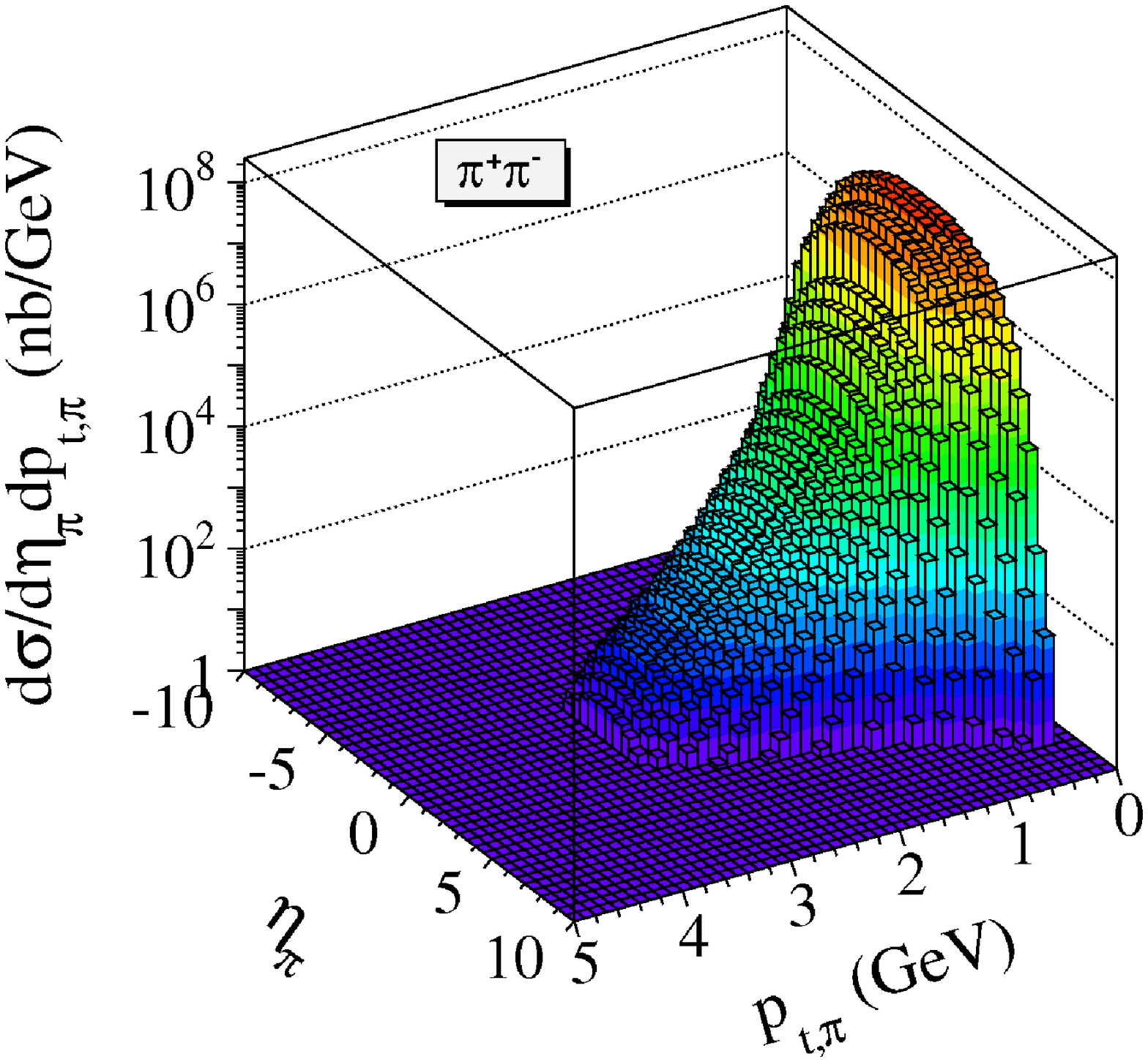}
\includegraphics[width=8cm]{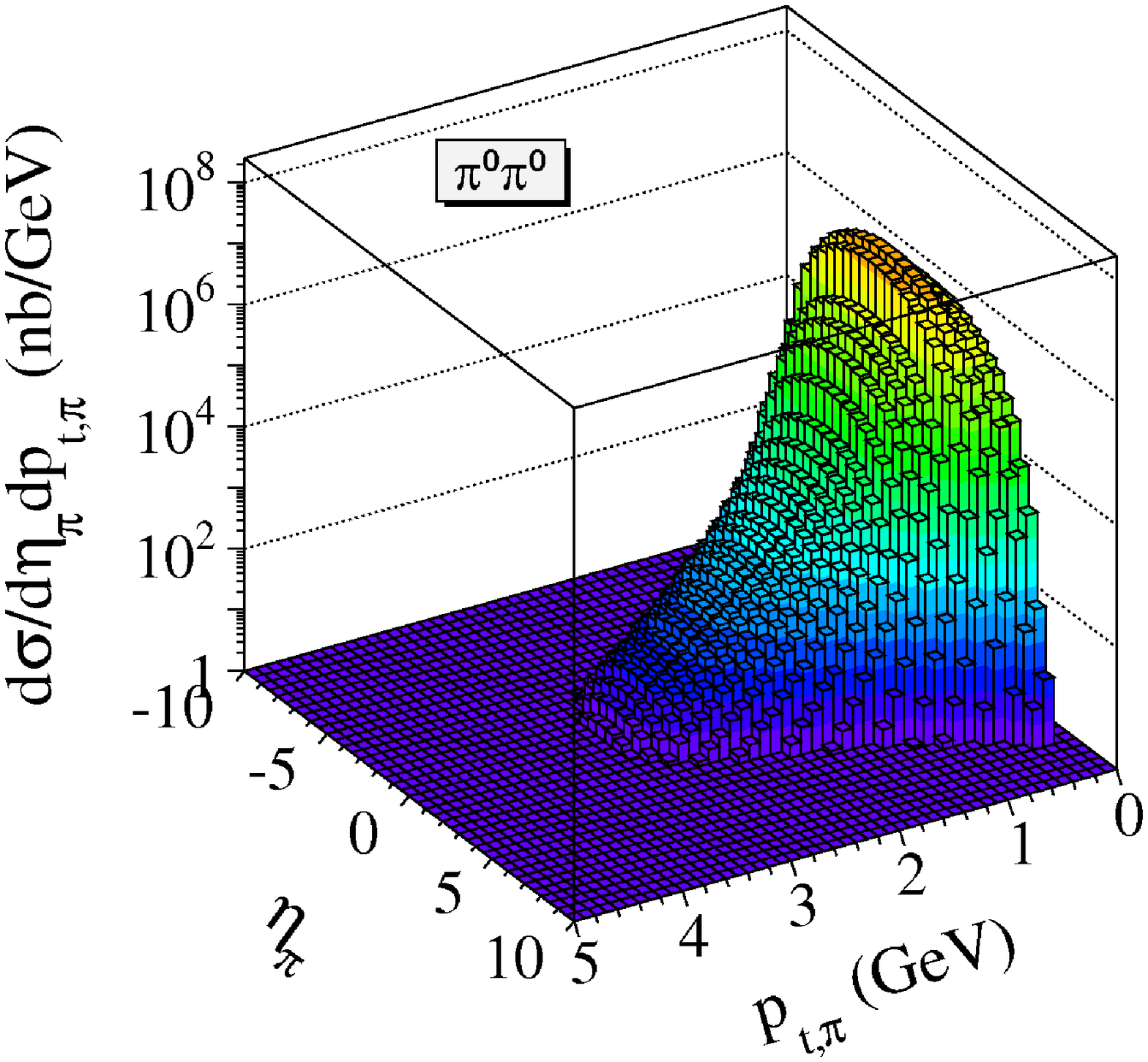}
   \caption{\label{fig:map_eta1pt}
   \small  $\frac{d\sigma}{d\eta_\pi dp_{t,\pi}}$ for 
   the $\mbox{$^{208}Pb$} + \mbox{$^{208}Pb$} \to \mbox{$^{208}Pb$} \mbox{$^{208}Pb$} \pi^+ \pi^-$ reaction (left panel) and
   the $\mbox{$^{208}Pb$} + \mbox{$^{208}Pb$} \to \mbox{$^{208}Pb$} \mbox{$^{208}Pb$} \pi^0 \pi^0$ reaction (right panel) 
   at the LHC energy $\sqrt{s_{NN}} =$ 3.5 TeV and \mbox{$p_{t,\pi} >$ 0.2 GeV}.}
\end{figure}

Fig.\ref{fig:map_eta1pt} shows two-dimensional distributions 
in pseudorapidity of charged pion (left panel)
or pseudorapidity of neutral pion (right panel)
and transverse momentum of one of the pions.
With higher $p_{t,\pi}$ values, 
the pseudorapidity distribution becomes somewhat narrower.
%
\begin{figure}[!h]             
\centering
\includegraphics[width=8cm]{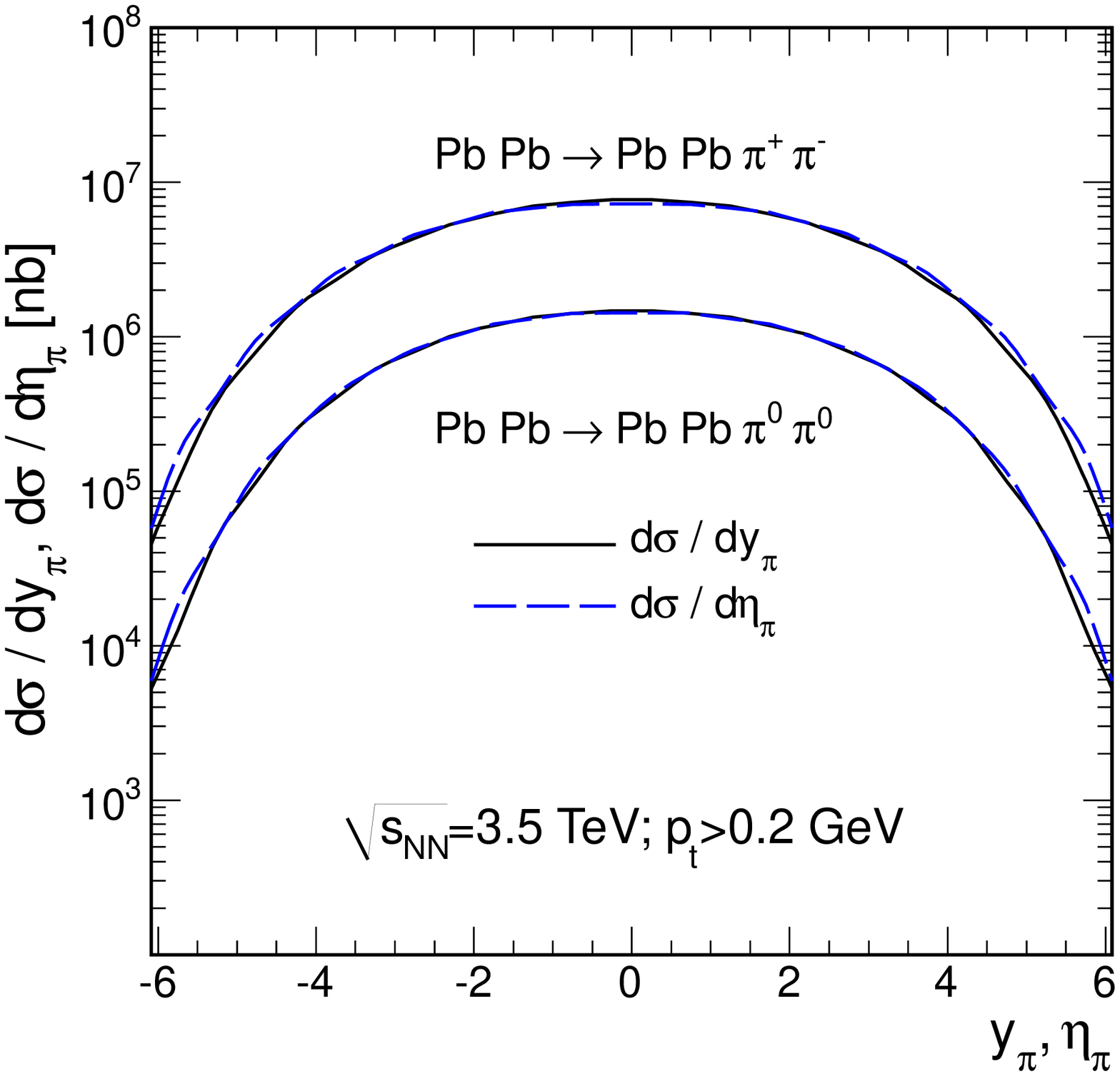}
\includegraphics[width=8cm]{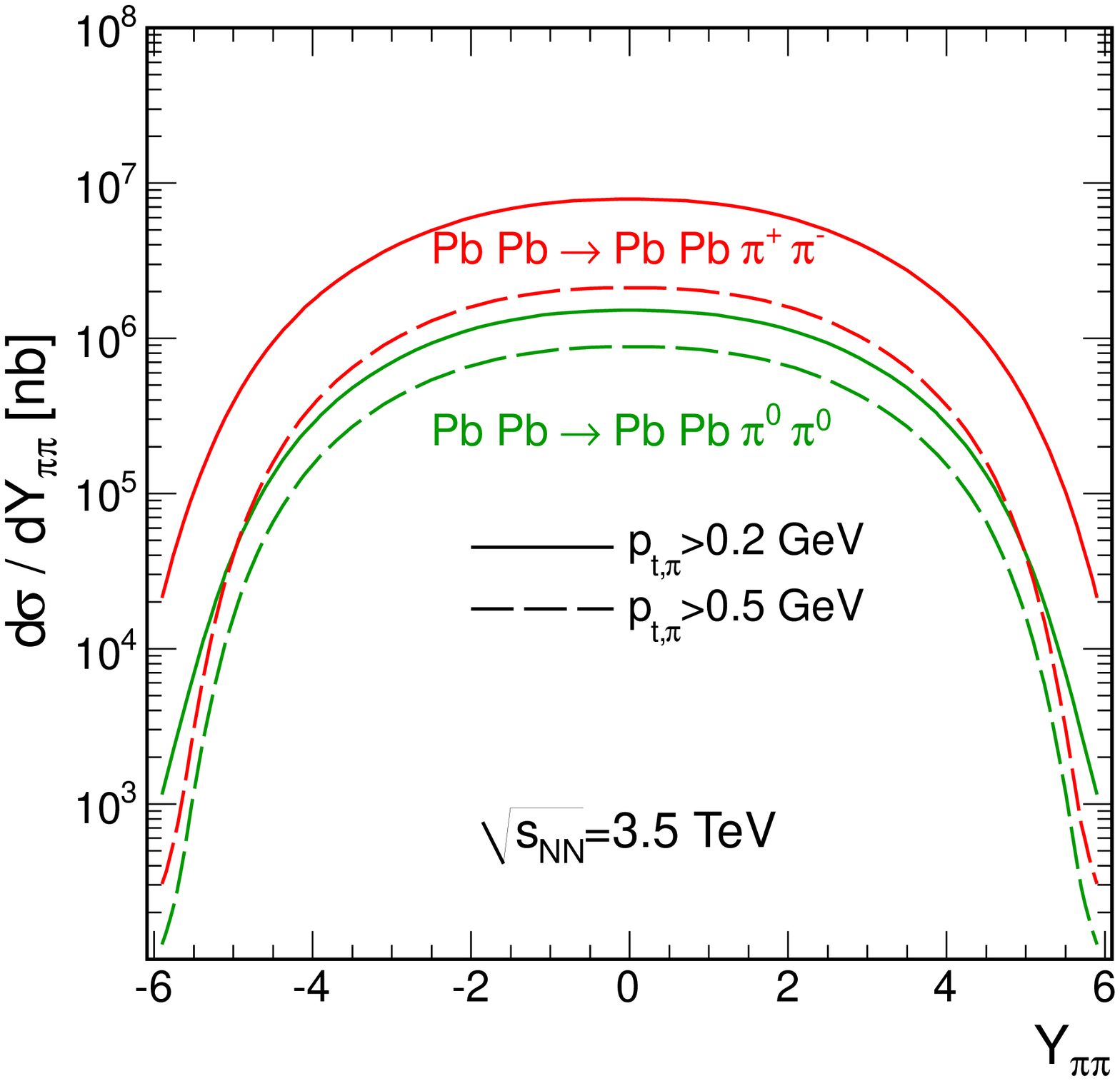}
   \caption{\label{fig:dsig_dy_deta}
   \small  $\frac{d\sigma}{dy_{\pi}}$ and $\frac{d\sigma}{d\eta_{\pi}}$ (left panel)
       and $\frac{d\sigma}{dY_{\pi\pi}}$ (right panel) for the LHC energy 
       $\sqrt{s_{NN}} =$ 3.5 TeV.}
\end{figure}

Let us start now presentation of theoretical differential distributions
which can be in principle measured.
In the left panel of Fig.\ref{fig:dsig_dy_deta} we show distributions 
in rapidity of individual pions from the pair (solid line) as well as 
distribution in pseudorapidity of the same pion (dashed line). 
Right panel illustrates nuclear cross section as the function of the pion pair rapidities.
Here we have imposed extra cuts on pion transverse momenta:
$p_{t,\pi}>$ 0.2 GeV (solid lines) and $p_{t,\pi}>$ 0.5 GeV (dashed lines).
In addition, we compare the nuclear cross section 
for $\pi^+\pi^-$ (upper curves) and for $\pi^0\pi^0$ (lower curves) production.
These distributions (in $y_\pi$, $\eta_\pi$ and $Y_{\pi\pi}$) look fairly similar.
%
\begin{figure}[!h]             
\centering
\includegraphics[width=8cm]{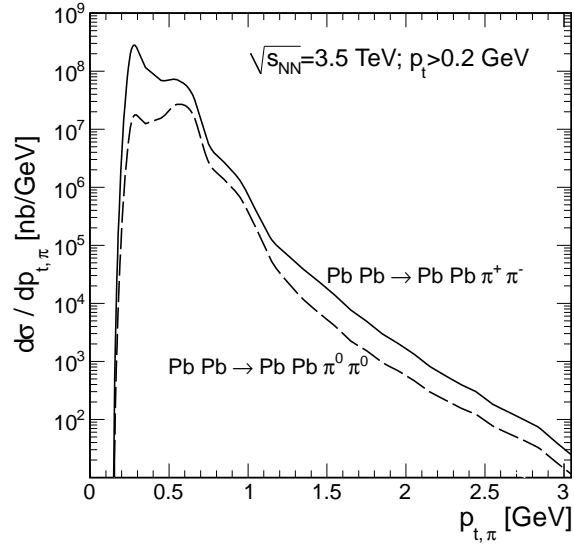}
   \caption{\label{fig:dsig_dpt}
   \small  $\frac{d\sigma}{dp_t}$ of one of pions for the LHC energy 
$\sqrt{s_{NN}} =$ 3.5 TeV and $p_{t,\pi} > $ 0.2 GeV.}
\end{figure}

The distribution in transverse momentum of one of the pions is shown 
in Fig.~\ref{fig:dsig_dpt} for the full phase space for 
$PbPb \to PbPb\pi^+\pi^-$ (solid line) and for
$PbPb \to PbPb\pi^0\pi^0$ (dashed line), respectively.
The distribution shown in Fig.\ref{fig:dsig_dpt} is 
for the full range of (pseudo)rapidities.
The ALICE detector can cover only a part of the whole rapidity range. 
%
\begin{figure}[!h]             
\centering
\includegraphics[width=8cm]{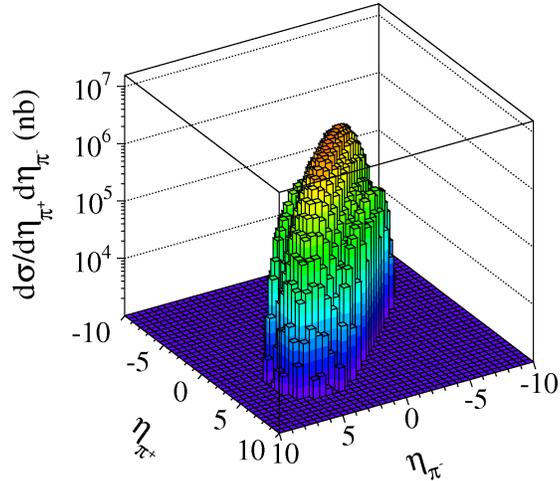}
   \caption{\label{fig:map_eta1eta2}
   \small  $\frac{d\sigma}{d\eta_{\pi^+} d\eta_{\pi^-}}$ for 
   the $\mbox{$^{208}Pb$}+ \mbox{$^{208}Pb$} \to \mbox{$^{208}Pb$} \mbox{$^{208}Pb$} \pi^+ \pi^-$ 
   reaction at the LHC energy $\sqrt{s_{NN}} =$ 3.5 TeV and $p_{t,\pi} > $ 0.2 GeV.}
\end{figure}

In Fig.\ref{fig:map_eta1eta2} we show a two-dimensional distribution
of pseudorapidities of pions. The cross section is concentrated along
the diagonal $\eta_1 = \eta_2$. The range covered by the ALICE detector
(a square (-0.9,0.9)$\times$(-0.9,0.9) centered around $\eta_1$ = $\eta_2$ = 0) is rather small.
The pseudorapidity distribution of one of the pions is shown in 
Fig.\ref{fig:dsig_dy_deta} (left panel) by the dashed line.

\subsection{$\rho^0(770)$ and $\rho^0(1450)$ photoproduction 
in PbPb collisions}

The $\pi^+ \pi^-$ pairs produced in ultraperipheral photon-photon
(sub)collisions constitutes a background for, e.g., exclusive
production of the $\rho^0$ meson. This reaction was studied 
experimentally at RHIC \cite{rho0_RHIC}. Theoretical calculations 
for LHC energies predict very large cross section of the order of a few barns 
(see e.g. \cite{KN99,FSZ,GM11,RSZ2011}), i.e., more than two orders of
magnitude larger than the contribution of the mechanism discussed here
(see also Table \ref{tab:values_rho}).
The $\rho^0$ spectrum is, however, concentrated in pion-pion
invariant mass around the $\rho$-resonance position.
It is an open question how far from the resonance position one 
can trust a rather naive Breit-Wigner form.
In photoproduction of $\rho^0$ meson on protons one often fits 
the following form:
\begin{equation}
\frac{d \sigma (Pb Pb \to Pb Pb \rho^0)}{d M_{\pi \pi}} = 
\left| A_{\rho^0}\frac{\sqrt{M_{\pi \pi} m_{\rho^0} \Gamma(M_{\pi \pi})}}{M_{\pi \pi}^2-m_{\rho^0}^2+i m_{\rho^0} \Gamma(M_{\pi \pi})} + A_{\pi\pi} \right|^2 \; ,
\label{eq:cross_rho}
\end{equation}
where $A_{\rho^0}$ is the amplitude for the Breit--Wigner function, 
$A_{\pi\pi}$ is the amplitude for the direct $\pi^+\pi^-$ production
and $\Gamma(M_{\pi \pi})$ is the momentum-dependent width of the $\rho^0$ resonance.
In our presentation below the values of parameters are taken from Ref.\cite{rho_STAR}
and the integrated total cross section was assumed to be 5 barn
(see Table \ref{tab:values_rho}).
%
\begin{table}[!h]  
\caption{Cross sections for coherent $\rho$ and $\rho'$ production for LHC.}
\begin{center}
\begin{tabular}{|r|r|r|r|}
\hline
	Reference/model  			&  $\sqrt{s_{NN}}$ [TeV]	&  $\sigma_{tot}^{\rho^0(770)}$ [b] &  $\sigma_{tot}^{\rho^0(1450)}$ [b] \\
\hline
    KN \cite{KN99}				& 5.6				&   5.2    							&			\\ \hline
    FSZ \cite{FSZ}          	& 5.5      			&   9.538  							& 	2.216	\\ \hline
    RSZ \cite{RSZ2011}			& 5.5				&   9.706							&			\\ \hline
    IKS \cite{IKS} /KST-R		& 5.5				&  	4.9 							&			\\
                   /KST			&					&	4.36 							&			\\
                   /GBW			&					&	3.99 							&			\\
                   /VDM			&					&	10.03 							&	 		\\ \hline
    GM \cite{GM}             	& 5.5      			&   10.069    						&			\\
\hline
\end{tabular}
\end{center}
\label{tab:values_rho}
\end{table}
%

\begin{figure}[!h]             
\centering
\includegraphics[width=8cm]{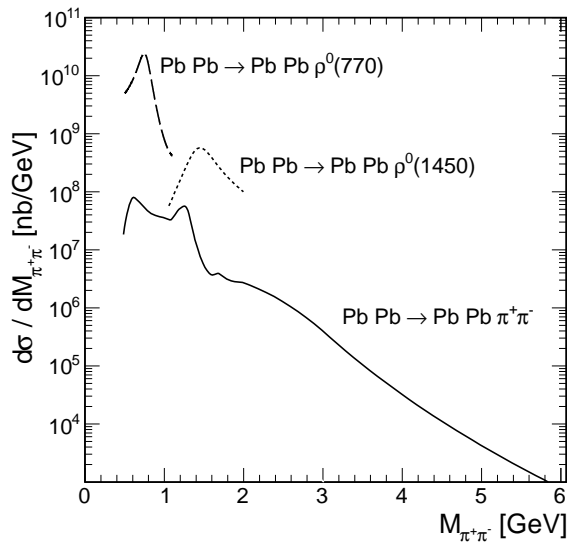}
   \caption{\label{fig:dsig_dW_rho}
   \small  Invariant mass distribution of $\pi^+\pi^-$ from the decay of 
   $\rho^0(770)$ and
   $\rho^0(1450)$ photo-production (resonance contributions represented by 
   the dashed and dotted lines) and our $\gamma\gamma$ fusion (solid line) 
   in ultraperipheral Pb-Pb collisions at $\sqrt{s_{NN}}=$ 3.5 TeV.}
\end{figure}

In Fig.\ref{fig:dsig_dW_rho} we show the contribution of the $\rho^0(770)$
meson integrated over the whole phase space, without any extra cuts.
While the predictions for the $\rho^0$ production are rather uncertain (within factor two or three), 
our predictions for the $\gamma \gamma$ process should be fairly precise
(a few percent precision).
A comparison with real data should be therefore interesting. 
At larger dipion invariant masses, other radial excitation may appear in addition.
The $\rho^0(1450)$ state is a first candidate of this type. Since the continuum 
single-photon photoproduction $\pi^+ \pi^-$ background
in this region was not studied, we have no information about its interference with
the $\rho^0(1450)$ resonance. Therefore in the following we shall show only the resonance
contribution and leave the question of the background for future analyses,
including perhaps the ALICE experimental data. Here we parametrize 
the $\rho^0(1450)$ contribution as:
\begin{equation}
\frac{d \sigma (Pb Pb \to Pb Pb \rho^0(1450))}{d M_{\pi \pi}} = 
\left| A_{\rho^0(1450)}\frac{\sqrt{M_{\pi \pi} m_{\rho^0(1450)} \Gamma(M_{\pi \pi})}}{M_{\pi \pi}^2-m_{\rho^0(1450)}^2+i m_{\rho^0(1450)} \Gamma(M_{\pi \pi})} 
 \right|^2 \; .
\label{eq:cross_rhoprime}
\end{equation}
The cross section for the photoproduction of $\rho^0(1450)$ was calculated
only in Ref.\cite{FSZ}. This calculation finds that the cross
section for the $\rho^0(1450)$ resonance is about a factor 5 smaller than for
$\rho^0(770)$. Here we are discussing the signal in the $\pi^+ \pi^-$ channel,
therefore a corresponding branching fraction has to be included in addition.
Unfortunately this branching fraction is not well known.
The Crystal Ball Collaboration has measured only the ratio of the cross
section for two and four pion channels \cite{Crys_Barrel}.
They have found $Br(2 \pi) / Br(4 \pi)$ = 0.37 $\pm$ 0.1. In this situation we can only
calculate an upper limit for the two-pion branching fraction:
\begin{equation}
Br\left(\rho^0(1450)\to\pi^+\pi^- \right) =
\frac{P_{2\pi}}{P_{2\pi}+P_{4\pi}+P_{other}} 
< \frac{P_{2\pi}}{P_{2\pi}+P_{4\pi}} 
= \frac{P_{2\pi}/P_{4\pi}}{P_{2\pi}/P_{4\pi}+1}\approx 0.27 \; .
\label{bf_etimate}
\end{equation}
%
%
%

In Fig.\ref{fig:dsig_dW_rho} the resonance contribution for the
$\rho^0(1450)$ is 5/0.27 times smaller than for $\rho^0(770)$ as suggested by the
calculation \cite{FSZ} and the above upper limit for the branching
fraction into pions.
It is therefore clear that this will be an upper estimate
for the $\rho^0(1450)$ contribution. The relative contribution of 
$\rho^0(1450)$ is still somewhat larger than observed for instance in the
electroproduction of these states at HERA \cite{ZEUS}.
In Ref.\cite{NNZ94} an anomalous (strongly nucleus mass and photon virtuality
dependent) production of $\rho^0(1450)$ has been predicted for
incoherent production in the $\gamma^* A \to \rho^{0}(1450) X$ reaction. 
We wish to emphasize in addition that our estimate does not take 
into account experimental cuts of any concrete experiment.

The photoproduction contribution cross sections are above our
photon-photon fusion one. It is not clear in the moment how the
kinematical cuts may change the proportions of the two mechanisms.
This will be studied elsewhere.

In contrast, the $Pb Pb \to Pb Pb \pi^0 \pi^0$ reaction is free
of the photoproduction mechanisms. However, it is not clear for us if, in this
case, a measurement is possible.

\section{Conclusions}

In the present paper we have discussed first 
$\gamma \gamma \to \pi^0 \pi^0$ and $\gamma \gamma \to \pi^+ \pi^-$
reactions and we have shown that different mechanisms
contribute to such processes below $\sqrt{s_{\gamma\gamma}} <$ 6 GeV.

These are: pion exchange for the $\pi^+ \pi^-$ channel, 
several resonances such as $f_0(600)$, $f_0(980)$, $f_2(1270)$, 
$f_0(1500)$, $f'_2(1525)$, $f_2(1565)$, $f_0(1710)$, $f_2(1950)$ 
and $f_4(2050)$ and pQCD mechanisms proposed first by Brodsky and Lepage 
as well as hand-bag mechanism proposed by Diehl, Kroll and Vogt. 
We have included also pion-pion rescattering which leads to 
a coupling between the $\pi^+\pi^-$ and $\pi^0\pi^0$ channels.
We have shown that the inclusion of such processes can help in understanding
energy dependence of the $\gamma \gamma \to \pi \pi$ processes
as well as angular distributions only at very low energies. 
We have not tried to get a perfect fit to the data
as several details of the model amplitudes are not known.
We have slightly adjusted some parameters to get a reasonable
description of the data.

We have found that the decay width $\sigma(600) \to \gamma \gamma$ is
much smaller than that found in other partial wave analyses.
The data (total cross section as well as angular distributions) close 
to the position of $f_2(1270)$ peak can be described
assuming the dominance of the amplitude\\
\begin{center}
${\cal M}_{\lambda_1 \lambda_2}^{f_2(1270)} 
\propto Y_{2,\lambda_1 - \lambda_2}(\theta,\phi) \cdot
\left( \delta_{\lambda_1-\lambda_2,-2} + \delta_{\lambda_1-\lambda_2, 2} \right)$ . 
\end{center}
We have included also the tensor $f_2(1565)$ resonance, considering
two models of the corresponding $\gamma \gamma \to \pi \pi$ amplitude.
A better description of the data has been obtained using
\begin{center}
${\cal M}_{\lambda_1 \lambda_2}^{f_2(1565)} 
\propto Y_{2, \lambda_1 - \lambda_2}(\theta,\phi) \cdot
\delta_{\lambda_1 - \lambda_2,0}$ , 
\end{center}
i.e., space, different than for $f_2(1270)$.
This may mean that the structure of this resonance is completely
different than that for $f_2(1270)$, $f'_2(1525)$ or $f_2(1950)$ ($q \bar q$ states).

Inclusion of the $f_4(2050)$ spin-4 resonance improves the spectrum 
and angular distributions around the resonance position. 
By fitting to the Belle Collaboration data we have found
$\Gamma_{f_4(2050) \to \gamma \gamma}$ = 0.7 $\pm$ 0.1 keV.

At still higher energies the situation is even less clear.
While in the $\pi^+ \pi^-$ channel the Brodsky-Lepage mechanism with
distribution amplitude discussed recently in the context of the BABAR data
describes the data, in the ''smaller'' $\pi^0 \pi^0$ channel more processes 
may contributed in addition to the BL pQCD mechanism.
Another possibility is the hand-bag mechanism proposed 
some time ago by Diehl, Kroll and Vogt. 
The hand-bag mechanism may be important only at energies $W >$ 3 GeV.

At high energies the angular distributions show an enhancement at large 
$|\cos\theta|$ for the $\gamma \gamma \to \pi^0 \pi^0$. Such 
an enhancement is predicted both, by the Brodsky-Lepage 
and by the hand-bag approaches. A proper mixture of both processes
provides a better description of the experimental data than 
each of them separately, as done usually in the literature.

Having fixed the details of the amplitude we have calculated
total cross sections and angular distribution as a function 
of $\gamma \gamma$ energy for both 
$\gamma \gamma \to \pi^+ \pi^-$ and $\gamma \gamma \to \pi^0 \pi^0$
processes. These two subprocess energy-dependent 
cross sections have been used next in Equivalent Photon 
Approximation in the impact parameter space to calculate 
for the first time corresponding production rate in ultraperipheral
ultrarelativistic heavy ion reactions. In this calculation we have taken into account
realistic charge distributions in colliding nuclei.

We have calculated both total cross sections at LHC energy, 
as well as distributions in rapidity and transverse momentum of pions
and dipion invariant mass. The calculation of distributions of 
individual pions is slightly more complicated in the b-space EPA. 
The distributions in dipion invariant mass have been compared
with the contribution of exclusive $\rho^0 \to \pi^+ \pi^-$ production
in photon-pomeron (pomeron-photon) mechanism taken from the literature.
Close to the $\rho^0$ resonance the 
$\gamma \gamma \to \pi^+ \pi^-$ mechanism yields only a small
contribution. The $\gamma \gamma$ contribution could be, perhaps, measured
outside of the $\rho^0$ resonance window. A detailed comparison with the
absolutely normalized ALICE experimental data should allow a quantitative
test of our predictions. Imposing several experimental cuts may 
enhance the $\gamma\gamma\to\pi\pi$ contribution.

In the present paper we have analysed exclusive production of two pions.
In principle exclusive production of four pions could be also interesting.
First experimental study has been presented in \cite{rho0_RHIC}.
Interesting theoretical study has been performed in \cite{PSSW2008}.

\appendix
\section{A comment on partial wave decomposition}

In our model, the amplitude for the $\gamma \gamma \to \pi \pi$ process
is a coherent sum of different contributions (amplitudes) for 
different mechanisms:
\begin{equation}
{\cal M}_{\lambda_1,\lambda_2}(W) = 
\sum_k {\cal M}^{(k)}_{\lambda_1,\lambda_2}(W) \; ,
\label{summing_amplitudes}
\end{equation}
where $\lambda_1$, $\lambda_2$ are initial photon helicities.
There are only four combinations of helicities defining the, in general,
complex amplitude: (-1,1),(1,-1),(1,1),(-1,-1).

As a consequence the differential distributions for the 
$\gamma \gamma \to \pi \pi$ reaction can be written in 
terms of spherical harmonics numbered with pion-pion angular momentum 
projection ($M=\lambda_1-\lambda_2$)
\begin{eqnarray}
\frac{d\sigma}{d \Omega} &=& 
| C_{0,0}(W) Y_{0,0}(z,\phi) + C_{2,0}(W) Y_{2,0}(z,\phi) + 
  C_{4,0}(W) Y_{4,0}(z,\phi) + (...) |^2 
\nonumber \\
&+& | C_{2,2}(W) Y_{2,2}(z,\phi) + C_{4,2}(W) Y_{4,2}(z,\phi) + (...) |^2
\nonumber \\
&+& | C_{2,-2}(W) Y_{2,-2}(z,\phi) + C_{4,-2}(W) Y_{4,-2}(z,\phi) + (...) |^2
\; .
\label{partial_wave_expansion}
\end{eqnarray}
The expansion coefficients $C_{L,M}$ are strongly energy dependent.
In general, resonance contributions (L = 0, 2, 4) interfere with 
slowly-energy-dependent backgrounds (L = 0, 2, 4,...) of different origin 
discussed in previous sections. 

Recently the Belle Collaboration \cite{Belle_Uehara09, Belle_Uehara08} has 
performed a simplified fit to the experimental angular distributions of 
the type:
\begin{eqnarray}
\frac{d \sigma}{d \Omega} &=&
| C_{0,0}(W) Y_{0,0}(z) |^2 + | C_{2,0}(W) Y_{2,0}(z) |^2 + | C_{4,0}
  Y_{4,0}(z)|^2 
\nonumber \\
&+& | C_{2,2}(W) Y_{2,2}(z) |^2 + | C_{4,2}(W) Y_{4,2}(z) |^2 \; .
\label{Belle_expansion}
\end{eqnarray}
Our model shows that such a simplified partial wave expansion 
has not always model justification and may therefore 
lead sometimes to unphysical solutions.
However, the correct partial wave decomposition 
(see (Eq.\ref{partial_wave_expansion})) of experimental
angular distributions is in practice too difficult, if not impossible.

\vspace{1cm}

{\bf Acknoweledgments}

We are indebted to Nikolai Achasov and Wolfgang Sch\"afer for
interesting exchange of informations and Christoph Mayer for a discussion of
ALICE experiment and exchange of a few experimental details
as well as careful reading of our manuscript.
This work was partially supported by the 
Polish grant N~N202 078735, N N202 236640 and DEC-2011/01/B/ST2/04535.
A big part of the calculations within this analysis was carried out 
with the help of cloud computer system (Cracow Cloud One\footnote{cc1.ifj.edu.pl}) 
of Institute of Nuclear Physics (PAN).



\end{document}